\documentclass[12pt]{article}

\usepackage{amsmath,amssymb}
\usepackage{graphics}
\usepackage{caption}
\usepackage{subcaption}
\usepackage{latexsym,ifthen}
\usepackage{booktabs}
\usepackage{epsfig,here}
\usepackage{authblk}

\newcommand{\Section}[1]{\section{#1}\setcounter{figure}{0}\setcounter{table}{0}\setcounter{equation}{0}}

\newtheorem{theo}{Theorem}[section]

\newtheorem{rem}{Remark}[section]

\title{A numerical method for the quasi-incompressible Cahn-Hilliard-Navier-Stokes equations for variable density flows with a discrete energy law}
\date{}
\author[1,2]{Z. Guo}\author[1]{P. Lin
}\author[3]{J. S. Lowengrub}
\affil[1]{Department of Mathematics, University of Dundee, Dundee, DD1
4HN, Scotland, United Kingdom}
\affil[2]{Department of Applied Mathematics and Mechanics, University of Science and Technology Beijing, Beijing, 100083, China}
\affil[3]{Department of Mathematics, University of California, Irvine, CA 92697, USA}

\begin{document}
\maketitle
\begin{abstract}
In this paper, we investigate numerically a diffuse interface model for the Navier-Stokes equation with fluid-fluid interface when the fluids have different densities \cite{Lowengrub1998}. Under minor reformulation of the system, we show that there is a continuous energy law underlying the system, assuming that all variables have reasonable regularities. It is shown in the literature that an energy law preserving method will perform better for multiphase problems. Thus for the reformulated system, we design a $C^0$ finite element method and a special temporal scheme where the energy law is preserved at the discrete level. Such a discrete energy law (almost the same as the continuous energy law) for this variable density two-phase flow model has never been established before with $C^0$ finite element. A Newton's method is introduced to linearise the highly non-linear system of our discretization scheme. Some numerical experiments are carried out using the adaptive mesh to investigate the scenario of coalescing and rising drops with differing density ratio. The snapshots for the evolution of the interface together with the adaptive mesh at different times are presented to show that the evolution, including the break-up/pinch-off of the drop, can be handled smoothly by our numerical scheme.  The discrete energy functional for the system is examined to show that the energy law at the discrete level is preserved by our scheme.
\end{abstract}
\section{Introduction}
Multiphase flows play an increasingly important role in many current scientific and engineering applications. In recent years, there have been both extensive theoretical and experimental studies, yet the area remains an active interdisciplinary research field. Improved numerical algorithms have resulted in direct numerical simulations of multiphase phenomena leading to improvement in predicting the behaviour of multiphase flows. The available numerical methods for solving multiphase problems can roughly be divided into two categories: interface tracking and interface capturing methods. In interface tracking methods, the position of the interface is explicitly tracked. It requires meshes that track the interfaces and are updated as the flow evolves. Boundary integral methods (see the review \cite{Hou2001}), front-tracking methods (see the review \cite{Tryggvason2001} and \cite{Glimm1998,Hua2008,Unverdi1992} for examples) and immersed boundary methods (see the review \cite{Mittal2005} and \cite{Li2003, Peskin2002} for examples) are examples of this type. In interface capturing methods, on the other hand, the interface is not tracked explicitly, but instead is implicitly defined through an interface function (e.g. level-set, color or phase-field function). Typically these methods are applied on fixed spatial domains, which eliminates the problem of updating the meshes encountered in interface tracking methods. Volume-of-fluid (VOF) methods (see the review \cite{Scardovelli1999} and \cite{Brackbill1992,Hirt1981} for examples), level-set methods (\cite{Osher2001, Osher1988, Sussman1994}) and phase-field methods (see the review\cite{Anderson1998,Emmerich2008}) are examples of interface capturing algorithms.
\\\\
Phase-field methods, or diffuse interface methods, have now emerged as a powerful method to simulate many types of the multiphase flows, including drop coalescence and break-up (\cite{Baldalassi2004,Boyer2002,Ding2007,Hua2011,Jacqmin1999,Kim2005,Lee2001,Lee2002,Yue2004}), and dynamics of interfaces with Marangoni effects ({\cite{Borcia2003,Guo2013,Sun2009,Verschueren2001, Teigena2011,Sman2006}), where the surface tension gradient is induced by the inhomogeneous distribution of the temperature or surfactants that can be absorbed at the liquid/gas or liquid/liquid interfaces. The basic idea for phase-field methods is to employ a conserved order parameter, the so-called phase variable, which varies continuously over thin interfacial layers and is mostly uniform in the bulk phases. Sharp interfaces are then replaced by thin but non-zero thickness transition regions where the interfacial forces are smoothly distributed. The corresponding evolution equations can be solved over the entire computational domain without any \emph{a} \emph{priori} knowledge of the location of the interfaces. When the width of the transition region approaches zero, the appropriately scaled phase-field model asymptotically reduces to the classical sharp-interface model under certain conditions, in which the jump conditions at the sharp interface can be recovered (\cite{Abels2012,Anderson1998,Anderson2000,LiuShen2002,Lowengrub1998}).
\\\\
The classical phase-field model, in the case of two incompressible, viscous Newtonian fluids, is the so-called Model H {\cite{Hohenberg1977}}, which couples fluid flow with Cahn-Hilliard diffusion for a conserved parameter. Model H has been successfully used to simulate complicated mixing flows involving incompressible components with matched densities {\cite{Chella1996}}. Gurtin $et$ $al.$ \cite{Gurtin1996} showed thermodynamic consistency of model H. This model, however, cannot be used if the two incompressible fluid components have different densities. To treat problems with small density ratios, a Boussinesq approximation is often used, where the small density difference is neglected except that in the gravitational force. The corresponding model remains thermodynamically consistent. This approach however is no longer valid for large density ratios. Several generalizations of Model H for the case of different densities were presented and discussed by Lowengrub and Truskinovsky \cite{Lowengrub1998}, Boyer \cite{Boyer2002}, Ding $et$ $al.$ \cite{Ding2007}, Shen and Yang \cite{shen2010}, and Abels et al. \cite{Abels2012}, in which, the models in \cite{Abels2012, shen2010} consider phenomenological modifications of the momentum equation, Further, the model in \cite{shen2010}
 is not frame-indifferent to rotations in the coordinate system. In \cite{Lowengrub1998}, a thermodynamically consistent extension of Model H for the case of different densities is derived. The sharp interface is replaced by a narrow transition layer, across which the two incompressible components of different densities can mix and be compressible, such fluids are termed as quasi-incompressible. The appropriate thermodynamic description is in terms of a Gibbs free energy, and the pressure instead of the density is employed as an independent variable. The resulting system couples the Navier-Stokes equations and the Cahn-Hilliard equations for variable density flows. It is thermodynamically and mechanically consistent, so that the fast changes of the diffuse interface can occur smoothly, and there is an energy law that is naturally associated with the model. (See $\S$ 2 and $\S$ 3). The NSCH system was studied theoretically by Abels \cite{Abels2009} who proved the existence of weak solutions for the non-stationary NSCH system. Very recently, another model of quasi-incompressible fluids for the phase transition simulation was developed by Giesselmann and Pryer \cite{Giesselmann2013}, where a discontinuous Galerkin finite element method is used and studied in \cite{Aki2014}. The model considered differs from the quasi-incompressible NSCH system developed in \cite{Lowengrub1998} in that the volume fraction, rather than the mass concentration, is used as the phase variable. In addition, the two models are derived with different energy functional.\\\\
Solving the quasi-incompressible NSCH system is more challenging than solving model H numerically because the velocity field is no longer divergence-free and because of the strong coupling between the Navier-Stokes and the Cahn-Hilliard equations. A simplified version of this quasi-incompressible Navier-Stokes-Cahn-Hilliard (NSCH) system has been used for numerical studies \cite{Kim2004, Kim2005}. However, there are, to the best of the authors’ knowledge, no discrete schemes available which are based on the full model. 
\\\\
Several numerical methods have been used to solve the phase-field model including spectral methods (\cite{Dong2012,Lee2002, LiuShen2002,shen2010,Yang2006}), adaptive moving mesh methods (\cite{Beckett2006, Zhang2010,Zhang2007}), finite element methods and adaptive finite element methods (\cite{Chen2000,Du2008,Feng2006,Giesselmann2013,Hua2011,Yue2006,Zhang2009}) and finite difference methods (\cite{Kim2004, Kim2005}), and we refer to \cite{Feng2014} and references therein for some recent discussions on adaptive methods for the long time simulation of the Cahn-Hilliard equation. In particular, the phase-field model can be derived from an energy variational approach. With thermodynamic consistency, an energy law is naturally associated with the well-posed nonlinearly coupled system. Note that the energy law can be preserved in the fully discretized system. It is highly desirable to have a numerical scheme that preserves the accurate energy law at the discrete level, in order to ensure the stability of the numerical algorithm and the accuracy of the solution, especially when a rapid change or a singularity occurs in the solution, such as occurs in non-Newtonian hydrodynamic systems (\cite{Lin2006,Lin2007}). When the underlining energy law is preserved in the fully discretized system, it is also possible to use a relatively coarse mesh in the simulation. Consequently such a method can reduce the cost of computations while resolving the prominent features of the flow.  Recently, Hua $et$ $al.$ \cite{Hua2011} introduced a $C^0$ finite elements formulation for a phase-field model in which the energy law of the system was well preserved at the discrete level. They obtained more accurate results for interface motion on coarser grids than another scheme that does not obey the energy law at the discrete level.\\\\
In this paper, a numerical method for the quasi-incompressible NSCH system with variable density, which respects the energy law is developed in this paper. A $C^0$ finite element formulation is introduced, under which the energy law of the system can be preserved at the discrete level. We emphasise $C^{0}$ finite elements because they are much simpler than finite elements with a higher degree of smoothness and are available in most existing finite element software packages. For all the examples computed in this paper, we assume the interface have no contact with the boundary of the domain. In the case that the interface contacts with the boundary of the domain, extra difficulties would arise from complicated interface/boundary interacting conditions and should be dealt with separately. Here we refer to \cite{Aland2010,XiaoPing2012,Qian2006,Teigen2011} and the discussions therein.\\\\
The rest of the paper is organised as follows. $\S$2 presents the established model for quasi-incompressible NSCH system with variable density. In $\S$3 we present a standard weak form of the system and formally rederive the corresponding continuous energy law. In $\S$4, we reformulate the equations of the system and present the temporal scheme with $C^0$ finite elements. We show that the energy law at the discrete level can be derived under our discretization. In $\S$5, we briefly introduce the Newton's method, which is employed to linearise the corresponding highly non-linear system. Several numerical experiments are carried out in $\S$6. $\S$ 7 presents conclusions and a discussion of future work.
\section{Quasi-Incompressible NSCH System}
We consider the following non-dimensional Navier-Stokes Cahn-Hilliard (NSCH) system with variable density governing the motion of a two-phase fluid:
\begin{align}
&\nabla \cdot \bold{u}=\cfrac{\alpha}{Pe}\Delta \mu,~~~~\operatorname{or}~~~~\dot{\rho}=-\rho\nabla \cdot \bold{u},\tag{2.1a,~2.1b}\label{qua1}\\
&\rho\dot{\bold{u}}=-\cfrac{1}{M}\big(\nabla p + C\nabla \cdot (\rho \nabla c \otimes \nabla c )\big)+\nabla \cdot (\cfrac{1}{Re} \nabla \bold{u})+\nabla\big(\cfrac{1}{3Re}(\nabla \cdot \bold{u}) \big)-\cfrac{1}{Fr^2}(\rho-\rho_0)\hat{\bold{j}},\tag{2.2}\label{qua2}\\
&\rho\dot{c}=\nabla \cdot (\cfrac{1}{Pe}\nabla \mu),\tag{2.3}\label{qua3}\\
&\mu= f(c)-\cfrac{\partial \rho}{\partial c}\cfrac{ p}{\rho^2} -\cfrac{C}{\rho}\nabla \cdot (\rho\nabla c ) -\cfrac{M\rho_0 \alpha}{Fr^2} y,  \tag{2.4}\label{qua4}
\end{align}
where
\begin{center}
\begin{tabular}{lll}
\toprule
$\dot{x}=x_t+(\bold{u}\cdot \nabla) x$ \hspace{5mm} & the material derivative,\\
$\bold{u}$ \hspace{5mm} & the velocity of mixtures,\\
$p$ \hspace{5mm} & the pressure,\\
$c$  \hspace{5mm} &  phase variable ($c=1$: fluid 1; $c=0$: fluid 2),\\
$\rho=\rho(c)$ \hspace{5mm} & the variable density of the mixture, a function of $c$,\\
$\mu$ \hspace{5mm} & the chemical potential,\\
$\rho_0$\hspace{5mm}& a constant representing the "background" or reference density,\\
$\hat{\bold{j}}$\hspace{5mm}& the vertical component of the unit vector in \\
&Cartesian coordinate system,\\ 
$y$ \hspace{5mm}&the vertical coordinate,\\
$F(c)=c^2(c-1)^2/4$ \hspace{5mm}&double-well potential (free energy),\\
$f(c)$ \hspace{5mm} & the derivative of $F$,\\
$\nabla c\otimes\nabla c$ \hspace{5mm} & extra reactive stress,\\
$C$\hspace{5mm} & the capillary number measuring the thickness of the interface,\\ 
$M$\hspace{5mm} &  an analogue of the Mach number measuring the relative strength\\
 &of the surface tension and chemical energies,\\
 $Re$ \hspace{5mm} & the Reynolds number,\\
 $Fr$\hspace{5mm} & the Froude number measuring the relative strengths of \\
 &inertial and gravitational forces,\\
$Pe$\hspace{5mm} & the diffusional Peclet number measuring the relative strengths\\
&of (chemical) diffusion and advection.\\
\bottomrule
\end{tabular}
\end{center}
Note that, as the gravitational effects are considered here, the system (2.1)-(\ref{qua4}) we investigate in this paper is modified from the quasi-incompressible NSCH system in {\cite{Lowengrub1998}}. Comparing to the system of equations in {\cite{Lowengrub1998}}, a gravitational force $-(\rho-\rho_0)\hat{\bold{j}}/Fr^2$ is added to our momentum equation (\ref{qua2}). Further, in order to maintain the thermodynamical consistency of our system, and an extra term $-\rho_0 g\alpha y/Fr^2$ corresponding to the gravitational forces is added to the chemical potential equation (\ref{qua4}) (See Remark 2.1 for detail).\\
Following Lowengrub and Truskinovsky \cite{Lowengrub1998}, the harmonic interpolation for the variable density model of the mixture of two incompressible fluids is adopted,
\begin{align}
\cfrac{1}{\rho(c)}=\cfrac{1}{\rho_1}c+\cfrac{1}{\rho_2}(1-c),\notag
\end{align}
where $c$ is the mass concentration (phase variable), and $\rho_1$, $\rho_2$ are constant representing the density of the two incompressible fluids respectively. Further we have
\begin{align}
\cfrac{\partial \rho}{\partial c}=-\alpha\rho^2,\tag{2.5}\label{partialho}
\end{align}
where $\alpha=(\rho_2-\rho_1)/\rho_2\rho_1$ is a constant. Note that the above variable density model is different from the more traditional models with linear interpolation and assume volume are preserved under mixing, see \cite{Aktin1976} and \cite{Bedford1983} for comprehensive reviews. Moreover, by considering Eq.(\ref{qua3}), we obtain
\begin{align}
\nabla \cdot \bold{u}=-\cfrac{1}{\rho} \cfrac{\partial \rho}{\partial c} \dot{c}=\cfrac{\alpha}{Pe}\Delta \mu. \tag{2.6}\label{der--chemi}
\end{align}
It can be observed that the two continuous equations in Eq.(2.1) are equivalent
\begin{align}
\dot{\rho}&=-\rho\nabla \cdot \bold{u}~~~~\Longleftrightarrow~~~~\nabla \cdot \bold{u}=\cfrac{\alpha}{Pe}\Delta \mu. \tag{2.7}\label{quasi--divv}
\end{align}
It is well known that for the incompressible flows, the velocity field satisfies the divergence free condition, 
\begin{align}
\nabla \cdot \bold{u}=0.\tag{2.8}\label{cond--divfree}
\end{align}
In this variable density model, however, this condition is no longer valid. Due to the variation of $c$, a mixture of two incompressible fluids with different densities can be compressible across the interface. In \cite{Lowengrub1998}, such binary fluids satisfying Eq.(2.1) are termed as quasi-incompressible. Note that the energy law preserving numerical method is designed in this paper corresponding to either Eq.(2.1a) or Eq.(2.1b), where in practice the method with Eq.(2.1a) is chosen for our computations. By presenting the distribution of $\nabla \cdot \bold{u}$, the compressibility of quasi-incompressible flows along the interface is demonstrated in $\S 6$ (Fig. {\ref{fig--rising1to2--divv}} and Fig. {\ref{fig--rising1to50--divv}}).\\
The system is supplied with the initial conditions:
\begin{align}
\bold{u}|_{t=0}=\bold{u}_0,~~c|_{t=0}=c_0,\notag
\end{align}
the Dirichlet boundary conditions for the velocity: $\bold{u}(=\bold{b_u})$ and the Neumann boundary conditions for the phase field $c$ and $\mu$: $\nabla c \cdot \hat{\bold{n}} =0$ and $\nabla \mu \cdot \hat{\bold{n}} =0$, where $\hat{\bold{n}}$ is the unit outward normal vector of the boundary.
\begin{rem}
Based on entropy production, the quasi-incompressible NSCH system {\cite{Lowengrub1998}} was derived through an energetic variational procedure. It is thermodynamically consistent with an energy functional $E$,
\begin{align}
&E=\int_{\Omega}\bigg(\cfrac{1}{2}\rho|\bold{u}|^2+\cfrac{1}{M}\rho F(c)+ \cfrac{C}{2M}\rho|\nabla c|^2  \bigg)dx,\tag{2.9}\label{original-energy}
\end{align}
where $\rho|\bold{u}|^2/2$ is the kinetic energy, and $\rho F(c)/M+C\rho |\nabla c|^2/2M$ is the Cahn-Hilliard free energy. Note that the potential energy is not considered in the original system.\\
In this paper, as the effects of gravity are considered, we consider the modified system (2.1)-(\ref{qua4}), where the corresponding non-dimensional total energy $\hat{E}$ for the system (2.1)-(\ref{qua4}) now can be defined as
\begin{align}
\widehat{E}&=E+E_g,\notag\\
&=\int_{\Omega}\bigg(\cfrac{1}{2}\rho|\bold{u}|^2+\cfrac{1}{M}\rho F(c)+ \cfrac{C}{2M}\rho|\nabla c|^2 +\cfrac{1}{Fr^2}\rho  y \bigg)dx\tag{2.10}\label{total--en}
\end{align}
in which $E$ is from the original system (Eq.(\ref{original-energy})) and $E_g$ is the gravitational potential energy,
\begin{align}
E_g=\int_{\Omega}\bigg(\cfrac{1}{Fr^2}\rho  y\bigg)dx,\tag{2.11}
\end{align}
where $\Omega$ is a bounded domain and $y$ is the coordinate in vertical direction.\\
We now rederive the system of equations with respect to the total energy $\hat{E}$ (Eq.(\ref{total--en})). Differentiating $E_g$ with respect to time (associated with the the variational procedure) we obtain
\begin{align}
\cfrac{d E_g}{dt} =\int_{\Omega}\bigg(-\cfrac{1}{Fr^2}\nabla \cdot (\rho  \bold{u})y \bigg)dx,\tag{2.12}\label{der--potential--1}
\end{align}
where Eq.(2.1b) is used. Note that, for the sake of convenience, a reference density $\rho_0$ is subtracted off from the gravitational forces in the NS equation (\ref{qua2}). In order to derive the gravitational force $-(\rho-\rho_0)\hat{\bold{j}}/Fr^2$ within a thermodynamically consistent framework, we multiply Eq.(2.1a) by $-\rho_0 y/ Fr^2$ to obtain
\begin{align}
-\int_{\Omega}\bigg(\cfrac{1}{Fr^2}\rho_0  y \nabla \cdot \bold{u}-\cfrac{1}{Fr^2}\rho_0  y \alpha \nabla \cdot (\cfrac{1}{Pe}\nabla \mu)\bigg)dx=0.\tag{2.13}\label{der--potential--2}
\end{align}
Adding Eq.(\ref{der--potential--2}) to Eq.(\ref{der--potential--1}) and using integration by parts, we obtain
\begin{align}
\cfrac{d E_g}{dt}&=-\int_{\partial \Omega}\bigg(\cfrac{1}{Fr^2}(\rho-\rho_0)  y \bold{u} -\cfrac{\alpha  y \rho_0}{Fr^2Pe}\nabla \mu\bigg) \cdot \hat{\bold{n}} ~dx\notag\\
&\hspace{10mm}+\int_{\Omega}\bigg(\cfrac{1}{Fr^2}(\rho-\rho_0)  \bold{u}-\cfrac{\rho_0 \alpha}{Fr^2Pe}\nabla \mu\bigg)\cdot \hat{\bold{j}}~dx,
\tag{2.14}\end{align}
 where $\hat{\bold{n}}$ is the unit outward normal vector of the boundary $\partial{\Omega}$ and
\begin{align}
\hat{\bold{j}}=\nabla y=\tbinom{0}{1}\tag{2.15}\label{jj}
\end{align} 
is the vertical component of the unit vector in Cartesian coordinate system. With the homogeneous boundary conditions that the interface is assumed to have no intersections with the boundary $\partial \Omega$, we arrive at
\begin{align}
\cfrac{d E_g}{dt}=\int_{\Omega}\bigg(\cfrac{1}{Fr^2}(\rho-\rho_0)   \hat{\bold{j}}\cdot\bold{u}-\cfrac{1}{MPe}\nabla \mu\cdot\nabla (\cfrac{M\rho_0  \alpha  y}{Fr^2} )\bigg)dx,\tag{2.16}\label{potentialrelation}
\end{align}
where, we notice that, with the reference density, not only the gravity but also the chemical potential contribute to the time rate of the change of the potential energy. Based on the entropy production and the variational procedure that used in \cite{Lowengrub1998}, the first term at the right hand side will result in the gravitational forces $-(\rho-\rho_0)\hat{\bold{j}}/Fr^2$ in the NS equation (\ref{qua2}), acting as the external forces of our the system, and the second term at the right hand side will contribute to the entropy production and will result in $-M\rho_0 \alpha y/Fr^2$ in the chemical potential (\ref{qua4}) of our system. We thus obtain a new chemical potential (\ref{qua4}), which differs in the original model derived in \cite{Lowengrub1998} that an extra term $-M\rho_0 \alpha y/Fr^2 $ appears to account for the potential energy associated with the gravitational force. In what follows ($\S 3$), we will derive the energy law with respect to the total energy $\widehat{E}$ (Eq.(\ref{total--en})) to show that the thermodynamic consistency is maintained in our new system. 
\end{rem}\label{remm}
\Section{Energy Law Preserving Weak Form}
For simplicity, we only consider two-dimensional case in this paper. The results can be straightforwardly extended to three dimensions. Let $\Omega$ be a bounded domain. We denote the boundary of $\Omega$ by $\partial\Omega$ and suppose that $\partial\Omega$ is sufficiently smooth (for example, Lipschitz-continuous). Denote the following spaces as $\bold{W}^{1,3}(\Omega)=(W^{1,3}(\Omega))^2$, $\bold{W}^{1,3}_{\bold{b}}(\Omega)=\{\bold{u} \in \bold{W}^{1,3}(\Omega), \bold{u}=\bold{b}$ on $\partial\Omega \}$. Note that in order to obtain a meaningful weak form, we require $\rho$ is positive and $\rho \in L^{ \infty } (\Omega)$, and $\rho\bold{u} \in \bold{W}^{1,3}(\Omega)$. A direct variational or weak form may be derived straightforwardly by multiplying Eq.(2.1a) with a test function $q$, Eq.(\ref{qua2}) with $\bold{v}$, Eq.(\ref{qua3}) with $\psi$ 
and Eq.(\ref{qua4}) with $\chi$. Using integration by parts, we obtain the weak
form as the following: Find $\bold{u} \in \bold{W}^{1,3}_{\bold{b}}(\Omega)$, $p \in W^{1,3/2}(\Omega)$, $c \in W^{1,3}(\Omega)$, $\mu \in W^{1,3}(\Omega)$ such that
\begin{align}
&\int_{\Omega}\bigg(\bold{u}\cdot\nabla  q-\cfrac{\alpha}{Pe} \nabla\mu \cdot\nabla q\bigg)dx=0, \hspace{48mm} \forall q \in W^{1,3/2}(\Omega),\label{wqua1}\\
&\int_{\Omega}\bigg(\rho \dot{\bold{u}}\cdot \bold{v}+\cfrac{1}{M} \bold{v}\cdot \nabla p  - \cfrac{C}{M} (\rho \nabla c \otimes \nabla c ) : \nabla \bold{v}+\cfrac{1}{Re} \nabla \bold{u} : \nabla\bold{v}+\cfrac{1}{3Re} (\nabla \cdot \bold{u} ) (\nabla \cdot \bold{v} )\notag\\
&\hspace{8mm}+\cfrac{1}{Fr^2}(\rho-\rho_0)\hat{\bold{j}}\cdot \bold{v}\bigg)dx=0,\hspace{54mm}\forall \bold{v} \in \bold{W}^{1,3}_{0}(\Omega),\label{wqua2}\\
&\int_{\Omega}\bigg(\rho\dot{c}\psi+\cfrac{1}{Pe}\nabla \mu \cdot \nabla\psi \bigg)dx=0,\hspace{51mm}\forall \psi \in W^{1,3/2}(\Omega), \label{wqua3}\\
&\int_{\Omega}\bigg(\mu\chi- f(c)\chi  +\cfrac{\partial \rho}{\partial c} \cfrac{p}{\rho^2} \chi+\cfrac{C}{\rho} (\nabla \rho \cdot \nabla c ) \chi-C\nabla c \cdot \nabla \chi+\cfrac{M\rho_0  \alpha}{Fr^2} y \chi\bigg)dx=0,\notag\\
&\hspace{114mm}\forall \chi \in W^{1,3}(\Omega).\label{wqua4}
\end{align}
At most, only first order derivatives of $\bold{u}$, $p$, $c$, $\mu$, $\bold{v}$, $q$, $\psi$ and $\chi$ are required so that the $C^0$ (conforming) finite element method can be used to solve the problem in this variational form. In the case of homogeneous boundary conditions, there is an energy law that underlies the system \cite{Lowengrub1998}. We now formally rederive the continuous energy law using the weak form (\ref{wqua1})-(\ref{wqua4}) (with homogeneous boundary conditions for $\bold{u}$), which is useful in deriving the discrete energy law preserving numerical scheme later.\\
For the momentum equation (\ref{wqua2}), by setting $\bold{v}=\bold{u}$ and using integration by parts, we obtain
\begin{align}
\int_{\Omega}\bigg(\rho \cfrac{d}{dt}\cfrac{1}{2}(\bold{u}\cdot \bold{u})-\nabla \cdot (\rho\bold{u})\cfrac{1}{2}(\bold{u}\cdot \bold{u})\bigg)dx&=\int_{\Omega}\bigg(\cfrac{1}{M} p (\nabla \cdot \bold{u}) + \cfrac{C}{M} (\rho \nabla c \otimes \nabla c ) : \nabla \bold{u}\notag\\
&\hspace{-30mm }-\cfrac{1}{Fr^2}(\rho-\rho_0)\hat{\bold{j}}\cdot \bold{u}\bigg)dx-\cfrac{1}{Re} ||\nabla \bold{u}||^2_{L^2} -\cfrac{1}{3Re}||\nabla \cdot \bold{u}||^2_{L^2},\label{enlawfv}
\end{align}
where we used the following identity under homogeneous boundary conditions
\begin{align}
\int_{\Omega}\bigg(\rho(\bold{u}\cdot \nabla )\bold{u}\cdot \bold {u}\bigg)dx=\int_{\Omega}\bigg(-\nabla \cdot (\rho\bold{u})\cfrac{1}{2}(\bold{u}\cdot \bold{u})\bigg)dx\hspace{5mm}\operatorname{for}~ \bold{u}|_{\partial \Omega}=0.\notag
\end{align}
For the Cahn-Hilliard equation (\ref{wqua3}), we set $\psi= \mu/M-\alpha (\rho-\rho_0) y/Fr^2 -\alpha \rho \bold{u}\cdot \bold{u}/2 - \alpha \rho F(c)/M-\alpha \rho C (\nabla c\cdot \nabla c )/2M -\alpha p/M$ to obtain
\begin{align}
&\int_{\Omega}\bigg(\cfrac{1}{M}\rho\dot{c}\mu+\cfrac{\rho_0  \alpha}{Fr^2} y\rho\dot{c}+\cfrac{1}{Fr^2} y\dot{\rho}+\cfrac{1}{2}\bold{u}\cdot \bold{u} \dot{\rho} +\cfrac{1}{M}F(c)\dot{\rho} +\cfrac{C}{2M}\nabla c\cdot \nabla c \dot{\rho}+\cfrac{1}{M}\cfrac{p}{\rho}\dot{\rho}\bigg)dx\notag\\
&=\int_{\Omega}\cfrac{\alpha}{Pe}\nabla \mu \cdot \nabla\bigg(\cfrac{\rho-\rho_0}{Fr^2} y+\cfrac{1}{2}\rho\bold{u}\cdot \bold{u}  +\cfrac{1}{M}\rho F(c) +\cfrac{C}{2M}\rho\nabla c\cdot \nabla c+\cfrac{p}{M} \bigg)dx-\cfrac{1}{M} \cfrac{1}{Pe}||\nabla \mu||^2_{L^2}.\label{enlawfc}
\end{align}
For the chemical potential equation (\ref{wqua4}), we set $\chi=\rho\dot{c}/M$ to obtain
\begin{align}
&\int_{\Omega}\bigg(\cfrac{1}{M} \rho \cfrac{d}{dt}F(c)-\cfrac{1}{M}\nabla \cdot  (\rho\bold{u})F(c)+\cfrac{C}{2M}\rho \cfrac{d}{dt}(\nabla c \cdot \nabla c)-\cfrac{C}{2M}\nabla \cdot (\rho \bold{u} )(\nabla c \cdot \nabla c )\bigg)dx\notag\\
&\hspace{28mm}=\int_{\Omega}\bigg( \cfrac{1}{M}\cfrac{p}{\rho}\dot{\rho}-\cfrac{C}{M}(\rho \nabla c \otimes \nabla c) : \nabla \bold{u}+\cfrac{1}{M}\rho\dot{c}\mu+\cfrac{\rho_0\alpha}{Fr^2} y\rho \dot{c}\bigg)dx,\label{enlawfm}
\end{align}
where, for the term $\int_{\Omega}( \rho f(c)\dot{c}/M)dx$, we have
\begin{align}
\int_{\Omega}\bigg(\cfrac{1}{M}\rho f(c)\dot{c}\bigg)dx&=\int_{\Omega}\bigg( \cfrac{1}{M}\rho\cfrac{d}{dt}F(c)+\cfrac{1}{M}\rho (\bold{u}\cdot \nabla ) F(c)\bigg)dx\notag\\
&=\int_{\Omega}\bigg( \cfrac{1}{M}\rho\cfrac{d}{dt}F(c)-\cfrac{1}{M}\nabla \cdot  (\rho\bold{u})F(c)\bigg)dx.
\end{align}
And for the term $\int_{\Omega}(C\rho \nabla c \cdot\nabla \dot{c}/M )dx$, we have
\begin{align}
&\int_{\Omega}\bigg(\cfrac{C}{M}\rho \nabla c \cdot\nabla \dot{c} \bigg)dx\notag\\
&=\int_{\Omega}\bigg( \cfrac{C}{2M}\rho \cfrac{d}{dt}(\nabla c \cdot \nabla c)+\cfrac{C}{M}\rho \nabla c \cdot\nabla\big( (\bold{u}\cdot\nabla)c\big) \bigg)dx\notag\\
&=\int_{\Omega}\bigg( \cfrac{C}{2M}\rho \cfrac{d}{dt}(\nabla c \cdot \nabla c)-\cfrac{C}{2M}\nabla \cdot(\rho\bold{u})(\nabla c \cdot \nabla c ) +\cfrac{C}{M}(\rho \nabla c \otimes \nabla c) : \nabla \bold{u}\bigg)dx,\label{highre}
\end{align}
in which the following identity is used,
\begin{align}
\nabla\cdot(\rho\nabla c \otimes \nabla c )=\nabla\cdot ( \rho \nabla c ) \nabla c  +\cfrac{1}{2} \rho \nabla (\nabla c \cdot \nabla c ),\label{ctimesc}
\end{align}
such that 
\begin{align}
&\int_{\Omega}\bigg(\cfrac{C}{M}\rho\nabla c \cdot \nabla \big((\bold{u}\cdot \nabla )c\big)\bigg)dx\notag\\
&\hspace{0mm}=\int_{\Omega}\bigg(-\cfrac{C}{M}\nabla \cdot (\rho\nabla c) (\bold{u}\cdot \nabla )c-\cfrac{C}{2M}\rho\bold{u}\cdot\nabla (\nabla c \cdot \nabla c )+\cfrac{C}{2M}\rho \bold{u}\cdot\nabla (\nabla c \cdot \nabla c ) \bigg)dx\notag\\
&\hspace{0mm}=\int_{\Omega}\bigg(-\cfrac{C}{M}\nabla \cdot (\rho\nabla c\otimes \nabla c )\cdot \bold{u} -\cfrac{C}{2M}\nabla \cdot (\rho\bold{u}) (\nabla c \cdot \nabla c ) \bigg)dx\notag\\
&\hspace{0mm}=\int_{\Omega}\bigg(\cfrac{C}{M}(\rho\nabla c\otimes \nabla c ) : \nabla \bold{u} -\cfrac{C}{2M}\nabla \cdot (\rho\bold{u}) (\nabla c \cdot \nabla c ) \bigg)dx.
\end{align}
By adding Eqs.(\ref{enlawfv})-(\ref{enlawfm}) together, we obtain
\begin{align}
&\cfrac{d}{dt}\bigg(\cfrac{1}{2}||\sqrt{\rho}\bold{u}||^2_{L^2} + \cfrac{C}{2M}||\sqrt{\rho}~\nabla c||^2_{L^2}+\int_{\Omega}\big(\cfrac{1}{M}\rho F(c)\big)dx\bigg)\notag\\
&=-\cfrac{1}{Re} ||\nabla \bold{u}||^2_{L^2} -\cfrac{1}{3Re}||\nabla \cdot \bold{u}||^2_{L^2}-\cfrac{1}{MPe} ||\nabla \mu||^2_{L^2}\notag\\
&-\int_{\Omega}\bigg(\cfrac{1}{Fr^2}(\rho-\rho_0)\hat{\bold{j}}\cdot \bold{u}+\cfrac{\rho_0  \alpha}{Fr^2Pe} \hat{\bold{j}}\cdot \nabla \mu+(\nabla \cdot  \bold{u})\big(\cfrac{1}{2}\rho\bold{u}\cdot \bold{u}+\cfrac{1}{M}\rho F(c)+\cfrac{C}{2M}\rho (\nabla c \cdot \nabla c)+\cfrac{1}{M}p~\big)\notag\\
&+\cfrac{\alpha}{Pe}\nabla \mu \cdot \nabla\big(\cfrac{1}{2}\rho\bold{u}\cdot \bold{u}  +\cfrac{1}{M}\rho F(c) +\cfrac{C}{2M}\rho\nabla c\cdot \nabla c+\cfrac{1}{M}p \big)\bigg)dx.\label{enlawf123}
\end{align}
For the continuous equation (\ref{wqua1}), we set $q=\rho\bold{u}\cdot \bold{u}/2 + \rho F(c)/M+\rho C(\nabla c\cdot \nabla c )/2M + p/M+(\rho -\rho_0)y/Fr^2$, through integration by parts we obtain
\begin{align}
0&=-\int_{\Omega}\bigg((\nabla \cdot  \bold{u})\big(\cfrac{1}{2}\rho\bold{u}\cdot \bold{u}+\cfrac{1}{M}\rho F(c)+\cfrac{C}{2M}\rho (\nabla c \cdot \nabla c)+\cfrac{1}{M}p+\cfrac{\rho-\rho_0}{Fr^2} y~\big)\notag\\
&-\cfrac{\alpha}{Pe}\nabla \mu \cdot \nabla\big(\cfrac{1}{2}\rho\bold{u}\cdot \bold{u}  +\cfrac{1}{M}\rho F(c) +\cfrac{C}{2M}\rho\nabla c\cdot \nabla c+\cfrac{1}{M}p +\cfrac{\rho-\rho_0}{Fr^2} y\big)\bigg)dx.\label{enlawfms}
\end{align}
By adding Eq.(\ref{enlawfms}) to Eq.(\ref{enlawf123}), having in mind Eq.(\ref{jj}) and the following identity under the homogeneous boundary conditions for $\bold{u}$
\begin{align}
\int_{\Omega}\bigg(\cfrac{1}{Fr^2}(\rho-\rho_0)\hat{\bold{j}}\cdot \bold{u}\bigg)dx=-\int_{\Omega}\bigg(\cfrac{1}{Fr^2}\nabla \cdot \big((\rho-\rho_0)\bold{u}\big)y\bigg)dx,
\end{align}
we finally obtain the continuous energy law for the quasi-incompressible NSCH system with respect to the total energy $\widehat{E}$ (defined in Eq.(\ref{total--en})),
\begin{align}
\cfrac{d\widehat{E}}{dt}=& \cfrac{d}{dt}\bigg(\cfrac{1}{2}||\sqrt{\rho}\bold{u}||^2_{L^2} + \cfrac{C}{2M}||\sqrt{\rho}~\nabla c||^2_{L^2}+\int_{\Omega}\big(\cfrac{1}{M}\rho F(c)+\cfrac{1}{Fr^2}\rho y\big)dx\bigg)\notag\\
=&-\cfrac{1}{Re} ||\nabla \bold{u}||^2_{L^2} -\cfrac{1}{3Re}||\nabla \cdot \bold{u}||^2_{L^2}-\cfrac{1}{MPe} ||\nabla \mu||^2_{L^2}.\label{energylaw11}
\end{align}
By using Eq.(\ref{potentialrelation}), we obtain
\begin{align}
\cfrac{dE}{dt}&= \cfrac{d}{dt}\bigg(\cfrac{1}{2}||\sqrt{\rho}\bold{u}||^2_{L^2} + \cfrac{C}{2M}||\sqrt{\rho}~\nabla c||^2_{L^2}+\int_{\Omega}\big(\cfrac{1}{M}\rho F(c)\big)dx\bigg)\notag\\
&=-\cfrac{1}{Re} ||\nabla \bold{u}||^2_{L^2} -\cfrac{1}{3Re}||\nabla \cdot \bold{u}||^2_{L^2}-\cfrac{1}{MPe} ||\nabla \mu||^2_{L^2}\notag\\
&-\int_{\Omega}\bigg(\cfrac{1}{Fr^2}(\rho-\rho_0)\hat{\bold{j}}\cdot \bold{u}+\cfrac{\rho_0  \alpha}{Fr^2Pe} \hat{\bold{j}}\cdot \nabla \mu\bigg)dx,\label{energylaw}
\end{align}
for the original energy $E$ (Eq.(\ref{original-energy})), without the potential energy. Note that two energy laws (\ref{energylaw11}) and (\ref{energylaw}) are equivalent in continuous case, whereas, at the discrete level, this equivalence relation depends on the discretization of the numerical methods. In $\S4$, two energy law preserving numerical methods are developed corresponding to the energy law (\ref{energylaw11}) and (\ref{energylaw}) respectively. The method (based on Eq.(2.1a)) with respect to the energy law (\ref{energylaw11}) is used for our computations, and the other method (based on Eq.(2.1b)) that preserves the energy law (\ref{energylaw}) at the discrete level is presented in Remark 4.2.\\\\
From Eqs.(\ref{enlawfc}), ({\ref{highre}}) and (\ref{enlawfms}) we notice that, in deriving the energy law, the terms $C\rho\nabla c \cdot \nabla \big((\bold{u}\cdot \nabla)c\big)/M$ and $-\alpha C\nabla \mu \cdot \nabla (\rho\nabla c\cdot \nabla c)/2MPe$ are involved, implying that a higher regularity than $c \in W^{1,3}$ is needed and that more complicated $C^1$ finite elements are needed. To ensure the energy law with the lower regularity $c \in W^{1,3}$, we need to reformulate the equations of the system (2.1)-(\ref{qua4}), so that only first order derivatives of $c$ are required during the derivation so that an energy law preserving $C^0$ finite element method can be derived.
\Section{Reformulation and the Numerical Method that Accurately Preserve the Energy Law}
We first reformulate the momentum equation (\ref{qua2}) and the chemical potential equation (\ref{qua4}) in order that a rigorous energy law can be derived using lower regularity requirements so that the continuous ($C^0$) finite element can be used. We emphasise $C^{0}$ finite elements because they are much simpler than the finite elements with a higher degree of smoothness and are available in most existing finite element software packages, which reduces various complications. Based on the weak form of reformulated system, we then develop a special numerical scheme such that a discrete energy law can be rigorously obtained that is very similar to the continuous one (\ref{energylaw11}). This feature not only immediately implies the stability of the numerical scheme, but also ensures the accuracy of the solutions.\\\\
\subsection{New Formulation of the Momentum Equation}
In order to obtain an accurate discrete energy law, we first consider the continuous equation (2.1b)
\begin{align}
\rho_t+\nabla \cdot (\rho \bold{u}) = 0,\notag
\end{align}
by multiplying $\bold{u}/2$, we obtain 
\begin{align}
\cfrac{1}{2}\rho_t\bold{u}+\cfrac{1}{2}\nabla \cdot (\rho \bold{u})\bold{u} = 0.\notag
\end{align}
By adding this term to Eq.(\ref{qua2}), we obtain
\begin{align}
\sqrt{\rho}(\sqrt{\rho}\bold{u})_t+\rho(\bold{u}\cdot \nabla)\bold{u}+\cfrac{1}{2}\nabla \cdot (\rho \bold{u})\bold{u}&=-\cfrac{1}{M}\big(\nabla p + C\nabla \cdot (\rho \nabla c \otimes \nabla c )\big)\notag\\
&\hspace{-10mm}+ \nabla \cdot (\cfrac{1}{Re} \nabla \bold{u})+\nabla\big(\cfrac{1}{3Re}(\nabla \cdot \bold{u}) \big)-\cfrac{1}{Fr^2}(\rho-\rho_0)\hat{\bold{j}}.\label{nnqua2}
\end{align}
We next consider the chemical potential equation (\ref{qua4}) and rewrite it as
\begin{align}
\mu -f(c)+\cfrac{\partial \rho}{\partial c}\cfrac{ p}{\rho^2} +\cfrac{C}{\rho}\nabla \cdot (\rho\nabla c )+\cfrac{M\rho_0 \alpha}{Fr^2} y=0,  \notag
\end{align}
by multiplying $\rho \nabla c/M$ and using Eq.(\ref{ctimesc}), we obtain
\begin{align}
&\cfrac{1}{M}\rho \nabla c\mu - \cfrac{1}{M}\rho \nabla F(c)+\cfrac{1}{M}\cfrac{ p}{\rho}\nabla \rho + \cfrac{C}{M}\nabla \cdot (\rho\nabla c \otimes \nabla c)\notag\\
&\hspace{50mm} -\cfrac{C}{2M}\rho\nabla (\nabla c \cdot \nabla c )+\cfrac{\rho_0 \alpha}{Fr^2} y\rho \nabla c=0. \label{surfaceterm}
\end{align}
By introducing a new pressure $\hat{p}$,
\begin{align}
\hat{p}=p+\rho F(c) +\cfrac{C}{2}\rho |\nabla c|^2\label{newnewp},
\end{align}
we obtain 
\begin{align}
p=\hat{p}-\rho F(c) -\cfrac{C}{2}\rho |\nabla c|^2.\label{newpressure}
\end{align}
Adding Eq.(\ref{surfaceterm}) to Eq.(\ref{nnqua2}) and using in mind of Eq.(\ref{newpressure}), we obtain a new formulation of the momentum equation
\begin{align}
&\sqrt{\rho}(\sqrt{\rho}\bold{u})_t+\rho(\bold{u}\cdot \nabla )\bold{u}+\cfrac{1}{2}\nabla \cdot (\rho \bold{u})\bold{u}=-\cfrac{1}{M}\rho\nabla( \cfrac{\hat{p}}{\rho})+\cfrac{1}{M}\rho \mu\nabla c\notag\\
&\hspace{30mm} +\nabla \cdot (\cfrac{1}{Re} \nabla \bold{u})+\nabla\big(\cfrac{1}{3Re}(\nabla \cdot \bold{u}) \big)+\cfrac{\rho_0 \alpha}{Fr^2} y\rho \nabla c-\cfrac{1}{Fr^2}(\rho-\rho_0)\hat{\bold{j}}.\label{newvelocity}
\end{align}
\subsection{New Formulation of the Chemical Potential Equation}
For the chemical potential equation (\ref{qua4}), by multiplying $\rho$ on both sides and substituting (\ref{newpressure}) we obtain
\begin{align}
\rho\mu&=\rho f(c)-\cfrac{\partial \rho}{\partial c} \cfrac{\hat{p}}{\rho}+\cfrac{\partial \rho}{\partial c} F(c)+\cfrac{\partial \rho}{\partial c} \cfrac{C}{2}(\nabla c \cdot \nabla c )- C \nabla \cdot (\rho\nabla c )-\cfrac{M\rho_0 \alpha}{Fr^2} y\rho.\notag
\end{align} 
Note that the reason we introduce a new pressure $\hat{p}$ and reformulate Eq.(\ref{qua2}) and Eq.(\ref{qua4}) is that, in the weak form presented later, we can derive an accurate discrete energy law while keeping the variables in appropriate functional spaces as we assume at the beginning of $\S$3.
\subsection{New System and Weak Form}
Now we obtain the new system for the quasi-incompressible NSCH system for variable density flows,
\begin{align}
&\nabla \cdot \bold{u}= \cfrac{\alpha}{Pe}\Delta \mu,\label{nnewmass}\\
&\sqrt{\rho}(\sqrt{\rho}\bold{u})_t+\rho(\bold{u}\cdot \nabla )\bold{u}+\cfrac{1}{2}\nabla \cdot (\rho \bold{u})\bold{u}=-\cfrac{1}{M}\rho \nabla( \cfrac{\hat{p}}{\rho})+\cfrac{1}{M}\rho \mu\nabla c\notag\\
&\hspace{20mm}+\nabla \cdot (\cfrac{1}{Re} \nabla \bold{u}) +\nabla\big(\cfrac{1}{3Re}(\nabla \cdot \bold{u}) \big)+\cfrac{\rho_0 \alpha}{Fr^2} y\rho \nabla c-\cfrac{1}{Fr^2}(\rho-\rho_0)\hat{\bold{j}},\label{nnewvelocity}\\
&\rho\dot{c}=\nabla \cdot (\cfrac{1}{Pe}\nabla \mu),\label{nnewqua3}\\
&\rho\mu=\rho f(c)-\cfrac{\partial \rho}{\partial c}\cfrac{\hat{p}}{\rho}+\cfrac{\partial \rho}{\partial c} F(c)+\cfrac{\partial \rho}{\partial c} \cfrac{C}{2} (\nabla c \cdot \nabla c) - C \nabla \cdot (\rho\nabla c )-\cfrac{M\rho_0 \alpha}{Fr^2} y\rho.\label{nnqua4}
\end{align} 
The corresponding new weak form reads: find $\bold{u} \in \bold{W}^{1,3}_{\bold{b}}(\Omega)$, $\hat{p} /\rho\in W^{1,3/2}(\Omega)$, $c \in W^{1,3}(\Omega)$ and $\mu \in W^{1,3}(\Omega)$ such that 
\begin{align}
&\int_{\Omega}\bigg(- \bold{u}\cdot \nabla q\bigg)dx=\int_{\Omega}\bigg(-\cfrac{\alpha}{Pe} \nabla\mu \cdot\nabla q\bigg)dx, \hspace{38mm} \forall q \in W^{1,3/2}(\Omega),\label{wnqua1}\\
&\int_{\Omega}\bigg(\sqrt{\rho}(\sqrt{\rho}\bold{u})_t\cdot \bold{v}+\rho(\bold{u}\cdot \nabla )\bold{u}\cdot \bold{v}+\cfrac{1}{2}\nabla \cdot (\rho \bold{u})\bold{u}\cdot \bold{v}\bigg)dx=\int_{\Omega}\bigg(-\cfrac{1}{M}\rho(\bold{v}\cdot \nabla) \cfrac{\hat{p}}{\rho} \notag\\
&-  \cfrac{1}{Re} \nabla \bold{u} : \nabla\bold{v}-\cfrac{1}{3Re}(\nabla \cdot \bold{u})(\nabla\cdot \bold{v})+\cfrac{1}{M}\rho (\bold{v}\cdot\nabla) c\mu +\cfrac{\rho_0 \alpha}{Fr^2} y\rho \bold{v}\cdot\nabla c\notag\\
&\hspace{10mm}-\cfrac{1}{Fr^2}(\rho-\rho_0)\hat{\bold{j}}\cdot \bold{v}\bigg)dx,\hspace{63mm}\forall \bold{v} \in \bold{W}^{1,3}_{0}(\Omega),\label{wnqua2}\\
&\int_{\Omega}\bigg(\rho\dot{c}\psi\bigg)dx=\int_{\Omega}\bigg(-\cfrac{1}{Pe}\nabla \mu \cdot \nabla\psi \bigg)dx,\hspace{44mm}\forall \psi \in W^{1,3/2}(\Omega), \label{wnqua3}\\
&\int_{\Omega}\bigg(\rho\mu\chi\bigg)dx=\int_{\Omega}\bigg( \rho f(c)\chi  - \cfrac{\partial \rho}{\partial c} \cfrac{\hat{p}}{\rho} \chi+\cfrac{\partial \rho}{\partial c} F(c)\chi+\cfrac{\partial \rho}{\partial c} \cfrac{C}{2} (\nabla c \cdot \nabla c)\chi +C\rho \nabla c \cdot \nabla \chi\notag\\
&\hspace{10mm}-\cfrac{M\rho_0 \alpha}{Fr^2} y\rho\chi\bigg)dx,\hspace{75mm}\forall \chi \in W^{1,3}(\Omega).\label{wnqua4}
\end{align}
If we let $\bold{v}=\bold{u}$, $\psi=\mu/M-(\rho-\rho_0)  \alpha y/Fr^2-\alpha \hat{p}/M $, $\chi=c_t/  M$ and $q=\hat{p}/M$, we can still obtain the continuous energy law 
\begin{align}
\cfrac{d\widehat{E}}{dt}=& \cfrac{d}{dt}\bigg(\cfrac{1}{2}||\sqrt{\rho}\bold{u}||^2_{L^2} + \cfrac{C}{2M}||\sqrt{\rho}~\nabla c||^2_{L^2}+\int_{\Omega}\big(\cfrac{1}{M}\rho F(c)+\cfrac{1}{Fr^2}\rho y\big)dx\bigg)\notag\\
=&-\cfrac{1}{Re} ||\nabla \bold{u}||^2_{L^2} -\cfrac{1}{3Re}||\nabla \cdot \bold{u}||^2_{L^2}-\cfrac{1}{MPe} ||\nabla \mu||^2_{L^2},\notag
\end{align}
which is consistent with the energy law (\ref{energylaw11}) derived with respect to the total energy $\widehat{E}$. Again, in order to get a meaningful weak form of the system, we require $\rho \in L^{ \infty } (\Omega)$ and $\rho\bold{u} \in \bold{W}^{1,3}(\Omega)$. Moreover, note that no second order derivative of $c$ is involved when deriving the energy law with the new weak form. Hence no higher regularity for variable $c$ is needed. Further, to have a meaningful weak form, all the integrals in Eq. (\ref{wnqua4}) should be finite or bounded in the current choice of functional spaces for $c\in W^{1,3}$ and $\chi \in W^{1,3}$. As it is obvious for other terms, we only need to check the third and fourth integral terms at the right hand side of (\ref{wnqua4}). From Sobolev's embedding theorem \cite{Adams1975}, we know that 
\begin{align}
||f||_{L^{q}(\Omega)} \le C||Df||_{L^p(\Omega)},~~~~\forall f \in W^{1,p},~~~~1 \le p \le q \le \cfrac{np}{n-p},\notag
\end{align}
where $C$ is a generic constant. By setting $f=c^{3}$, $n=2$, $p=1$ and $q=np/(n-p)$, we have
\begin{align}
||c^4||^{3}_{3/2} \le C||c||^{2}_{3}||\nabla c||_{3}~,\notag
\end{align}
where the H$\operatorname{\ddot{o}}$lder's inequality is used. Hence, in the third term at the right hand side, $F(c) \in L^{3/2}$. In the fourth term at the right hand side, it is obvious that $\nabla c \cdot \nabla c \in L^{3/2}$ when $c \in W^{1,3}$ by H$\operatorname{\ddot{o}}$lder's inequality. Having in mind $\rho \in L^{ \infty } (\Omega)$ and Eq.(\ref{partialho}), we obtain 
\begin{align}
 \cfrac{\partial \rho}{\partial c} F(c) \in L^{3/2}~~~~\operatorname{and}~~~~ \cfrac{\partial \rho}{\partial c}\nabla c \cdot \nabla c \in L^{3/2},
\end{align}
and thus both integrals are finite when $c\in W^{1,3}$ and $\chi \in W^{1,3}$.\\
Next we present a special temporal scheme where an accurate discrete energy law can be obtained. If $C^0$ finite elements are used and if time remains continuous, the finite element solution belongs to the functional spaces required in weak form (\ref{wnqua1})-(\ref{wnqua4}).
\subsection{A Special Temporal Scheme with an Accurate Discrete Energy Law}
We seek to solve the weak problem (\ref{wnqua1})-(\ref{wnqua4}) using a finite difference scheme in time and a conformal $C^0$ finite element method in space. Let
\begin{align}
\bold{W}=\bold{W}^{1,3}_{b}(\Omega)\times W^{1,3/2}(\Omega)\times W^{1,3}(\Omega)\times W^{1,3}(\Omega),\notag
\end{align}
and $\bold{W}^h=\bold{U}^{h}\times P^{h}\times H^{h}\times H^{h},
$ be a finite dimensional subspace of $\bold{W}$ given by a finite element discretization of $\Omega$. $\bold{W}^{h}_{0}$ represents the space $\bold{W}^h$ satisfying homogeneous Dirichlet boundary conditions. Let $\Delta t > 0$ represent the time step size and $(\bold{u}^{n}_{h},\hat{p}^{n}_{h},c^{n}_{h},\mu^{n}_{h}) \in \bold{W}^{h}$ be an approximation of $\bold{u}(t^n)=\bold{u}(n\Delta t)$, $\hat{p}(t^n)=p(\Delta t)$, $c(t^n)=c(\Delta t)$ and $\mu(t^n)=\mu(\Delta t)$. The approximation at time $t^{n+1}=(n+1)\Delta t$ is denoted as $(\bold{u}^{n+1}_{h},\hat{p}^{n+1}_{h},c^{n+1}_{h},\mu^{n+1}_{h}) \in \bold{W}^{h}$ and is computed by the following finite element scheme 
\begin{align}
&\int_{\Omega}\bigg(- (\sqrt{\rho}\bold{u})^{n+1}_{h} \cdot \nabla q\bigg)dx=\int_{\Omega}\bigg(-\cfrac{\alpha}{Pe}\nabla \mu^{n+\frac{1}{2}}_{h} \cdot \nabla q\bigg)dx,\label{dwqua11}\\
&\int_{\Omega}\bigg(\sqrt{\rho}^{n+\frac{1}{2}}_{h}(\sqrt{\rho}\bold{u})^{n+1}_{\bar{t}}\cdot \bold{v}+\rho^{n+\frac{1}{2}}_{h}\big((\sqrt{\rho}\bold{u})^{n+1}_{h}\cdot \nabla \big)(\sqrt{\rho}\bold{u})^{n+1}_{h}\cdot \bold{v}\notag\\
&\hspace{70mm}+\cfrac{1}{2}\nabla \cdot \big(\rho^{n+\frac{1}{2}}_{h} (\sqrt{\rho}\bold{u})^{n+1}_{h}\big)(\sqrt{\rho}\bold{u})^{n+1}_{h}\cdot \bold{v}\bigg)dx\notag\\
&=\int_{\Omega}\bigg(-\cfrac{1}{M} {\rho^{n+\frac{1}{2}}_h}(\bold{v}\cdot\nabla)\cfrac{ \hat{p}^{n+\frac{1}{2}}_h}{\rho^{n+\frac{1}{2}}_h}- \cfrac{1}{Re} \nabla (\sqrt{\rho}\bold{u})^{n+1}_{h} : \nabla  \bold{v}-\cfrac{1}{3Re}\big(\nabla \cdot(\sqrt{\rho}\bold{u})^{n+1}_{h}\big)\big)\big(\nabla\cdot \bold{v}\big)\notag\\
&-\cfrac{1}{M}\cfrac{1}{\alpha \rho^{n+\frac{1}{2}}_{h}}(\bold{v}\cdot \nabla)\rho^{n+\frac{1}{2}}_{h} \mu^{n+\frac{1}{2}}_{h}-\cfrac{\rho_0 \alpha}{Fr^2} y\cfrac{1}{\alpha \rho^{n+\frac{1}{2}}_{h}} \bold{v}\cdot\nabla \rho^{n+\frac{1}{2}}_{h}-\cfrac{1}{Fr^2}(\rho^{n+\frac{1}{2}}_{h}-\rho_0)\hat{\bold{j}}\cdot \bold{v}\bigg)dx,\label{dwqua21}\\
&\int_{\Omega}\bigg(-\cfrac{r(c^{n+1}_{h},c^n_{h})}{\alpha \rho^{n+\frac{1}{2}}_{h}}c^{n+1}_{\bar{t}}\psi+\cfrac{1}{2\rho^{n+\frac{1}{2}}_{h}}\big((\rho^{n+1}_h)^2(\sqrt{\rho}\bold{u})^{n+1}_{h}\cdot \nabla c^{n+1}_{h}+(\rho^{n}_h)^2(\sqrt{\rho}\bold{u})^{n+1}_{h}\cdot \nabla c^{n}_{h}\big)\psi\bigg)dx\notag\\
&\hspace{90mm}=\int_{\Omega}\bigg(-\cfrac{1}{Pe}\nabla \mu^{n+\frac{1}{2}}_{h} \cdot \nabla\psi \bigg)dx, \label{dwqua31}\\
&\int_{\Omega}\bigg( -\cfrac{r(c^{n+1}_{h},c^n_{h})}{\alpha \rho^{n+\frac{1}{2}}_{h}} \mu^{n+\frac{1}{2}}_{h}\chi\bigg)dx=\int_{\Omega}\bigg( \rho^{n+\frac{1}{2}}_{h} g(c^{n+1}_{h},c^n_{h})\chi  -\cfrac{1}{{\rho^{n+\frac{1}{2}}_{h}} }r(c^{n+1}_{h},c^n_{h}) \hat{p}^{n+\frac{1}{2}}_{h} \chi\notag\\
&+ \cfrac{1}{2} \big( F(c^{n+1}_{h})+F(c^{n}_{h})\big)r(c^{n+1}_{h},c^n_{h})\chi+\cfrac{C}{4}(\nabla c^{n+1}_{h} \cdot \nabla c^{n+1}_{h}+\nabla c^{n}_{h} \cdot \nabla c^{n}_{h})r(c^{n+1}_{h},c^n_{h})\chi\notag\\
&\hspace{62mm}+C \rho^{n+\frac{1}{2}}_{h} \nabla c^{n+\frac{1}{2}}_{h}\cdot \nabla   \chi+\cfrac{r(c^{n+1}_{h},c^n_{h})}{ \rho^{n+\frac{1}{2}}_{h}}\cfrac{M\rho_0 }{Fr^2} y\chi\bigg)dx,\label{dwqua41}
\end{align}
for all $(\bold{v}, q, \psi, \chi)\in \bold{W}^{h}_0$, where
\begin{footnotesize}
\begin{center}
\begin{tabular}{l@{\hspace{1mm}} l@{\hspace{1mm}}l}
\toprule
 \vspace{2mm}
 $\rho^{n+1}_{h}=\cfrac{\rho_1 \rho_2}{(\rho_2-\rho_1)c^{n+1}_{h}+\rho_1},$ &\hspace{15mm} $\rho^{n+\frac{1}{2}}_{h}=\cfrac{\rho^{n+1}_h+\rho^{n}_h}{2},$ &\hspace{15mm} $\sqrt{\rho}^{n+\frac{1}{2}}_{h}=\cfrac{\sqrt{\rho^{n+1}_h}+\sqrt{\rho^{n}_h}}{2},$\\
 \vspace{2mm}
 $\hat{p}^{n+\frac{1}{2}}_h=\cfrac{\hat{p}^{n+1}_{h}+\hat{p}^{n}_{h}}{2},$& \hspace{-10mm}$(\sqrt{\rho}\bold{u})^{n+1}_{h}=\cfrac{\sqrt{\rho^{n+1}_h}\bold{u}^{n+1}_h+\sqrt{\rho^{n}_h}\bold{u}^{n}_h}{\sqrt{\rho^{n+1}_h}+\sqrt{\rho^n_h}},$&$(\sqrt{\rho}\bold{u})^{n+1}_{\bar{t}}=\cfrac{\sqrt{\rho^{n+1}_{h}}\bold{u}^{n+1}_{h}-\sqrt{\rho^{n}_{h}}\bold{u}^{n}_{h}}{\Delta t},$\\
 \vspace{2mm}
$c^{n+1}_{\bar{t}}=\cfrac{c^{n+1}_{h}-c^{n}_{h}}{\Delta t},$ &\hspace{17mm}$c^{n+\frac{1}{2}}_{h}=\cfrac{c^{n+1}_{h}+c^{n}_{h}}{2},$&\hspace{25mm}$\mu^{n+\frac{1}{2}}_{h}=\cfrac{\mu^{n+1}_{h}+\mu^{n}_{h}}{2}.$\\
\bottomrule
\end{tabular}
\end{center}
\end{footnotesize}
Further,
\begin{align}
g(c^{n+1}_{h},c^n_{h})=\cfrac{1}{4}\big(c^{n+1}_{h}(c^{n+1}_{h}-1)+c^{n}_{h}(c^{n}_{h}-1)\big)(c^{n+1}_{h}+c^{n}_{h}-1)
\end{align}
is an approximation to the nonlinear function $F'(c)=f(c)=c(c-1)(c-1/2)$, where we note the identity,
\begin{align}
&F(c^{n+1}_h)-F(c^{n}_h)=g(c^{n+1}_{h},c^n_{h})(c^{n+1}_h-c^{n}_h).
\end{align}
And
\begin{align}
&r(c^{n+1}_h,c^{n}_h)=-\cfrac{\rho_1 \rho_2(\rho_2-\rho_1)}{\big((\rho_2-\rho_1\big)c^{n+1}_h+\rho_1)\big((\rho_2-\rho_1)c^{n}_h+\rho_1\big)}~,
\end{align}
is an approximation of the nonlinear function
\begin{align}
\cfrac{\partial \rho(c)}{\partial c}=-\alpha \rho^2(c)=-\cfrac{\rho_1 \rho_2(\rho_2-\rho_1)}{\big((\rho_2-\rho_1)c+\rho_1\big)^2}~,
\end{align}
where we note the identity
\begin{align}
&\rho(c^{n+1}_h)-\rho(c^{n}_h)=r(c^{n+1}_h,c^{n}_h)(c^{n+1}_h-c^{n}_h).
\end{align}
And we use
\begin{align}
-\cfrac{r(c^{n+1}_{h},c^n_{h})}{\alpha \rho^{n+\frac{1}{2}}_{h}},~~\cfrac{(\rho^{n+1}_{h})^2}{\rho^{n+\frac{1}{2}}_{h}}~~\operatorname{and}~\cfrac{(\rho^{n}_{h})^2}{\rho^{n+\frac{1}{2}}_{h}}
\end{align}
to approximate $\rho(c)$.
\begin{theo}
In the case of homogeneous boundary conditions, the solution of scheme  (\ref{dwqua11})-(\ref{dwqua41}) satisfies the following discrete energy law, which is analogous to that obtained in the continuous case in Eq.(\ref{energylaw11}),
\begin{align}
\widehat{E}^{n+1}_{\bar{t}}&=\bigg(\cfrac{1}{2}||\sqrt{\rho^{n+1}_h}\bold{u}^{n+1}_h||_{L^2}^2+\cfrac{C}{2M}||\sqrt{\rho^{n+1}_h}\nabla c^{n+1}_h||_{L^2}^2+\int_{\Omega}\big( \cfrac{1}{M}\rho^{n+1}_hF(c^{n+1}_h) +\cfrac{1}{Fr^2}\rho^{n+1}_h y\big)dx\bigg)_{\bar{t}}\notag\\
&= - \cfrac{1}{Re} ||\nabla (\sqrt{\rho}\bold{u})^{n+1}_{h} ||_{L^2}^2-\cfrac{1}{3Re}||\nabla \cdot (\sqrt{\rho}\bold{u})^{n+1}_{h} ||_{L^2}^2-\cfrac{1}{MPe}||\nabla \mu^{n+\frac{1}{2}}_{h}||_{L^2}^2.
\label{disenergylaw}
\end{align}
$Proof$. According to the continuous energy law, by setting $\bold{v}=(\sqrt{\rho}\bold{u})^{n+1}_{h}$ in Eq.(\ref{dwqua21}) and using integration by parts, we have
\begin{align}
\bigg(\cfrac{1}{2}||\sqrt{\rho^{n+1}_{h}}\bold{u}^{n+1}_{h}||^2_{L^2}\bigg)_{\bar{t}}&=\int_{\Omega}\bigg(\cfrac{1}{M}\cfrac{ \hat{p}^{n+\frac{1}{2}}_h}{\rho^{n+\frac{1}{2}}_h} \nabla \cdot \big({\rho^{n+\frac{1}{2}}_h}(\sqrt{\rho}\bold{u})^{n+1}_{h}\big)-\cfrac{1}{M}\cfrac{\rho_0 y}{\alpha\rho^{n+\frac{1}{2}}_{h}}(\sqrt{\rho}\bold{u})^{n+1}_{h}\cdot \nabla \rho^{n+\frac{1}{2}}_{h}\notag\\
&\hspace{-10mm}-\cfrac{1}{M}\cfrac{1}{\alpha\rho^{n+\frac{1}{2}}_{h}}(\sqrt{\rho}\bold{u})^{n+1}_{h}\cdot \nabla \rho^{n+\frac{1}{2}}_{h}\mu^{n+\frac{1}{2}}_{h} +\cfrac{1}{Fr^2}(\rho^{n+\frac{1}{2}}_{h}-\rho_0)\hat{\bold{j}}\cdot (\sqrt{\rho}\bold{u})^{n+1}_{h}\bigg)dx\notag\\
&\hspace{-10mm}- \cfrac{1}{Re} ||\nabla (\sqrt{\rho}\bold{u})^{n+1}_{h}||^2_{L^2}-\cfrac{1}{3Re}||\nabla \cdot (\sqrt{\rho}\bold{u})^{n+1}_{h}||^2_{L^2}.\label{ddwqua1}
\end{align}
Taking $\psi=-(\rho^{n+1/2}_{h}-\rho_0)\alpha y/Fr^2-\alpha \hat{p}^{n+1/2}_{h}/M+\mu^{n+1/2}_h/M$ in Eq.(\ref{dwqua31}) leads to 
\begin{align}
\bigg(\int_{\Omega}\big(\cfrac{1}{Fr^2}\rho^{n+1}y\big)dx\bigg)_{\bar{t}}&=\int_{\Omega}\bigg(\cfrac{\alpha}{Fr^2Pe}\nabla \mu^{n+\frac{1}{2}}_{h} \cdot \nabla(\rho^{n+\frac{1}{2}}_h y)-\cfrac{1}{Fr^2}y(\sqrt{\rho}\bold{u})^{n+1}_{h}\cdot \nabla \rho^{n+\frac{1}{2}}_{h}\notag\\
&\hspace{-10mm}+\cfrac{1}{Fr^2}\cfrac{\rho_0 y}{\rho^{n+\frac{1}{2}}_{h}}\rho^{n+1}_{\bar{t}}+\cfrac{1}{M}\cfrac{\rho_0 y}{\alpha\rho^{n+\frac{1}{2}}_{h}}(\sqrt{\rho}\bold{u})^{n+1}_{h}\cdot \nabla \rho^{n+\frac{1}{2}}_{h}-\cfrac{\alpha}{Fr^2Pe}\nabla \mu^{n+\frac{1}{2}}_{h} \cdot \nabla(\rho_0 y) \notag\\
&\hspace{-10mm}-\cfrac{1}{M}\cfrac{\hat{p}^{n+1/2}_{h}}{\rho^{n+\frac{1}{2}}_h}\rho^{n+1}_{\bar{t}}-\cfrac{1}{M}\cfrac{\hat{p}^{n+1/2}_{h}}{\rho^{n+\frac{1}{2}}_h}(\sqrt{\rho}\bold{u})^{n+1}_{h}\cdot \nabla \rho^{n+\frac{1}{2}}_{h}+\cfrac{\alpha}{MPe}\nabla \mu^{n+\frac{1}{2}}_{h} \cdot \nabla\hat{p}^{n+\frac{1}{2}}_h\notag\\
&\hspace{-10mm}+\cfrac{1}{M}\cfrac{\mu^{n+\frac{1}{2}}_{h}}{\alpha \rho^{n+\frac{1}{2}}_{h}}\rho^{n+1}_{\bar{t}}+\cfrac{1}{M}\cfrac{\mu^{n+\frac{1}{2}}_{h}}{\alpha\rho^{n+\frac{1}{2}}_{h}}(\sqrt{\rho}\bold{u})^{n+1}_{h}\cdot \nabla \rho^{n+\frac{1}{2}}_{h}\bigg)dx-\cfrac{1}{MPe}||\nabla \mu^{n+\frac{1}{2}}_h||_{L^2}^2. \label{ddwqua3}
\end{align}
By taking $\chi= c^{n+1}_{\bar{t}}/M$ in Eq.(\ref{dwqua41}), we obtain
\begin{align}
\bigg(\cfrac{C}{2M}||\sqrt{\rho^{n+1}_h}\nabla& c^{n+1}_h||_{L^2}^2+\cfrac{1}{M}\int_{\Omega}\big( \rho^{n+1}_hF(c^{n+1}_h) \big)dx\bigg)_{\bar{t}}=\notag\\
&\int_{\Omega}\bigg(-\cfrac{1}{M}\cfrac{1}{\alpha \rho^{n+\frac{1}{2}}_{h}}\rho^{n+1}_{\bar{t}}\mu^{n+\frac{1}{2}}_{h}+\cfrac{1}{M}\cfrac{\hat{p}^{n+1/2}_{h}}{\rho^{n+\frac{1}{2}}_h}\rho^{n+1}_{\bar{t}}-\cfrac{1}{Fr^2}\cfrac{\rho_0 y}{\rho^{n+\frac{1}{2}}_{h}}\rho^{n+1}_{\bar{t}}\bigg)dx.\label{ddwqua41}
\end{align}
Taking $q=\hat{p}^{n+1/2}_h/M+(\rho^{n+1/2}_h-\rho_0)y/Fr^2$ in Eq.(\ref{dwqua11}) and using integration by parts, we obtain
\begin{align}
0=\int_{\Omega}\bigg(-\nabla \cdot (\sqrt{\rho}\bold{u})^{n+1}_{h}&\big(\cfrac{1}{M}\hat{p}^{n+1/2}_h+\cfrac{1}{Fr^2}(\rho-\rho_0)y\big)\notag\\
&-\cfrac{\alpha}{Pe}\nabla \mu^{n+\frac{1}{2}}_{h} \cdot \nabla\big(\cfrac{1}{M}\hat{p}^{n+1/2}_h+\cfrac{1}{Fr^2}(\rho-\rho_0)y\big) \bigg)dx.\label{ddwqua4}
\end{align}
Combing Eqs.(\ref{ddwqua1})-(\ref{ddwqua4}) together, we finally obtain the discrete energy law (\ref{disenergylaw}) for the weak form (\ref{dwqua11})-(\ref{dwqua41}),
\begin{align}
\widehat{E}^{n+1}_{\bar{t}}&=\bigg(\cfrac{1}{2}||\sqrt{\rho^{n+1}_h}\bold{u}^{n+1}_h||_{L^2}^2+\cfrac{C}{2M}||\sqrt{\rho^{n+1}_h}\nabla c^{n+1}_h||_{L^2}^2+\int_{\Omega}\big( \cfrac{1}{M}\rho^{n+1}_hF(c^{n+1}_h) +\cfrac{1}{Fr^2}\rho^{n+1}_h y\big)dx\bigg)_{\bar{t}}\notag\\
&= - \cfrac{1}{Re} ||\nabla (\sqrt{\rho}\bold{u})^{n+1}_{h} ||_{L^2}^2-\cfrac{1}{3Re}||\nabla \cdot (\sqrt{\rho}\bold{u})^{n+1}_{h} ||_{L^2}^2-\cfrac{1}{MPe}||\nabla \mu^{n+\frac{1}{2}}_{h}||_{L^2}^2.\notag
\end{align}
This completes the proof. $\Box$
\end{theo}
Note that the numerical scheme (\ref{dwqua11})-(\ref{dwqua41}) is used for our computations later. Another energy-law preserving numerical method with respect to the continuous energy law (\ref{energylaw}) may be also designed, see Remark 4.2.
\begin{rem}
We now remark about how to implement the scheme (\ref{dwqua11})-(\ref{dwqua41}) and how to choose continuous finite element spaces based on the above derivation of the discrete energy law. From weak form (\ref{dwqua11})-(\ref{dwqua41}) we observe that $\bold{u}$ appears in the form of $\sqrt{\rho} \bold{u}$ and $\hat{p}$ appears in the form (or can be made in the form) of $\hat{p}/\rho$. We can thus introduce new variables $\tilde{\bold{u}}=\sqrt{\rho}\bold{ u}$ and $\tilde{p}= \hat{p}/\rho$, and denote the corresponding finite element spaces as $\tilde{\bold{U}}^h$ and $\tilde{P}^h$ respectively. 
\end{rem}
\begin{rem}
If we want to solve the model based on equation (2.1b) instead of (2.1a), we can also design a method which can preserve the energy law (\ref{energylaw}) at the discrete level. With $\bold{u}^{n+1}_{h}, p^{n+1}_{h}, c^{n+1}_{h}, \mu^{n+1}_{h}, \bold{v}, q, \psi, \chi$ in the appropriate spaces, we seek solutions $(\bold{u}^{n+1}_{h},p^{n+1}_{h},c^{n+1}_{h},\mu^{n+1}_{h})$ by computing the following finite element scheme 
\begin{align}
&\int_{\Omega}\bigg(\rho^{n+1}_{\bar{t}}q\bigg)dx=\int_{\Omega}\bigg(-\nabla\cdot\big(\rho^{n+\frac{1}{2}}_{h}(\sqrt{\rho}\bold{u})^{n+1}_{h} \big)q\bigg)dx,\label{dwqua1}\\
&\int_{\Omega}\bigg(\sqrt{\rho}^{n+\frac{1}{2}}_{h}(\sqrt{\rho}\bold{u})^{n+1}_{\bar{t}}\cdot \bold{v}+\rho^{n+\frac{1}{2}}_{h}\big((\sqrt{\rho}\bold{u})^{n+1}_{h}\cdot \nabla \big)(\sqrt{\rho}\bold{u})^{n+1}_{h}\cdot \bold{v}\notag\\
&\hspace{68mm}+\cfrac{1}{2}\nabla \cdot \big(\rho^{n+\frac{1}{2}}_{h} (\sqrt{\rho}\bold{u})^{n+1}_{h}\big)(\sqrt{\rho}\bold{u})^{n+1}_{h}\cdot \bold{v}\bigg)dx\notag\\
&=\int_{\Omega}\bigg(\cfrac{1}{M} p^{n+\frac{1}{2}}_{h}(\nabla \cdot \bold{v})- \cfrac{1}{Re} \nabla (\sqrt{\rho}\bold{u})^{n+1}_{h} : \nabla  \bold{v}-\cfrac{1}{3Re}\big(\nabla \cdot(\sqrt{\rho}\bold{u})^{n+1}_{h})\big)(\nabla\cdot \bold{v})\notag\\
&+\cfrac{1}{M}\rho^{n+\frac{1}{2}}_{h}\mu^{n+\frac{1}{2}}_{h} (\bold{v}\cdot\nabla) c^{n+\frac{1}{2}}_{h} + \cfrac{1}{2M} \big( F(c^{n+1}_{h})+F(c^{n}_{h})\big)\nabla\cdot(\rho^{n+\frac{1}{2}}_{h}\bold{v})\notag\\
&+\cfrac{1}{M}\cfrac{ p^{n+\frac{1}{2}}_{h}}{\rho^{n+\frac{1}{2}}_{h}}(\bold{v}\cdot \nabla)\rho^{n+\frac{1}{2}}_{h}+\cfrac{C}{4M}(\nabla c^{n+1}_{h} \cdot \nabla c^{n+1}_{h}+\nabla c^{n}_{h} \cdot \nabla c^{n}_{h})\nabla\cdot (\rho^{n+\frac{1}{2}}_{h}\bold{v})\notag\\
&\hspace{54mm}+\cfrac{\rho_0 \alpha}{Fr^2} y\rho^{n+\frac{1}{2}}_{h} \bold{v}\cdot\nabla c^{n+\frac{1}{2}}_{h}-\cfrac{1}{Fr^2}(\rho^{n+\frac{1}{2}}_{h}-\rho_0)\hat{\bold{j}}\cdot \bold{v}\bigg)dx,\label{dwqua2}\\
&\int_{\Omega}\bigg(\rho^{n+\frac{1}{2}}_{h} c^{n+1}_{\bar{t}}\psi+\rho^{n+\frac{1}{2}}_{h}\big((\sqrt{\rho}\bold{u})^{n+1}_{h}\cdot \nabla\big)c^{n+\frac{1}{2}}_{h}\psi\bigg)dx=\int_{\Omega}\bigg(-\cfrac{1}{Pe}\nabla \mu^{n+\frac{1}{2}}_{h} \cdot \nabla\psi \bigg)dx, \label{dwqua3}\\
&\int_{\Omega}\bigg( \rho^{n+\frac{1}{2}}_{h} \mu^{n+\frac{1}{2}}_{h}\chi\bigg)dx=\int_{\Omega}\bigg( \rho^{n+\frac{1}{2}}_{h} g(c^{n+1}_{h},c^n_{h})\chi  -\cfrac{1}{{\rho^{n+\frac{1}{2}}_{h}} }r(c^{n+1}_{h},c^n_{h}) p^{n+\frac{1}{2}}_{h} \chi\notag\\
&\hspace{68mm}+C \rho^{n+\frac{1}{2}}_{h} \nabla c^{n+\frac{1}{2}}_{h}\cdot \nabla   \chi-\cfrac{M\rho_0 \alpha}{Fr^2} y\rho^{n+\frac{1}{2}}_{h}\chi\bigg)dx,\label{dwqua4}
\end{align}
where, instead of Eq.(2.1a), Eq.(2.1b) is used for the weak form, $\rho^{n+1}_{\bar{t}}=(\rho^{n+1}_{h}-\rho^{n}_{h})/\Delta t,$ and $p^{n+1/2}_{h}=(p^{n+1}_{h}+p^{n}_{h})/2$.
Note that when taking $\bold{v}=(\sqrt{\rho}\bold{u})^{n+1}_{h},\psi=\mu^{n+1/2}_h/M+\rho_0  \alpha y/Fr^2,\chi= c^{n+1}_{\bar{t}}/M$ and $q=p^{n+1/2}_h/M\rho^{n+1/2}_{h} + \big(F(c^{n+1}_h)+ F(c^{n}_h)\big)/2M+ C(\nabla c ^{n+1}_h \cdot \nabla c^{n+1}_h+\nabla c ^{n}_h \cdot \nabla c^{n}_h)/4M$ in the weak form (\ref{dwqua1})-(\ref{dwqua4}) respectively, we can obtain the discrete energy law corresponding to the continuous energy law (\ref{energylaw}), 
\begin{align}
&E^{n+1}_{\bar{t}}=\bigg(\cfrac{1}{2}||\sqrt{\rho^{n+1}_h}\bold{u}^{n+1}_h||_{L^2}^2+\cfrac{C}{2M}||\sqrt{\rho^{n+1}_h}\nabla c^{n+1}_h||_{L^2}^2+\int_{\Omega}\big( \cfrac{1}{M}\rho^{n+1}_hF(c^{n+1}_h) \big)dx\bigg)_{\bar{t}}\notag\\
&\hspace{10mm}= - \cfrac{1}{Re} ||\nabla (\sqrt{\rho}\bold{u})^{n+1}_{h} ||_{L^2}^2-\cfrac{1}{3Re}||\nabla \cdot (\sqrt{\rho}\bold{u})^{n+1}_{h} ||_{L^2}^2-\cfrac{1}{MPe}||\nabla \mu^{n+\frac{1}{2}}_{h}||_{L^2}^2\notag\\
&\hspace{10mm}-\int_{\Omega}\bigg(\cfrac{1}{Fr^2}(\rho^{n+\frac{1}{2}}_{h}-\rho_0)\hat{\bold{j}}\cdot (\sqrt{\rho}\bold{u})^{n+1}_{h}+\cfrac{\rho_0  \alpha}{Fr^2Pe} \hat{\bold{j}}\cdot \nabla \mu^{n+\frac{1}{2}}_{h}\bigg)dx.\label{disenergylaw11}
\end{align}
\end{rem}
\Section{Implementation Issues}
As weak form (\ref{dwqua11})-(\ref{dwqua41}) is a highly nonlinear system consisting of the variable density Navier-Stokes equation with extra stress term, an advective Cahn-Hilliard equations, we employ Newton's method to linearise the system. We first briefly introduce a Newton's linearisation for time-dependent nonlinear equations associated with the unknowns, $\tilde{\bold{u}}$, $\tilde{p}$, $c$ and $\mu$ at the implicit time level,
\begin{align}
\bold{F}(\tilde{\bold{u}}^{n+1},\tilde{p}^{n+\frac{1}{2}},c^{n+1},\mu^{n+1})=0,\label{aaa}
\end{align}
where $\bold{F}$ is a vector function corresponding to the weak form equations (\ref{dwqua11})-(\ref{dwqua41}) respectively, $\tilde{\bold{u}}$ and $\tilde{p}$ are the new variables defined in Remark 4.1. and $\tilde{p}^{n+1/2}_h=\hat{p}^{n+1/2}_h/\rho^{n+1/2}_h$ (treating it as a solution for $\tilde{p}$ at the $(n+1/2)$th time step in our computations). The solutions at the $(n+1)$th time step ($(n+1/2)$th for $\tilde{p}$) ($\tilde{\bold{u}}^{n+1}, \tilde{p}^{n+\frac{1}{2}}, c^{n+1}, \mu^{n+1}$) are unknown, and ($\tilde{\bold{u}}^n, \tilde{p}^{n-\frac{1}{2}}, c^n, \mu^n$) at the $n$th~($(n-1/2)$th for $\tilde{p}$)  time step are obtained. If $n=0$ the initial conditions for $\tilde{\bold{u}}, \tilde{p}, c, \mu$ are taken as the initial guess ($\tilde{\bold{u}}^0, \tilde{p}^0, c^0, \mu^0$) for the nonlinear iteration with $\tilde{p}^0=\hat{p}^0$. For the iteration of Eq.(\ref{aaa}) at the $(n+1)$th time step, the Newton's method is
\begin{align}
0&=\bold{F}(\tilde{\bold{u}}_{s+1}^{n+1},\tilde{p}_{s+1}^{n+\frac{1}{2}},c_{s+1}^{n+1},\mu_{s+1}^{n+1})\notag\\
&\approx {\bold{F}}'_{\tilde{\bold{u}}^{n+1}}(\tilde{\bold{u}}_s^{n+1},\tilde{p}_s^{n+\frac{1}{2}},c_s^{n+1},\mu_s^{n+1})(\tilde{\bold{u}}_{s+1}^{n+1}-\tilde{\bold{u}}_s^{n+1})+{\bold{F}}'_{\tilde{p}^{n+\frac{1}{2}}}(\tilde{\bold{u}}_s^{n+1},\tilde{p}_s^{n+\frac{1}{2}},c_s^{n+1},\mu_s^{n+1})(\tilde{p}_{s+1}^{n+\frac{1}{2}}-\tilde{p}_s^{n+\frac{1}{2}})\notag\\
&\hspace{2mm}+{\bold{F}}'_{c^{n+1}}(\tilde{\bold{u}}_s^{n},\tilde{p}_s^{n+\frac{1}{2}},c_s^{n},\mu_s^{n})(c_{s+1}^{n+1}-c_s^{n+1})+{\bold{F}}'_{\mu^{n+1}}(\tilde{\bold{u}}_s^{n+1},\tilde{p}_s^{n+\frac{1}{2}},c_s^{n+1},\mu_s^{n+1})(\mu_{s+1}^{n+1}-\mu_s^{n+1})\notag\\
&\hspace{2mm}+\bold{F}(\tilde{\bold{u}}_s^{n+1},\tilde{p}_s^{n+\frac{1}{2}},c_s^{n+1},\mu_s^{n+1}).\notag
\end{align}
Further we can have
\begin{align}
&{\bold{F}}'_{\tilde{\bold{u}}^{n+1}}(\tilde{\bold{u}}_s^{n+1},\tilde{p}_s^{n+\frac{1}{2}},c_s^{n+1},\mu_s^{n+1})\tilde{\bold{u}}_{s+1}^{n+1}+{\bold{F}}'_{\tilde{p}^{n+\frac{1}{2}}}(\tilde{\bold{u}}_s^{n+1},\tilde{p}_s^{n+\frac{1}{2}},c_s^{n+1},\mu_s^{n+1})\tilde{p}_{s+1}^{n+\frac{1}{2}}\notag\\
&+{\bold{F}}'_{c^{n+1}}(\tilde{\bold{u}}_s^{n+1},\tilde{p}_s^{n+\frac{1}{2}},c_s^{n+1},\mu_s^{n+1})c_{s+1}^{n+1}+{\bold{F}}'_{\mu^{n+1}}(\tilde{\bold{u}}_s^{n+1},\tilde{p}_s^{n+\frac{1}{2}},c_s^{n+1},\mu_s^{n+1})\mu_{s+1}^{n+1}\notag\\
&={\bold{F}}'_{\tilde{\bold{u}}^{n+1}}(\tilde{\bold{u}}_s^{n+1},\tilde{p}_s^{n+\frac{1}{2}},c_s^{n+1},\mu_s^{n+1})\tilde{\bold{u}}_s^{n+1}+{\bold{F}}'_{\tilde{p}^{n+\frac{1}{2}}}(\tilde{\bold{u}}_s^{n+1},\tilde{p}_s^{n+\frac{1}{2}},c_s^{n+1},\mu_s^{n+1})\tilde{p}_s^{n+\frac{1}{2}}\notag\\
&+{\bold{F}}'_{c^{n+1}}(\tilde{\bold{u}}_s^{n+1},\tilde{p}_s^{n+\frac{1}{2}},c_s^{n+1},\mu_s^{n+1})c_s^{n+1}+{\bold{F}}'_{\mu^{n+1}}(\tilde{\bold{u}}_s^{n+1},\tilde{p}_s^{n+\frac{1}{2}},c_s^{n+1},\mu_s^{n+1})\mu_s^{n+1}\notag\\
&-\bold{F}(\tilde{\bold{u}}_s^{n+1},\tilde{p}_s^{n+\frac{1}{2}},c_s^{n+1},\mu_s^{n+1}),\label{newton}
\end{align}
where $(\tilde{\bold{u}}^{n+1}_{s},\tilde{p}^{n+\frac{1}{2}}_{s},c^{n+1}_{s},\mu^{n+1}_{s})$ are the solutions obtained after the $s$th iteration step at the $(n+1)$th time step. In the 2 dimensional case, we define the differentiation operator for the velocity $\bold{u}$ as
\begin{align}
{\bold{F}}'_{\tilde{\bold{u}}_{s+1}^{n+1}}(\tilde{\bold{u}}_{s+1}^{n+1}-\tilde{\bold{u}}_s^{n+1})=
{\bold{F}}'_{\tilde{u}_1} ({{\tilde{u_1}}}_{s+1}^{n+1} -\tilde{u_1}_s^{n+1})+{\bold{F}}'_{\tilde{u}_2}(\tilde{u_2}_{s+1}^{n+1}-\tilde{u_2}_s^{n+1}).
\end{align}
 In accordance with the local quadratic convergence theory of Newton's method, Eq.(\ref{newton}) should converge rapidly with good initial guesses. Note that the initial guesses of iteration at the $(n+1)$th time step are usually given as the solutions of linearised PDE at the previous $n$th time step 
\begin{align}
(\tilde{\bold{u}}^{n+1}_{0},\tilde{p}^{n+\frac{1}{2}}_{0},c^{n+1}_{0},\mu^{n+1}_{0})=(\tilde{\bold{u}}^{n},\tilde{p}^{n-\frac{1}{2}},c^{n},\mu^{n}).\notag
\end{align}
We then solve Eq.(\ref{newton}) by treating $(\tilde{\bold{u}}^{n+1}_{s+1},\tilde{p}^{n+\frac{1}{2}}_{s+1},c^{n+1}_{s+1},\mu^{n+1}_{s+1})$ as an approximation for the solution at the $(n+1)$th time step $(\tilde{\bold{u}}^{n+1},\tilde{p}^{n+\frac{1}{2}},c^{n+1},\mu^{n+1})$, where, in the end, we expect
\begin{align}
(\tilde{\bold{u}}^{n+1},\tilde{p}^{n+\frac{1}{2}},c^{n+1},\mu^{n+1})=\lim_{s \to \infty}(\tilde{\bold{u}}^{n+1}_{s+1},\tilde{p}^{n+\frac{1}{2}}_{s+1},c^{n+1}_{s+1},\mu^{n+1}_{s+1}).\notag
\end{align}
Note that the solutions of velocity $\bold{u}$ and pressure $\hat{p}$ at the $(n+1)$th time step can then be obtained by 
\begin{align}
\bold{u}^{n+1}=\cfrac{\tilde{\bold{u}}^{n+1}}{\sqrt{\rho^{n+1}}},~~~~\hat{p}^{n+\frac{1}{2}}=\tilde{p}^{n+\frac{1}{2}}{\rho^{n+\frac{1}{2}}}.\notag
\end{align}
In practice, we introduce a stopping criteria for the inner nonlinear iteration 
\begin{align}
\big(||\tilde{\bold{u}}^{n+1}_{s+1}-\tilde{\bold{u}}^{n+1}_{s}||^2_{H^1}+||\tilde{p}^{n+\frac{1}{2}}_{s+1}-\tilde{p}^{n+\frac{1}{2}}_{s}||^2_{H^1}+||c^{n+1}_{s+1}-c^{n+1}_{s}||^2_{H^1}+||\mu^{n+1}_{s+1}-\mu^{n+1}_{s}||^2_{H^1}\big)^{\frac{1}{2}}< tol.\label{tol}
\end{align}
With a sufficiently small tolerance (here we set $tol = 10^{-5}$ in Eq.(\ref{tol})) and proper initial guess, our numerical method for (\ref{dwqua11})-(\ref{dwqua41}) converges rapidly in practice. Several examples are presented in the next section to demonstrate the capability of our method.
\Section{Numerical Experiments}
With the help of the FreeFem++ platform \cite{Freefem} and MATLAB, the computations are carried out under the $P_2$ (piecewise quadratic) continuous finite element for the velocity $\tilde{\bold{u}}$, the phase variables $c$, $\mu$ and the pressure $\tilde{p}$. On the square domain $\Omega=[-1,1] \times [-1,1]$, we apply Dirichlet boundary conditions for the velocity ($\tilde{\bold{u}}=0$) and Neumann boundary conditions for the phase variables ($\nabla c \cdot \hat{\bold{n}}=0$, $\nabla \mu \cdot \hat{\bold{n}}=0$). Moreover the initial condition for the velocity is set to be zero ($\tilde{\bold{u}}|_{t=0}=0$). For the phase variable $c$, different initial conditions will be supplied as appropriate for the examples considered.\\
The solutions to the Cahn-Hilliard equations are nearly constant in the so called bulk region, which typically comprise the largest part of the domain. Between the bulk regions, the solutions exhibit thin transition layers, through which their values can change rapidly but continuously between their values in the bulk regions. In many cases, it is sufficient to finely resolve only the transition layers, and a fixed grid meshing represents a waste of computational resources. Thus, efficient adaptive mesh which resolves only the thin layers near the interface is desirable \cite{Kim2004}. In this paper, we adopt a variable metric/Delaunay automatic meshing algorithm for all the examples, which is built in the FreeFem++ platform (see \cite{Hecht1998} for details of this method). Snapshots for the adaptive mesh together with the evolution of the phase variable $c$ at different times are presented in Example 6.2 below.\\\\
$\bold{Example}$ $\bold{6.1}$  We study the coalescence of two kissing drops with different density ratios:  (a) 1 : 1 (density matched case, $\rho_1=\rho_2=1$); (b) 1 : 10 (variable case, $\rho_1=1$, $\rho_2=10$), where the heavier drop is set in a lighter medium. The effect of gravity is ignored in this experiment. Moreover we set
\begin{center}
\begin{tabular}{lll}
\toprule
$\epsilon=0.01$,& \hspace{10mm}$C=100\epsilon^{2},$\hspace{10mm}$M=1/(10\epsilon),$\hspace{10mm}$Pe=100/\epsilon,$\\
\bottomrule
\end{tabular}
\end{center}
and $\Delta t = 0.01$ for the time step. In \cite{Lowengrub1998}, the stationary solutions for the phase field and pressure were given, which can be used as the approximations for the initial conditions. Based on their work, we use the following as the initial condition for the phase variable $c$, 
\begin{align}
&c=\cfrac{1}{2}\operatorname{tanh}\bigg(\cfrac{-r+\sqrt{(x-a_x)^2+(y-a_y)^2}}{2\sqrt{2}\epsilon}\bigg)+\cfrac{1}{2}\operatorname{tanh}\bigg(\cfrac{-r+\sqrt{\big((x-b_x)^2+(y-b_y)^2}}{2\sqrt{2}\epsilon}\bigg),
\end{align}
where $r$ is the drop radius, $(a_x, a_y)$ and $(b_x, b_y)$ are the initial centre positions of two drops. Here we set $r=0.2\sqrt{2}$, $(a_x, a_y)=(-r/\sqrt{2}, r/\sqrt{2})$ and $(b_x, b_y)=(r/\sqrt{2}, -r/\sqrt{2})$. For simplicity, we set $p=0$ as the initial iteration of the Newton's method for the pressure. Here we refer \cite{Lowengrub1998} for detailed study of the pressure.
\\
Fig. \ref{figkiss} shows the evolution of the $c=0.5$ contours (black solid line) together with the velocity field for the density matched case and the variable density case from times $t=0.01$ to $t=260$ and $t=0.05$ to $t=70$, respectively. Because their interfaces overlap, the two drops coalesce into one larger drop for both cases with slightly visual differences.\\
In \cite{Lowengrub1998}, a sharp-interface asymptotic analysis was carried out for the quasi-incompressible NSCH system, where the interface was assumed to be normal to the $z$-axis and the fluid velocity, pressure and concentration are independent of time, x and y. At the equilibrium, they used the excess free energy to identify the surface tension $\sigma$ of the system, which can be given as
\begin{align}
\sigma=\cfrac{C}{M}\int_{+\infty}^{-\infty} \rho(c) c^{2}_{z}dz.
\end{align}
We see that the surface tension for the quasi-incompressible NSCH system is density dependent, where to be specific, for our example, the surface tension for the density matched case ($\rho_1=\rho_2=1$) is smaller than that for the variable density case  ($\rho_1=1$, $\rho_2=10$). As can be seen from Fig.\ref{figkiss}, our numerical example agrees with this analysis, where the effects of surface tension are stronger in the variable density case leading to a higher rate of interface coalescence, and the corresponding velocity fields are stronger as well. Moreover, the magnitude of the velocity is decreasing as the time passes for both cases, which states that the systems are approaching to the equilibrium. From Fig. {\ref{fig--volume}}, we observe that the volume of the drop in the entire domain is preserved well, where $\int_{\Omega} c dx = constant = 3.49321$ for the kissing drop examples in the whole time interval of computation.\\
We also examined the time evolution of the total energy (with the discrete energy law defined in Eq.(\ref{disenergylaw})) for both cases. The discrete energy functional is shown in Fig. \ref{figenlaw}. As the interface deforms, we observe that the total energy is decreasing monotonically as predicted by Theorem 4.1, and tends to a constant value corresponding to a single larger drop near the equilibrium. Moreover, the case for variable density has a larger total energy than that of the density matched case, which is consistent with the density dependence of the total energy of the quasi-incompressible NSCH system (Eq.(\ref{total--en})). We have also tested different mesh sizes and can conclude that the energy law (the decay of the energy) for this variable density model is indeed preserved very well by using our energy preserving numerical scheme.
\begin{figure}
\hspace{-5mm}
        \begin{subfigure}[h]{0.25\textwidth}
                \centering
                \includegraphics[width=\textwidth]{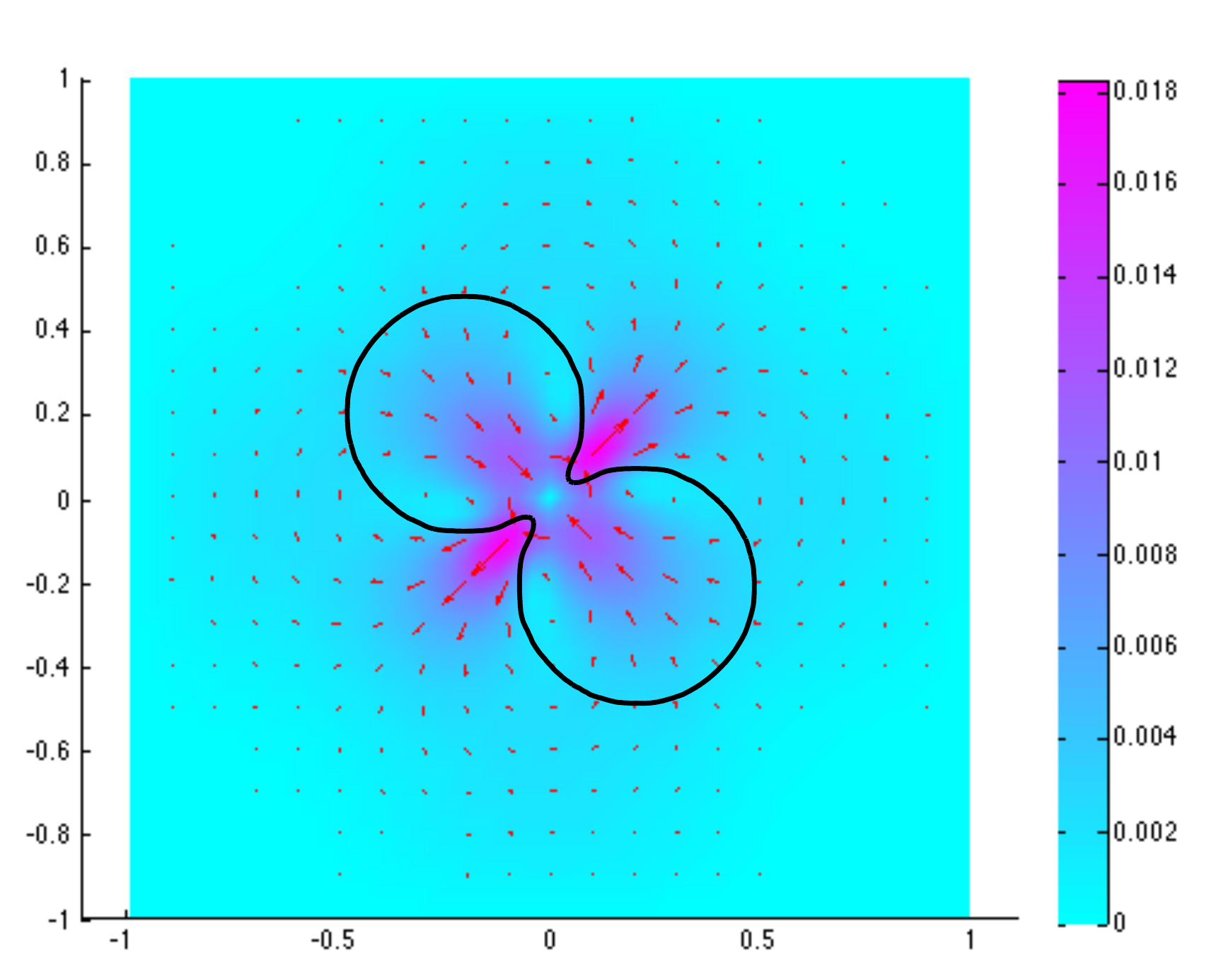}
                \caption*{t=0.01}
        \end{subfigure}%
\hspace{0mm}
        \begin{subfigure}[h]{0.25\textwidth}
                \centering
                \includegraphics[width=\textwidth]{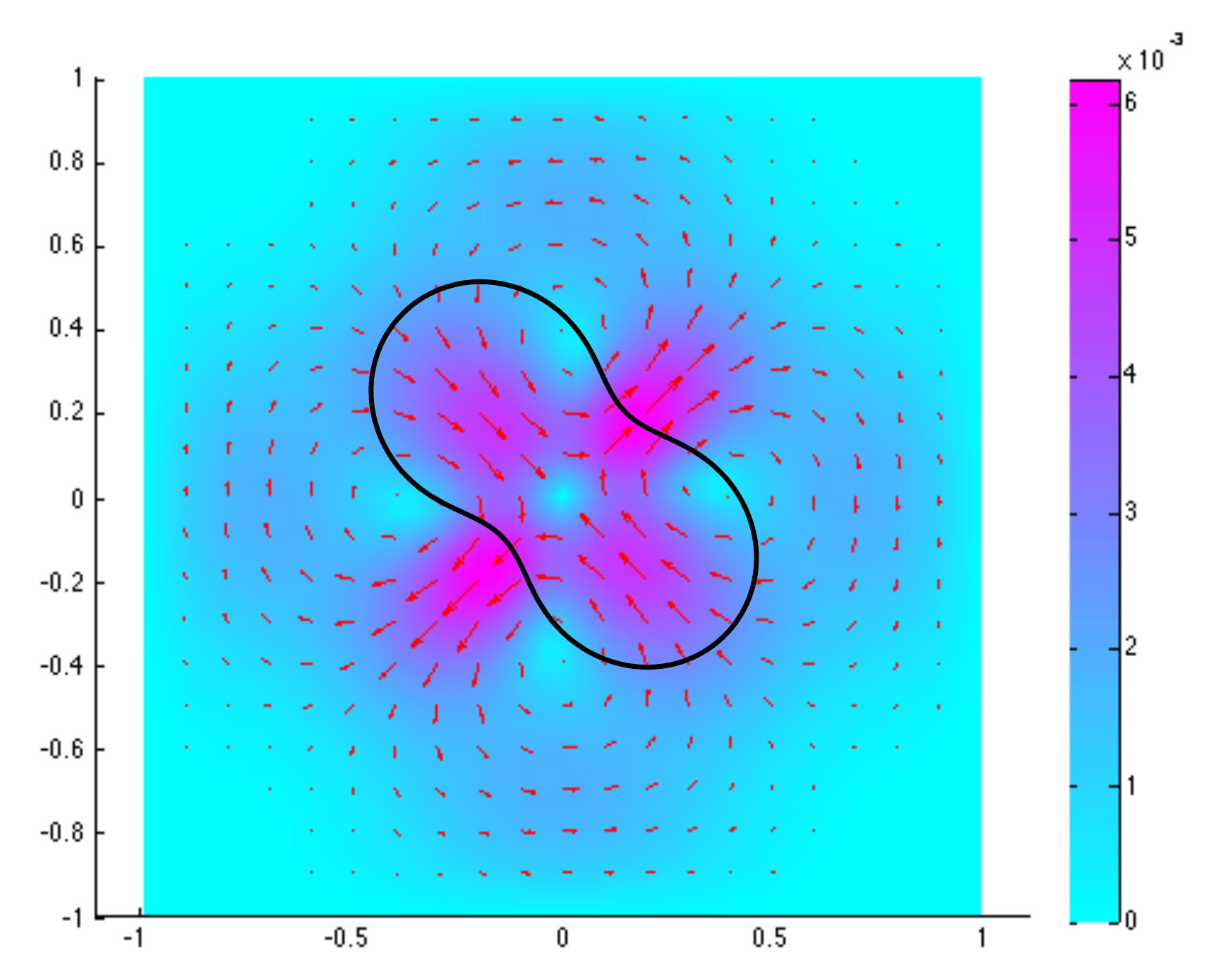}
                \caption*{t=8}
        \end{subfigure}%
\hspace{0mm}
        \begin{subfigure}[h]{0.25\textwidth}
                \centering
                \includegraphics[width=\textwidth]{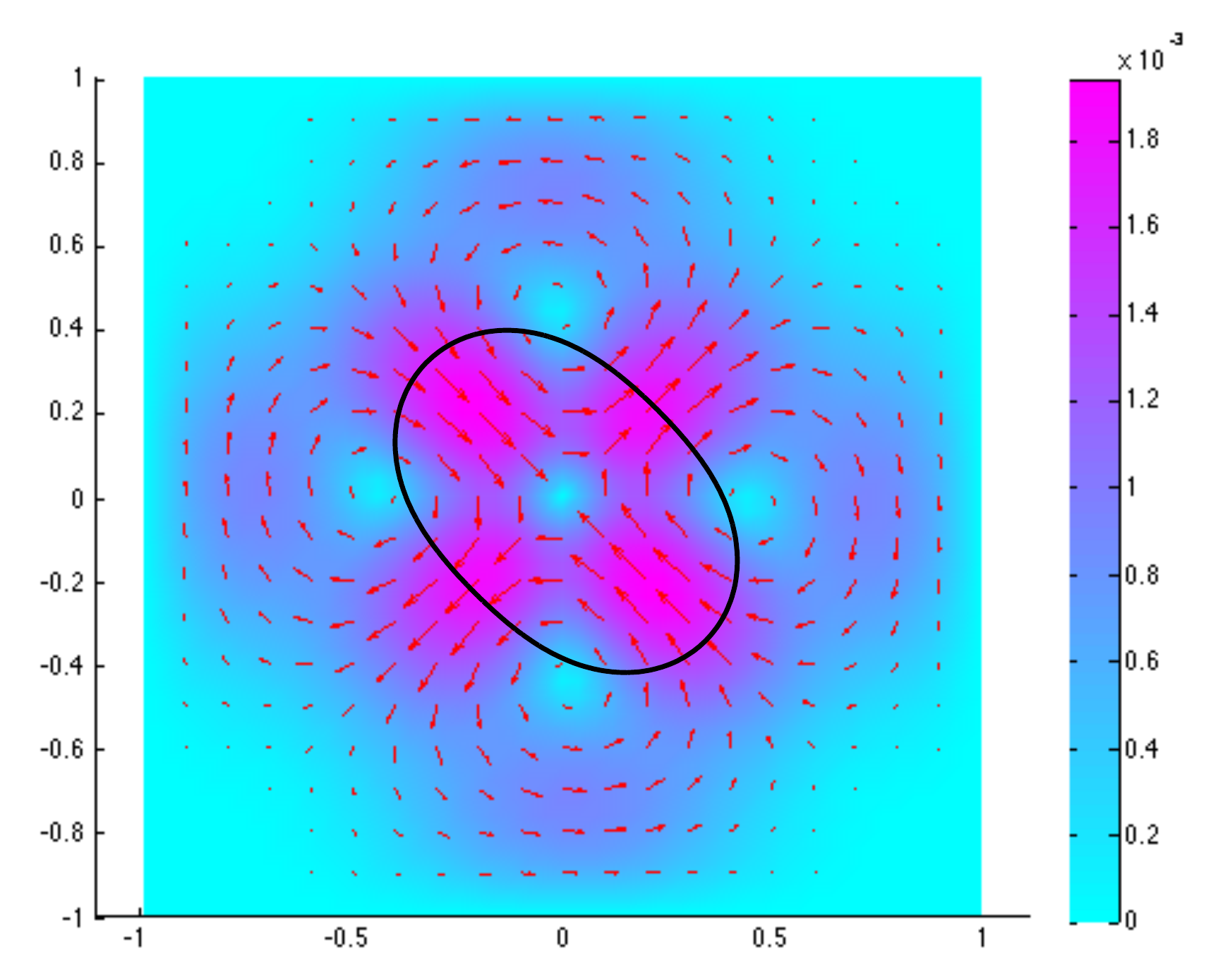}
                \caption*{t=40}
        \end{subfigure}%
\hspace{0mm}
        \begin{subfigure}[h]{0.25\textwidth}
                \centering
                \includegraphics[width=\textwidth]{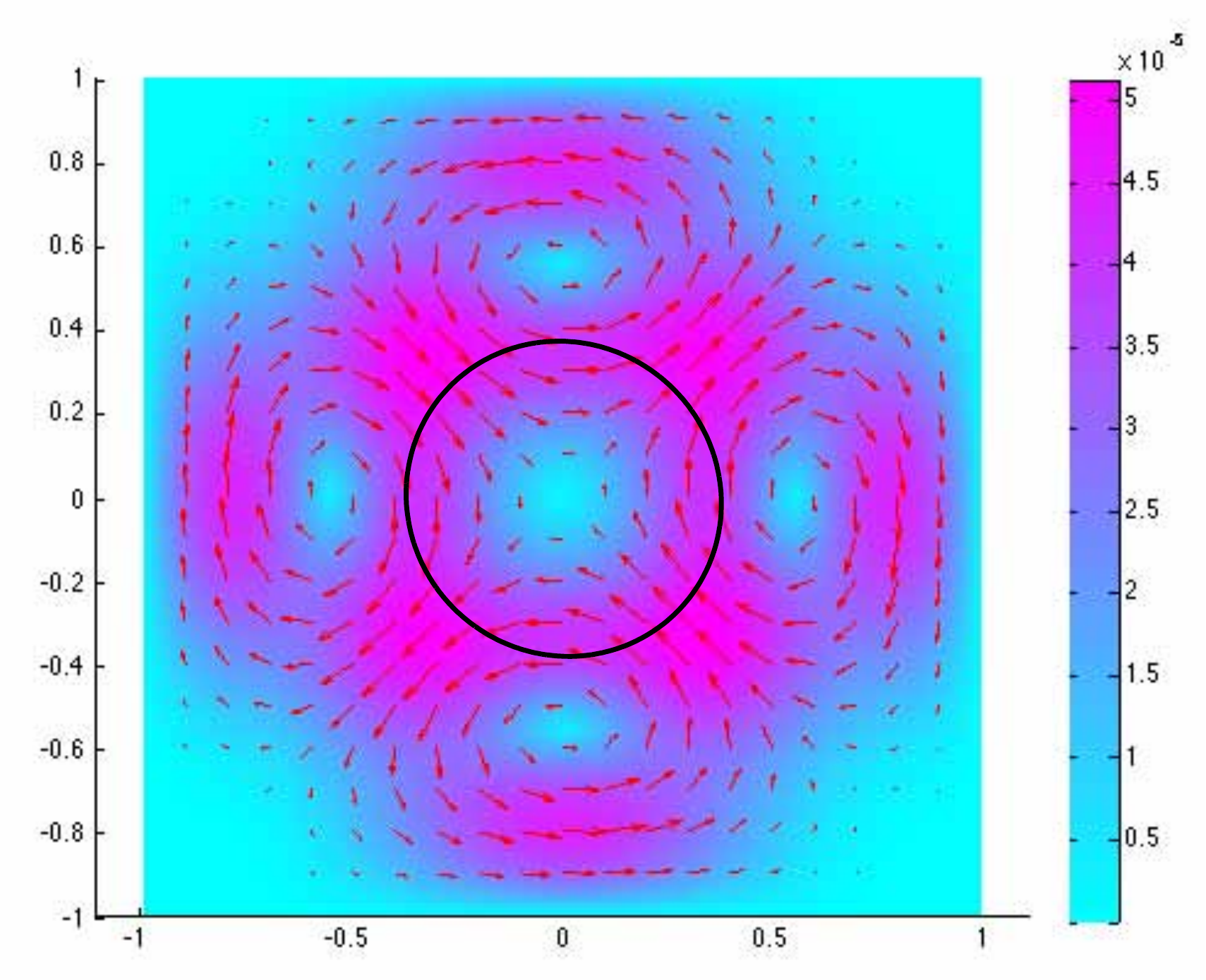}
                \caption*{t=260}
        \end{subfigure}%
                \caption*{(a)~~Density matched case ($\rho_1=\rho_2=1$)}\label{figkissa}
  
\hspace{-5mm}
        \begin{subfigure}[h]{0.25\textwidth}
                \centering
                \includegraphics[width=\textwidth]{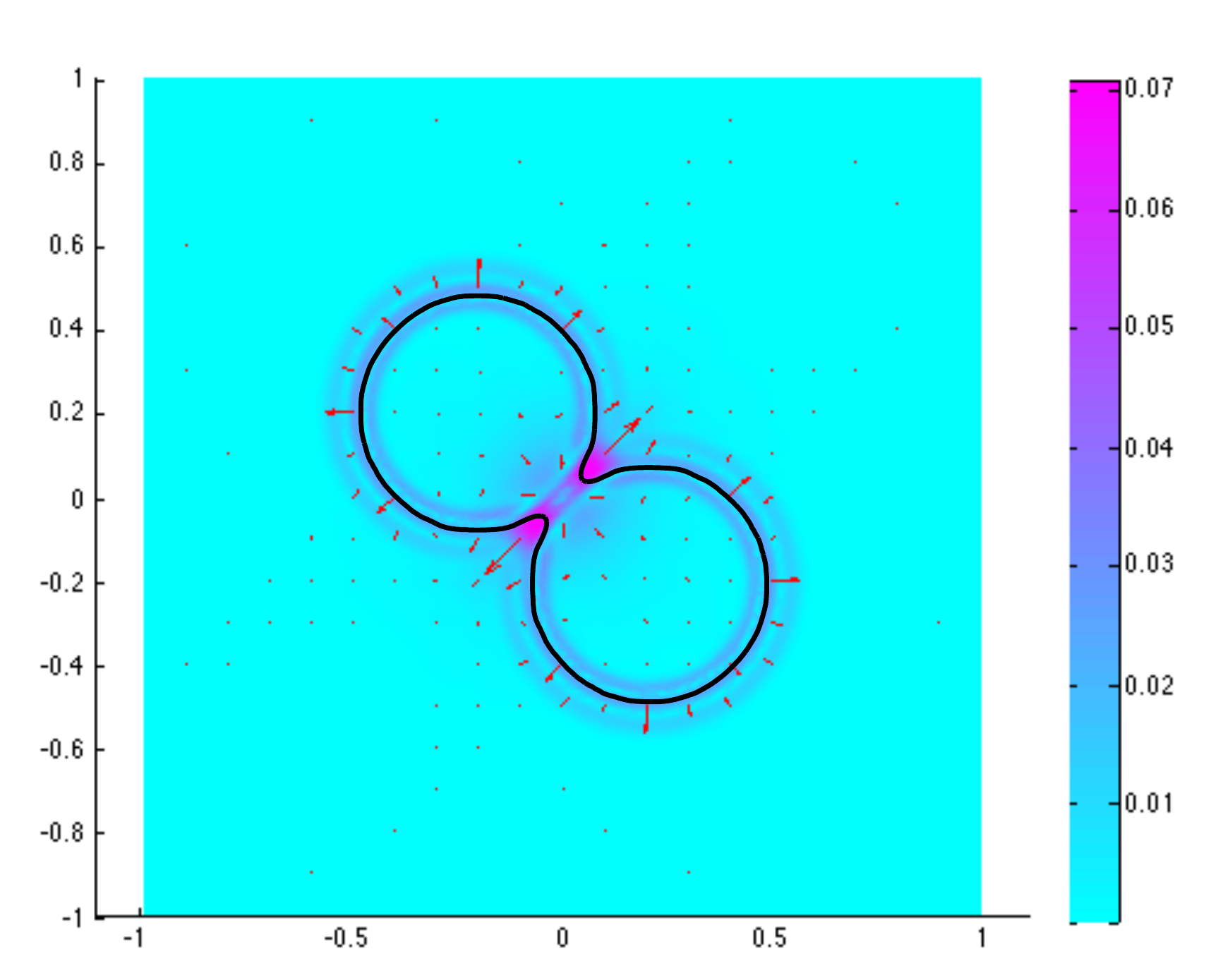}
                \caption*{t=0.05}
        \end{subfigure}%
\hspace{0mm}
        \begin{subfigure}[h]{0.25\textwidth}
                \centering
                \includegraphics[width=\textwidth]{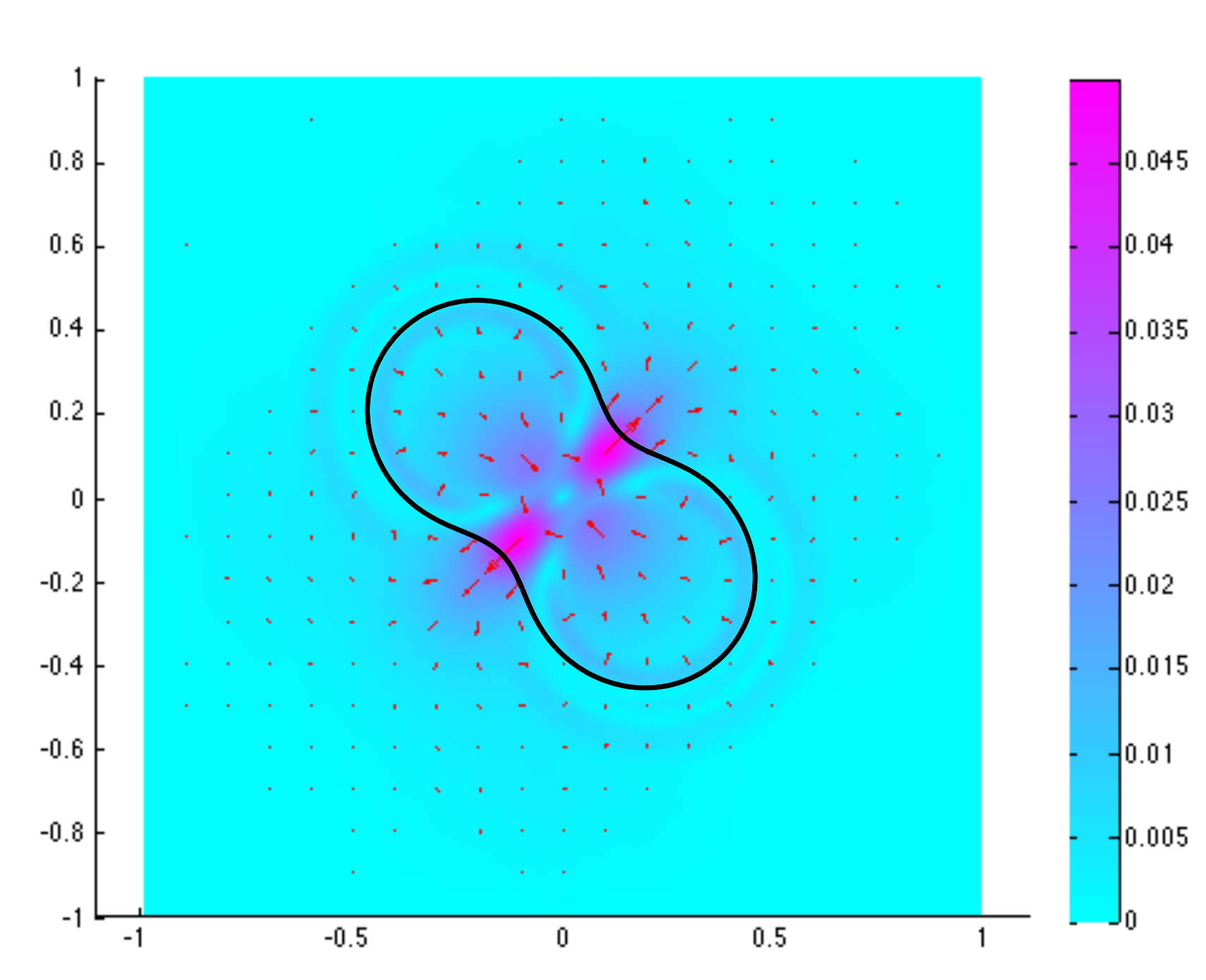}
                \caption*{t=4}
        \end{subfigure}%
\hspace{0mm}
        \begin{subfigure}[h]{0.25\textwidth}
                \centering
                \includegraphics[width=\textwidth]{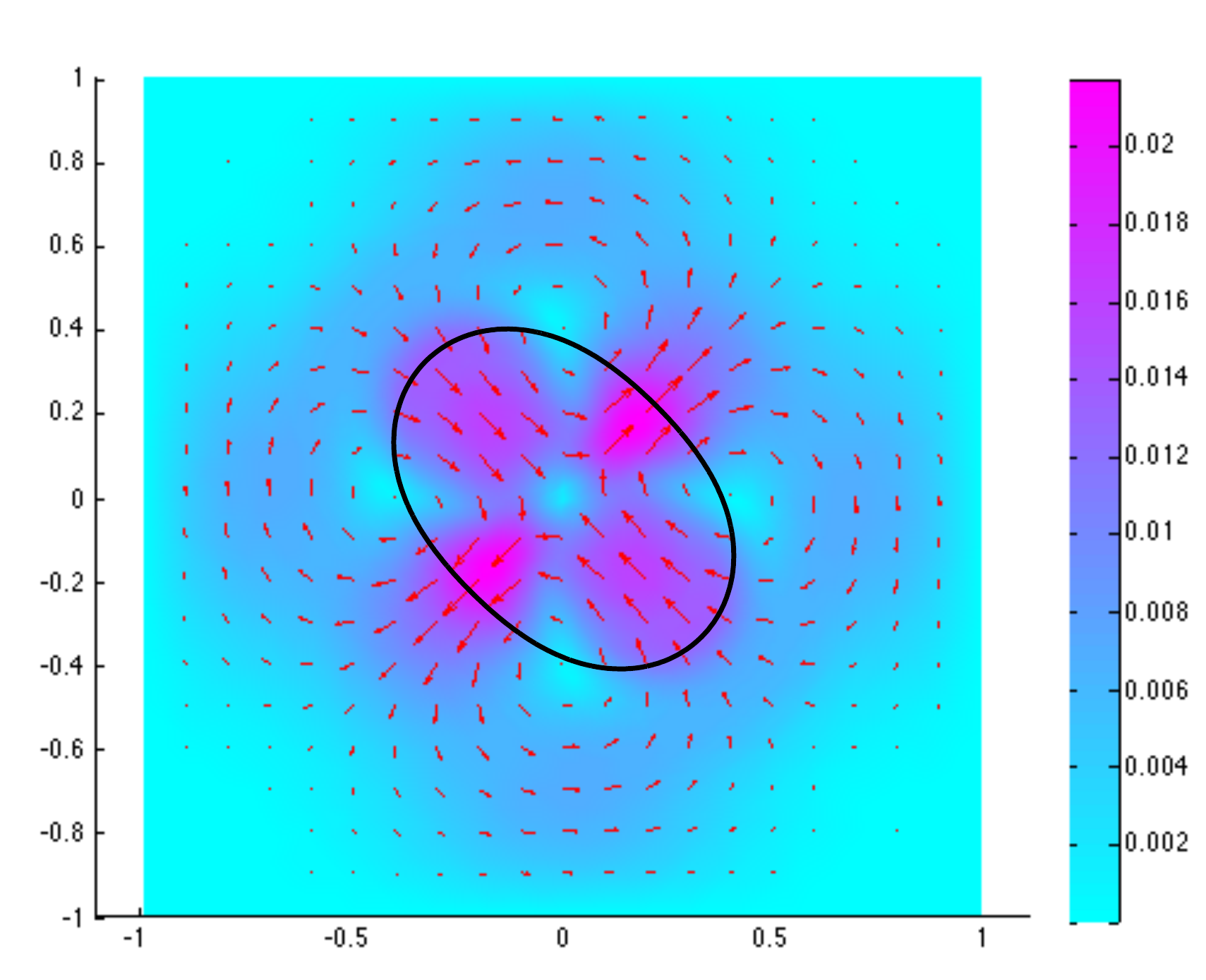}
                \caption*{t=13.5}
        \end{subfigure}%
\hspace{0mm}
        \begin{subfigure}[h]{0.25\textwidth}
                \centering
                \includegraphics[width=\textwidth]{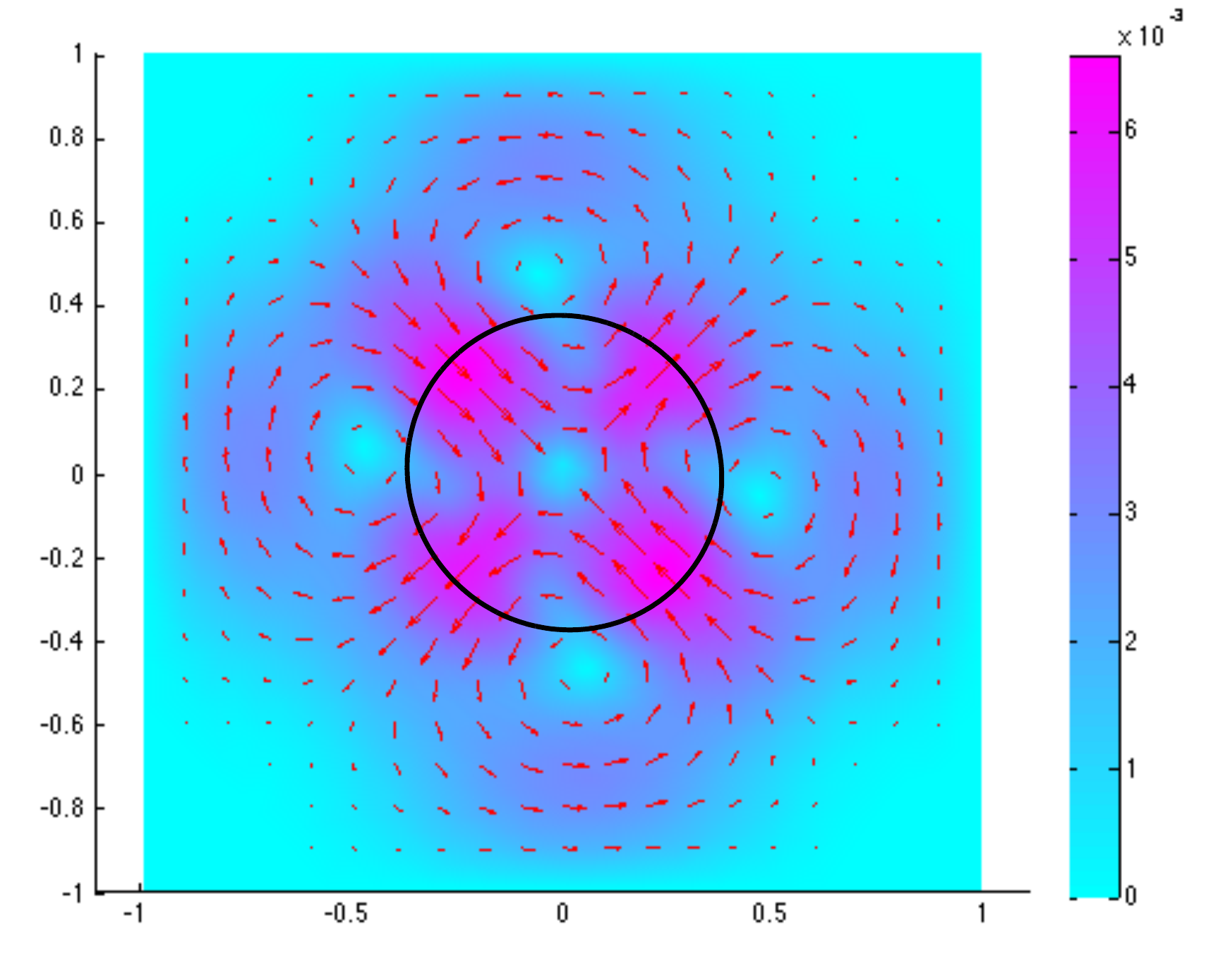}
                \caption*{t=70}
        \end{subfigure}%
                \caption*{(b)~~Variable density case ($\rho_1=1, \rho_2=10$)}\label{figkissb}
  
               \caption{The deformed drop interfaces (black solid line) with the velocity fields (with arrows representing the velocity vectors and color representing the norm value) for kissing drops. $\epsilon=0.01$, $C=100\epsilon^{2}$, $M=1/(10\epsilon)$, $Pe=100/\epsilon$ and $\Delta t = 0.01$.}\label{figkiss}
\end{figure}
\begin{figure}
 \vspace{-5mm}               \centering
                \includegraphics[width=0.6\textwidth]{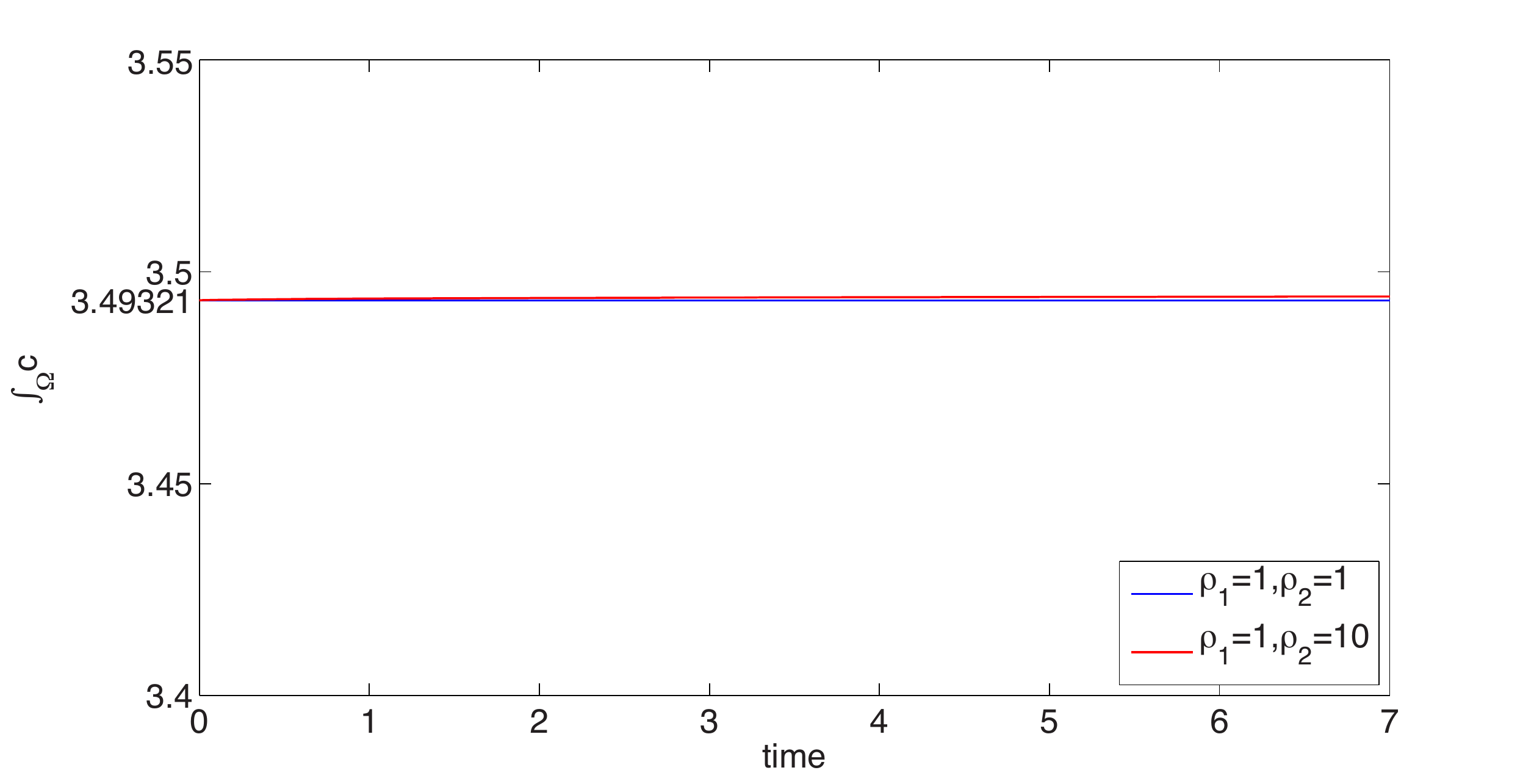}
\hspace{5mm}

  
               \caption{The integral of the volume fraction $c$ for kissing drop example in Fig.{\ref{figkiss}}.}\label{fig--volume}
\end{figure}

\begin{figure}
\hspace{5mm}
\centering
                \includegraphics[width=0.6\textwidth]{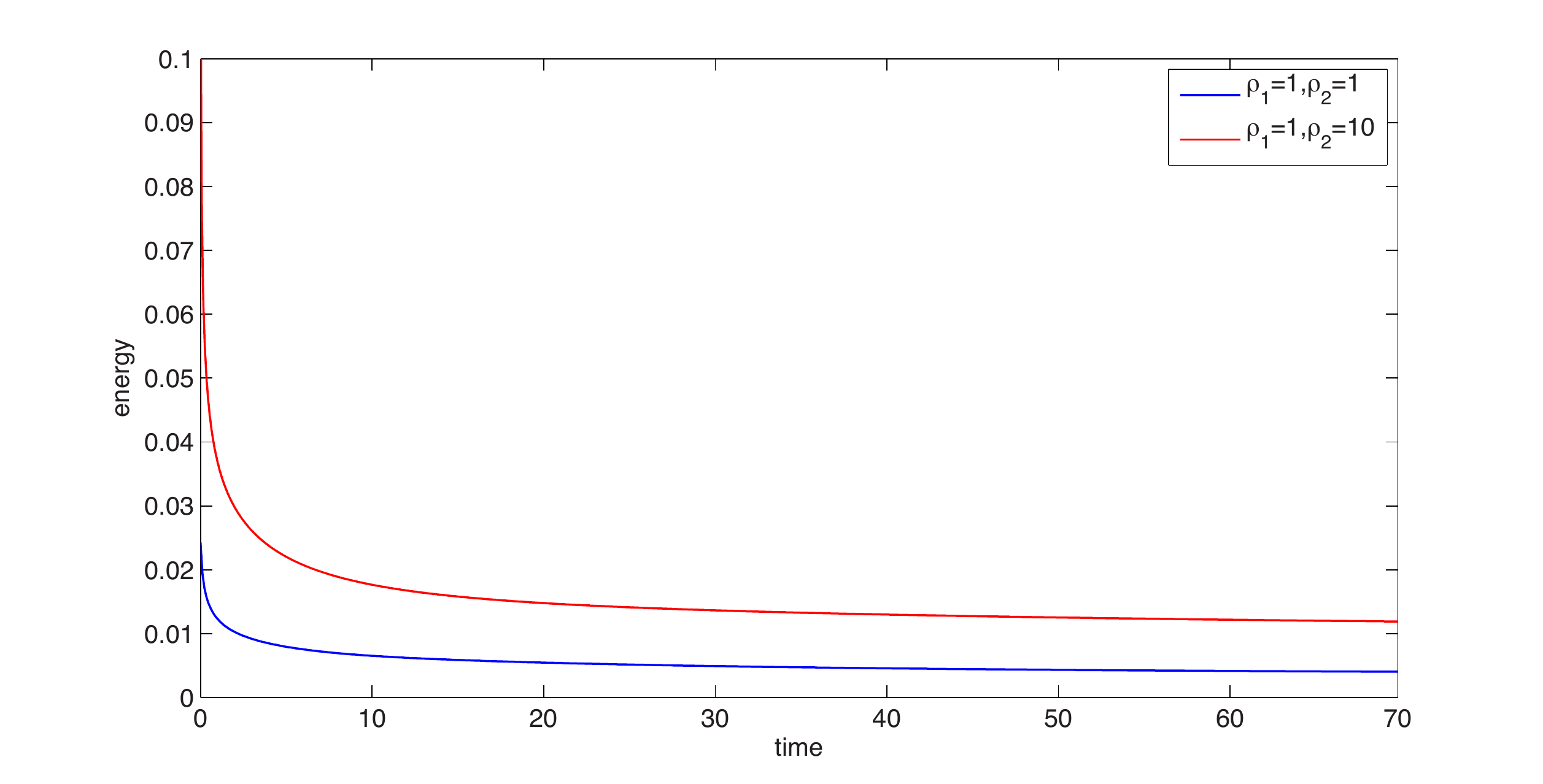}
                \caption{The energy for both cases in Fig.{\ref{figkiss}.}}
                \label{fig--enlaw}
 
\hspace{5mm}

  
\label{figenlaw}
\end{figure}
~\\
$\bold{Example}$ $\bold{6.2}$  We now consider a drop of lighter fluid rising in a heavier medium with different density ratios: (a) 1 : 2 ($\rho_1=1$, $\rho_2=2$) and (b) 1 : 50 ($\rho_1=1$, $\rho_2=50$).  Moreover we set
\begin{center}
\begin{tabular}{lll}
\toprule
$\epsilon=0.01$,& \hspace{10mm}$C=200\epsilon^{2},$\hspace{10mm}$M=1/(20\epsilon),$\hspace{10mm}$Pe=1000/\epsilon,$\hspace{10mm}$1/Fr^2=10.$\\
\bottomrule
\end{tabular}
\end{center}
And we set the time step to be $\Delta t = 0.001$ in (a) and $\Delta t = 0.00025$ in (b). The initial condition for the phase variable $c$
\begin{align}
&c=\cfrac{1}{2}\operatorname{tanh}\bigg(\cfrac{r-\sqrt{(x-c_x)^2+(y-c_y)^2}}{2\sqrt{2}\epsilon}\bigg)+\cfrac{1}{2}~,
\end{align}
where $r$ is the radius of the drop and $(c_x,c_y)$ is the initial position of the drop centre. Here we set $r=0.2$ and $(c_x,c_y)=(0,-0.6)$.\\
For a drop rising in a liquid, it is often desirable to study the long-term behaviour and the temporal evolution of the interface. The drops  may move a relatively long distance. Thus the computational domain must be correspondingly large which is likely to be computationally
infeasible, and this constraint may therefore prevent the desired long time simulation. A non-inertial reference frame \cite{Hua20088,Hua2008} that moves with the interface can be employed as a remedy for this problem. In this paper, we enlarge the effect of the gravity as an alternative, so that the dramatic deformation can be detected in a relatively short time period. $1/Fr^2$ is therefore set to be 10 in this example, whereas in \cite{Kim2005}, $1/Fr^2=1$.\\
The square domain $\Omega=[-1,1]\times[-1,1]$ is discretized using an adaptive mesh, where we use a variable metric/Delaunay automatic meshing algorithm built in the FreeFem++ platform.  
Since the values of phase variable $c$ can change rapidly but continuously across the interface, the mesh is adapted according to the value of $|\nabla c|$ after every 10 time steps, in order to track the features of the phase variable as the computation progresses. The shortest edge of all the grids is set to be $1/128<\epsilon$, so that at least one grid cell is located across the interface to ensure accuracy. An example of the dynamic adaptive mesh can be seen in Fig.{\ref{fig--mesh}}.\\
Snapshots of the deformed interface (black solid line) together with the $\nabla \cdot \bold{u}$ are presented in Fig. \ref{fig--rising1to2--divv} and Fig. \ref{fig--rising1to50--divv}, where we observe that the drop with a low density ratio deforms slowly, resulting in a mushroom shape by time $t=1.4$. On the other hand, the high density ratio drop rises much faster and has deformed into a mushroom shape by the time $t=0.225$. Our results are similar to those found in \cite{Sussman1994} when the effects of density ratio in rising drops were simulated using a level-set method. In the large density ratio case, the upwelling motion continues to deform the drops which attain a horseshoe shape around $t=0.4$. As the drop broadens, we observe drop pinch-off around $t=0.85$, as the tips of the drop roll up and smaller drops eventually detach.\\
The divergence of velocity $\nabla \cdot \bold{u}$ (color representing its value) are presented in Fig. \ref{fig--rising1to2--divv} and Fig. \ref{fig--rising1to50--divv}. Recall that the divergence-free condition does not hold for quasi-incompressible fluids with different densities because the fluids may mix slightly across the interface. The two incompressible fluids can be compressible across the interface where the two components are mixed. It can be observed that $\nabla \cdot \bold{u}=0$, such that the fluid is incompressible almost everywhere except along the moving interface. Near the interface, waves of expansion ($\nabla \cdot \bold{u}>0$) and compression ($\nabla \cdot \bold{u}<0$) are observed. Note that in the case of the larger density ratio, the distribution of $\nabla \cdot \bold{u}>0$ is found to be larger and more spatially localized than in the case with smaller density ratio. Further, the compression and expansion waves tend to trail the drop in the large density ratio case whereas in the smaller density ratio case, the waves tend to be ahead of the drop. The divergence free condition is satisfied in the bulk regions for each component during the pinch-off (from $t=0.85$ to $t=0.915$), where the distribution of $\nabla \cdot \bold{u}$ with non-zero value tends to appear around three drops. We expect that a smaller value for parameter $C$ and corresponding finer mesh grids causes a narrower distribution of non-zero values of $\nabla \cdot \bold{u}$ around the drop boundary.\\
The flow field and the vorticity contours are presented in Fig.\ref{fig--risdrop1to2--vor}-Fig.\ref{fig--risdrop1to50--vor2}, where we can observe that the velocity field for the second case with the lager density ratio are much more enhanced than that for the smaller matched case, with strong spinning motions of the fluid distributed across the drop boundary. Both cases have vorticity contours along the wall and the boundary of the drop, where the positive value stands for the clockwise rotation and the negative for the conterclockwise rotation. The rising of the drop seems to affect the shape of the streamwise vortices, where the concentration of vorticity becomes more apparent near the rear as the drop rises.\\
For the second case with larger density ratio, the vorticity contours are much more condensed with the absolute value being increased significantly. At $t=0.5$, we can observe that, as the drop broadens, a pair of co-rotating streamwise vortices begin to appear at the rounded bottom of the drop. They gain strength as they grow along the bottom boundary, where, on the other hand, another pair of vortices that are associated with the rotation of the elongating trailing arms of the drops tends to diminish the dominance. It is interesting to note that, during the pinch-off (from $t=0.85$ to $t=0.915$), two pairs of co-rotating vortices seem to balance and both concentrate near the rear of the larger drop.\\
The adaptive meshes for the rising drop of density ratio 1 : 50 are presented in Fig.{\ref{fig--mesh}}, where a grid structure that adapts to the locations of the moving interfaces is generated automatically and adapts around the isolated drops after break-up.\\
\begin{figure}
\hspace{-8mm}
        \begin{subfigure}[h]{0.275\textwidth}
                \includegraphics[width=\textwidth]{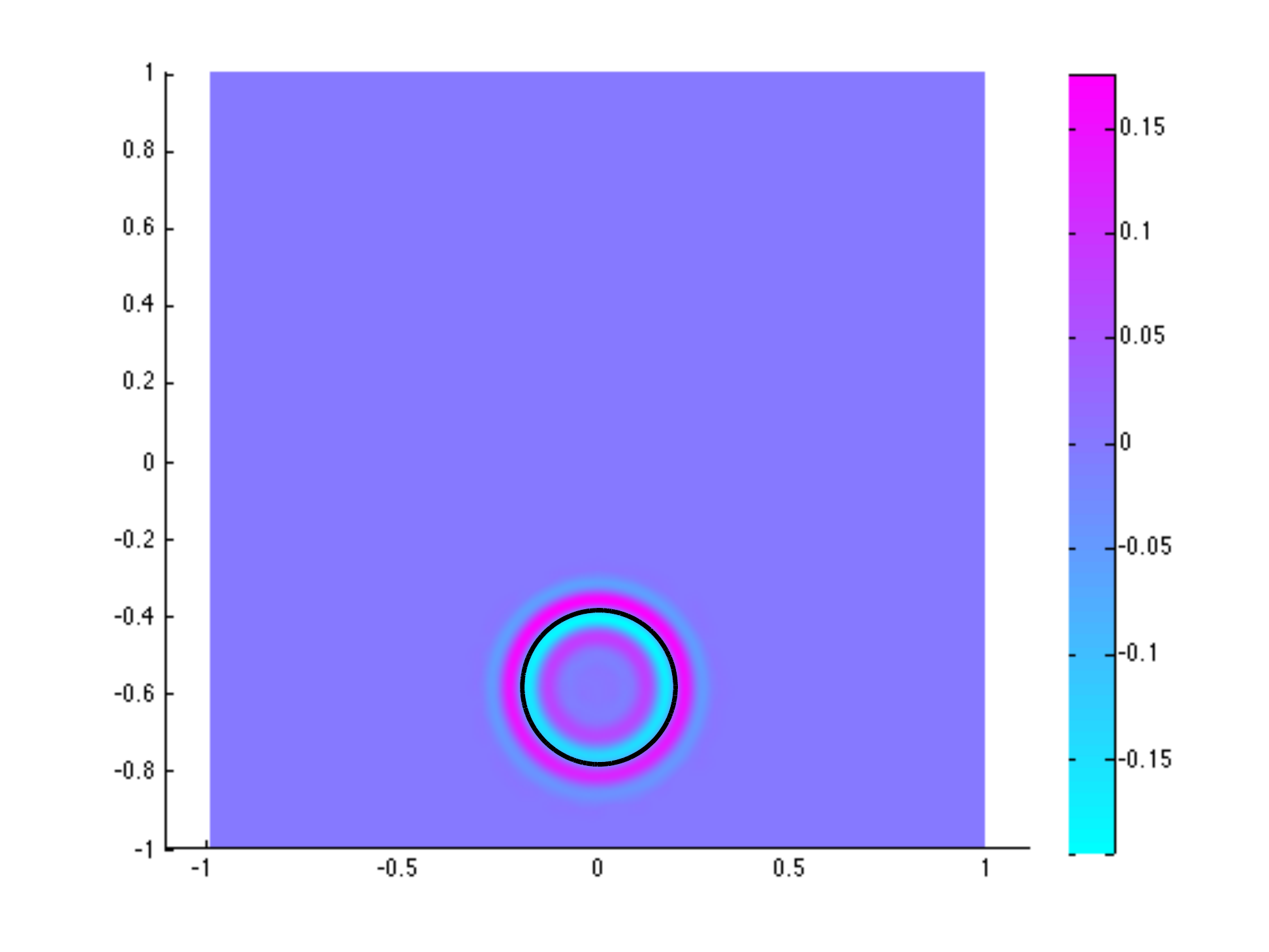}
                \caption*{t=0}
        \end{subfigure}%
\hspace{-4mm}
        \begin{subfigure}[h]{0.275\textwidth}
                \includegraphics[width=\textwidth]{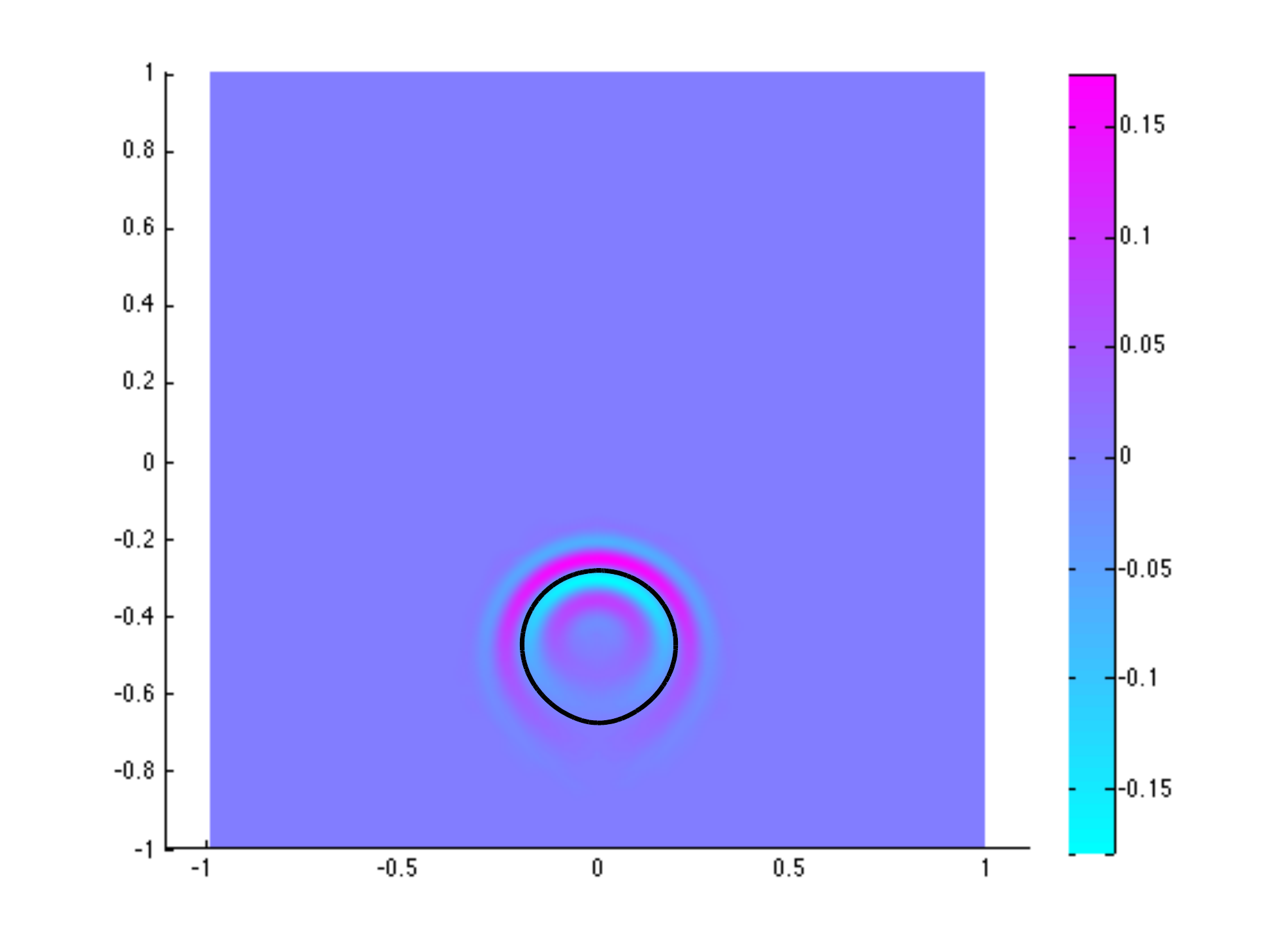}
                \caption*{t=0.4}
        \end{subfigure}%
\hspace{-4mm}
        \begin{subfigure}[h]{0.275\textwidth}
                \includegraphics[width=\textwidth]{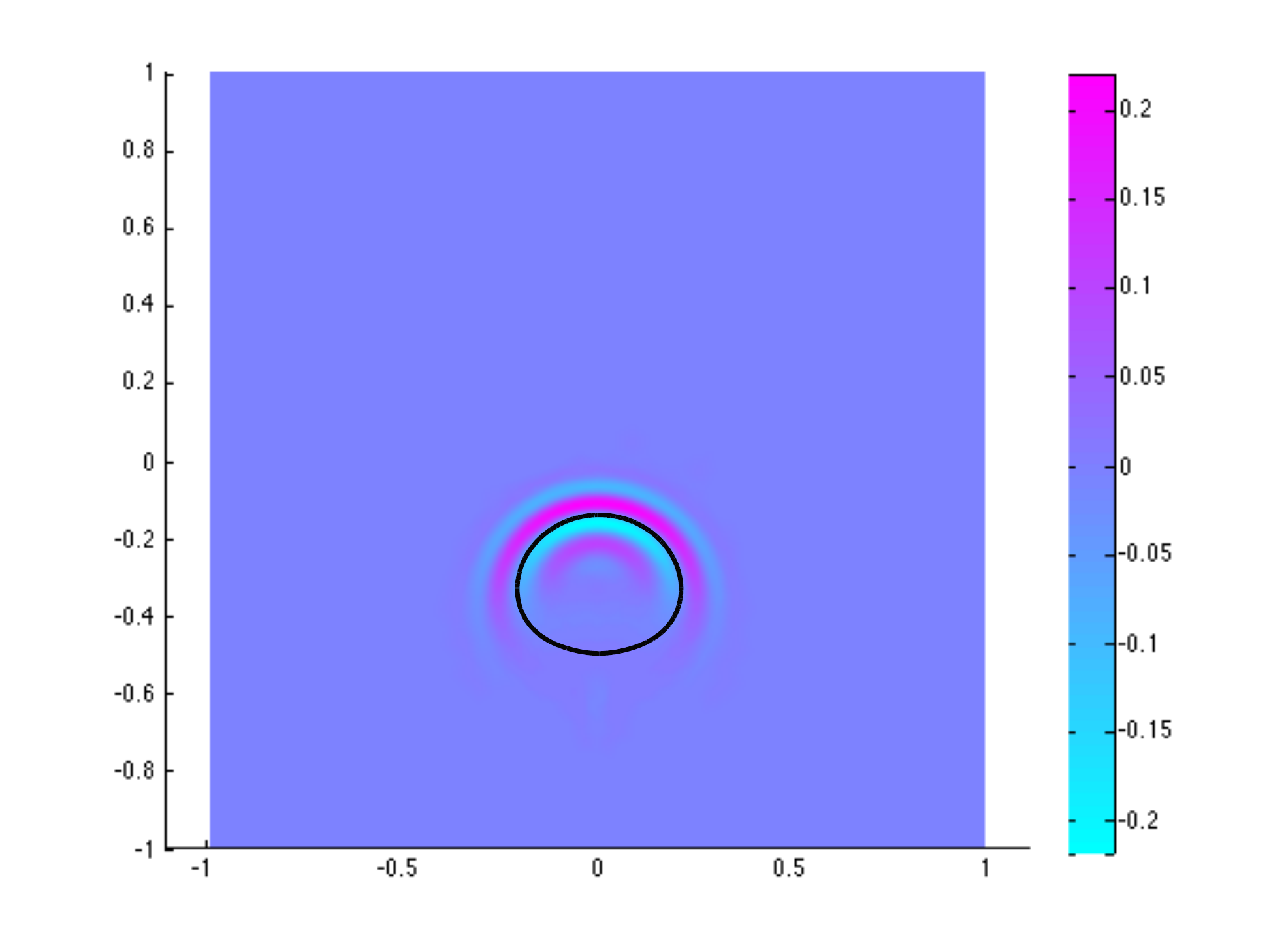}
                \caption*{t=0.7}
        \end{subfigure}%
\hspace{-4mm}
        \begin{subfigure}[h]{0.275\textwidth}
                \includegraphics[width=\textwidth]{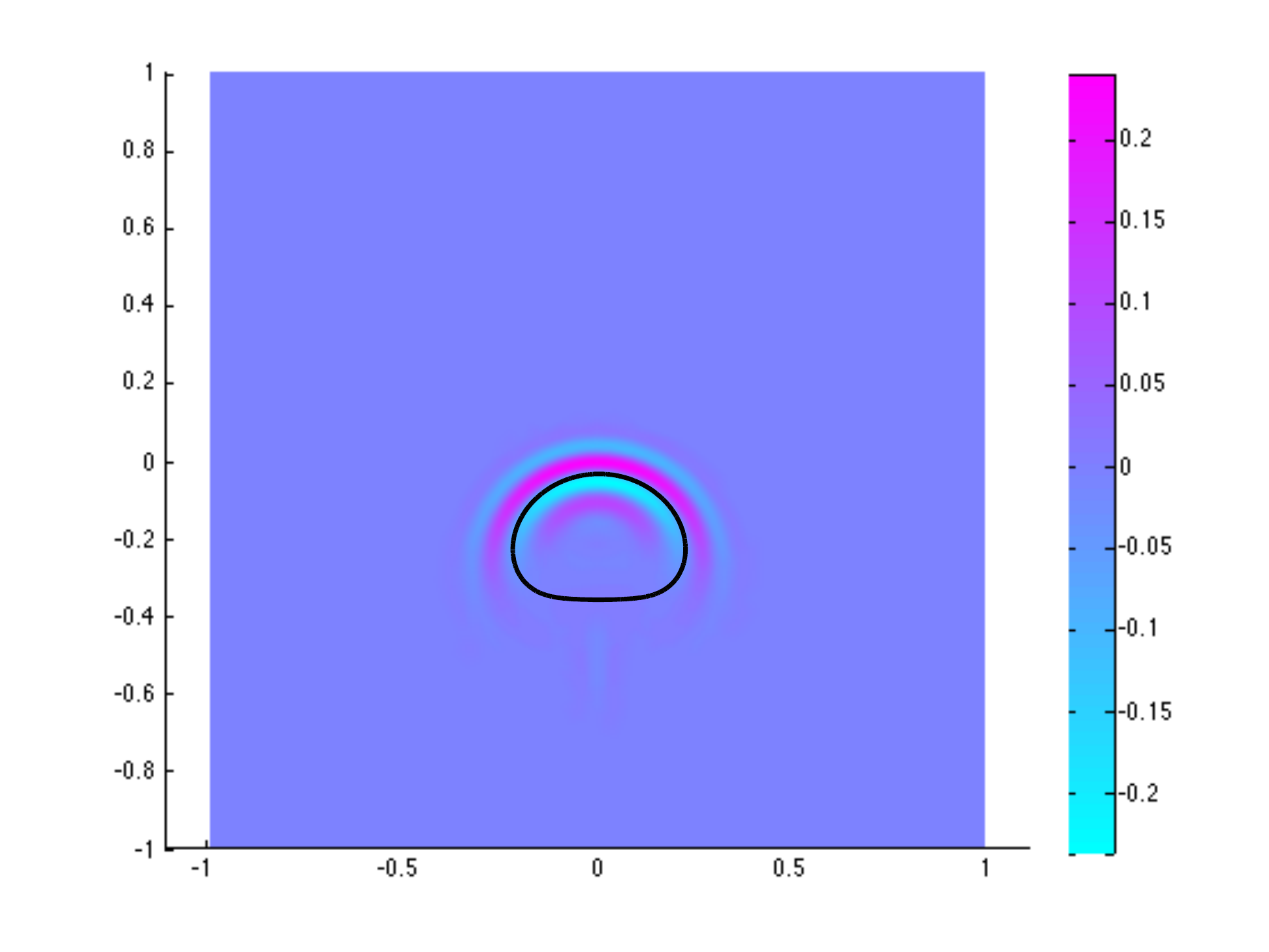}
                \caption*{t=0.9}
        \end{subfigure}
        \caption*{}
        \vspace{-8mm}
\hspace{-8mm}
        \begin{subfigure}[h]{0.275\textwidth}
                \includegraphics[width=\textwidth]{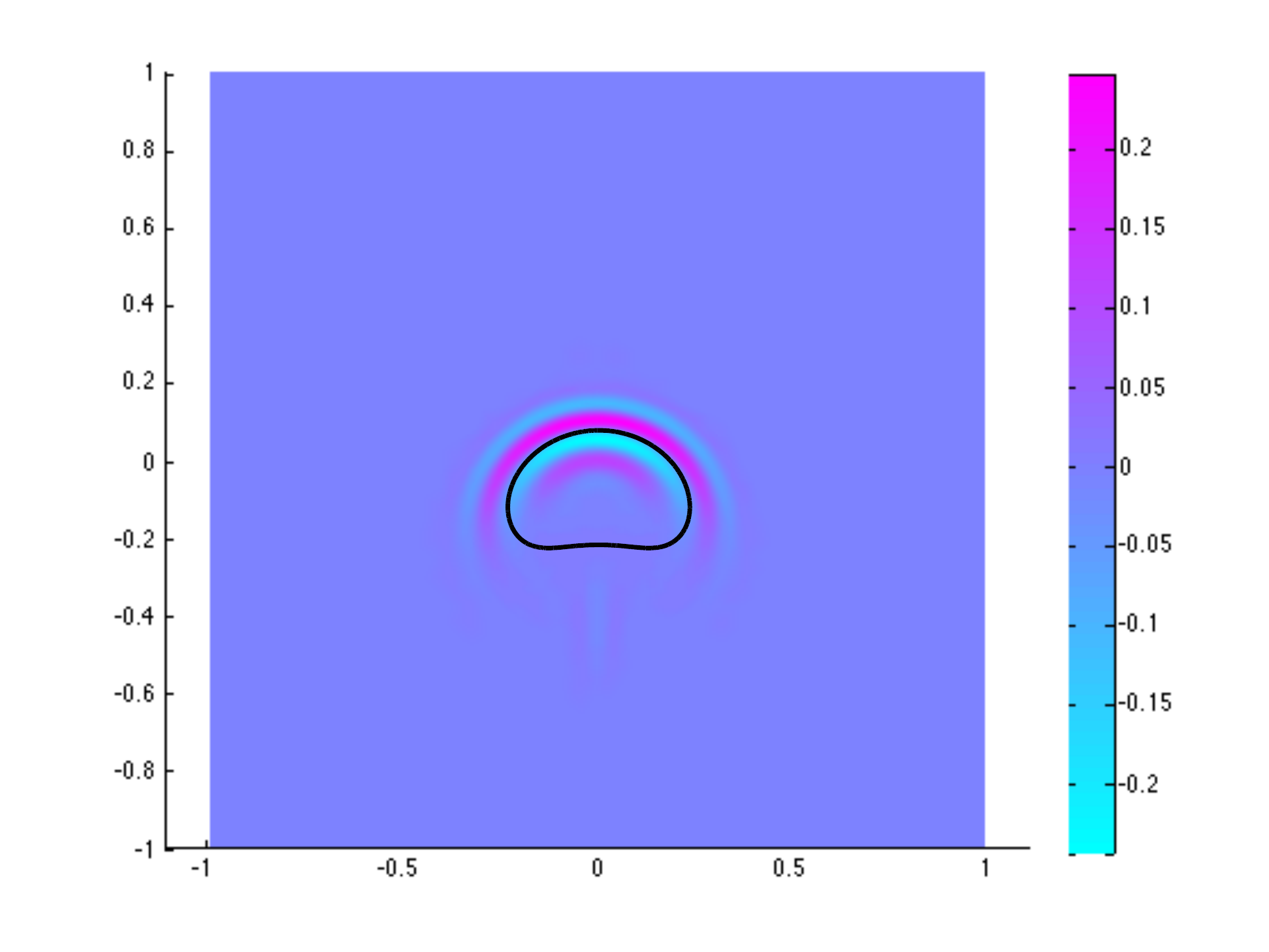}
                \caption*{t=1.1}
        \end{subfigure}%
\hspace{-4mm}
        \begin{subfigure}[h]{0.275\textwidth}
                \includegraphics[width=\textwidth]{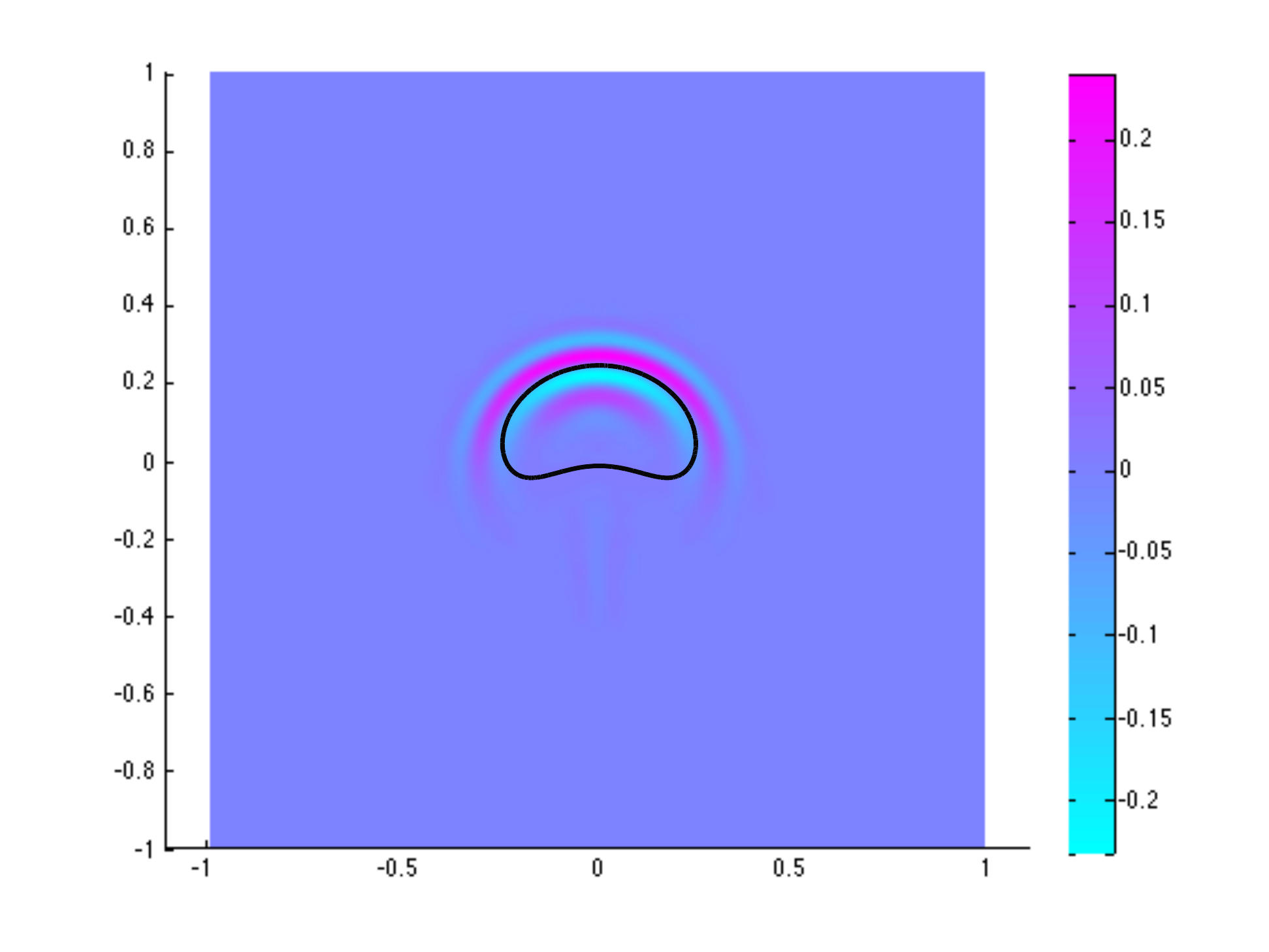}
                \caption*{t=1.4}
        \end{subfigure}%
\hspace{-4mm}
        \begin{subfigure}[h]{0.275\textwidth}
               \includegraphics[width=\textwidth]{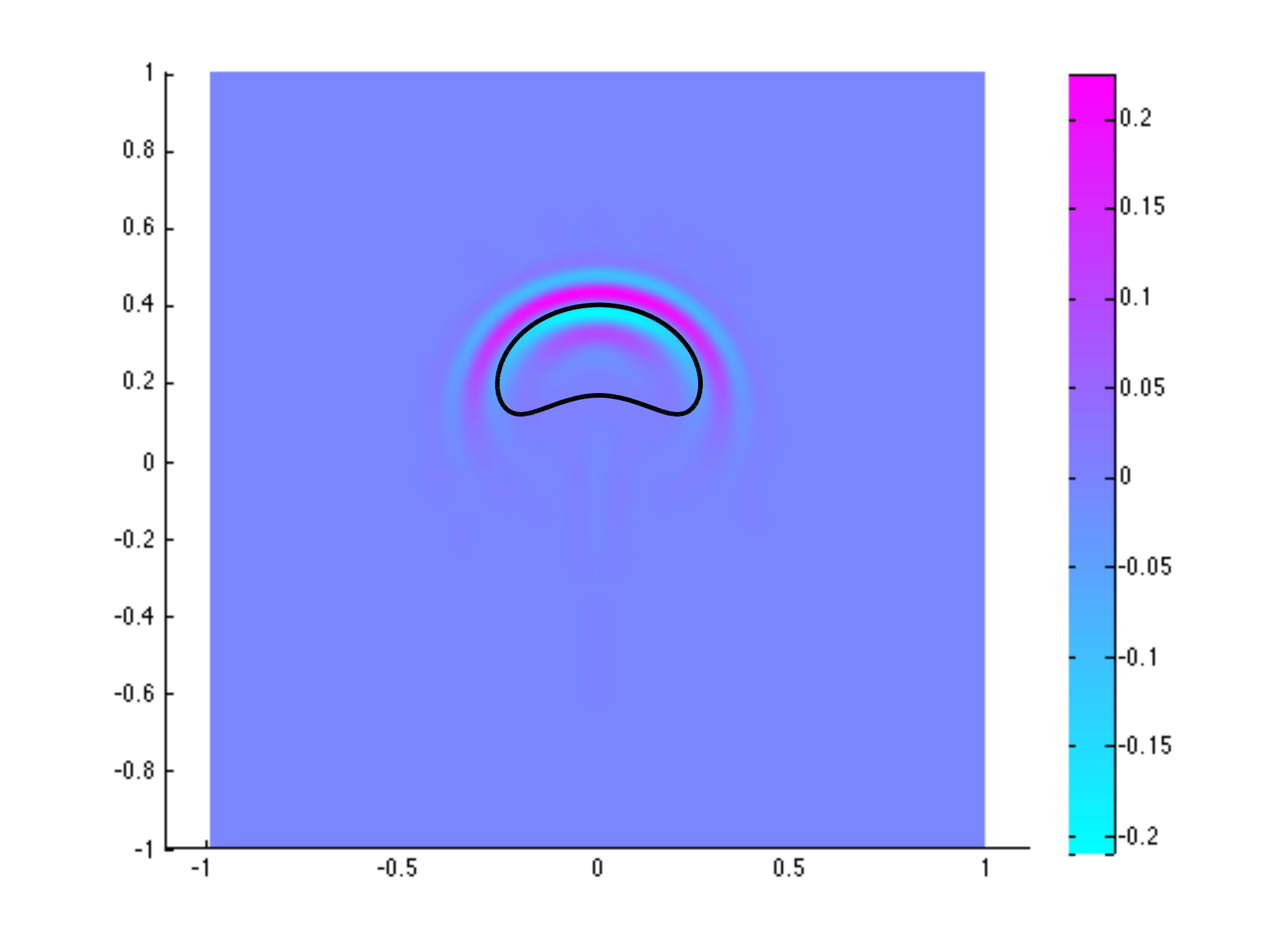}
                \caption*{t=1.7}
        \end{subfigure}%
\hspace{-4mm}
        \begin{subfigure}[h]{0.275\textwidth}
                \includegraphics[width=\textwidth]{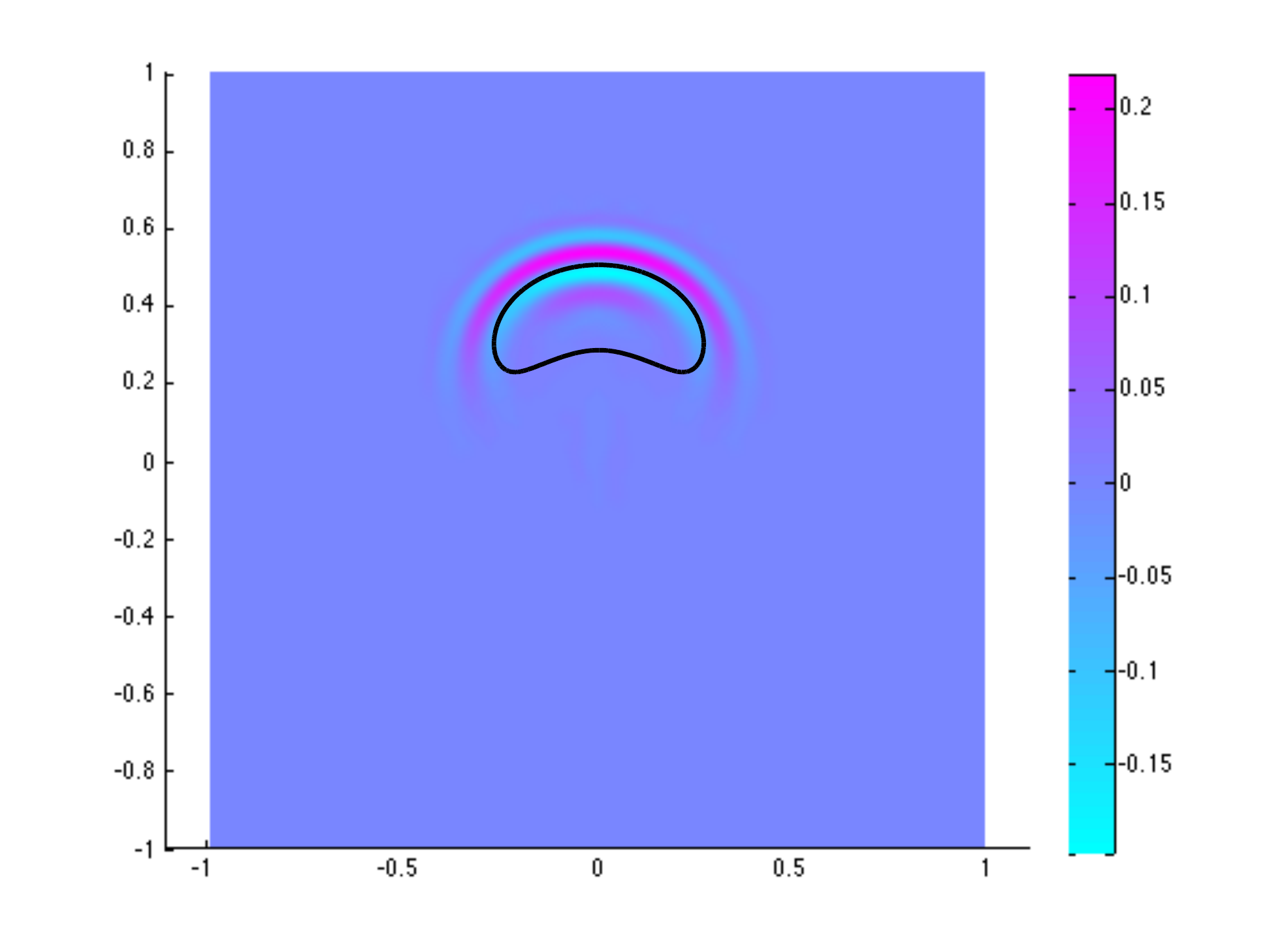}
                \caption*{t=1.9}
        \end{subfigure}%
                 \caption{The drop interfaces (black solid line) with $\nabla \cdot \bold{u}$ (with color representing the value distribution) for the rising drop with density ratio 1 : 2, $\rho_1=1, \rho_2=50$, $\epsilon=0.01$, $C=200\epsilon^{2}$, $M=1/(20\epsilon)$, $Pe=1000/\epsilon$, $1/Fr^2=10$ and $\Delta t = 0.00025$.}\label{fig--rising1to2--divv}
\end{figure}
\begin{figure}
\hspace{-8mm}
        \begin{subfigure}[h]{0.275\textwidth}
                \centering
                \includegraphics[width=\textwidth]{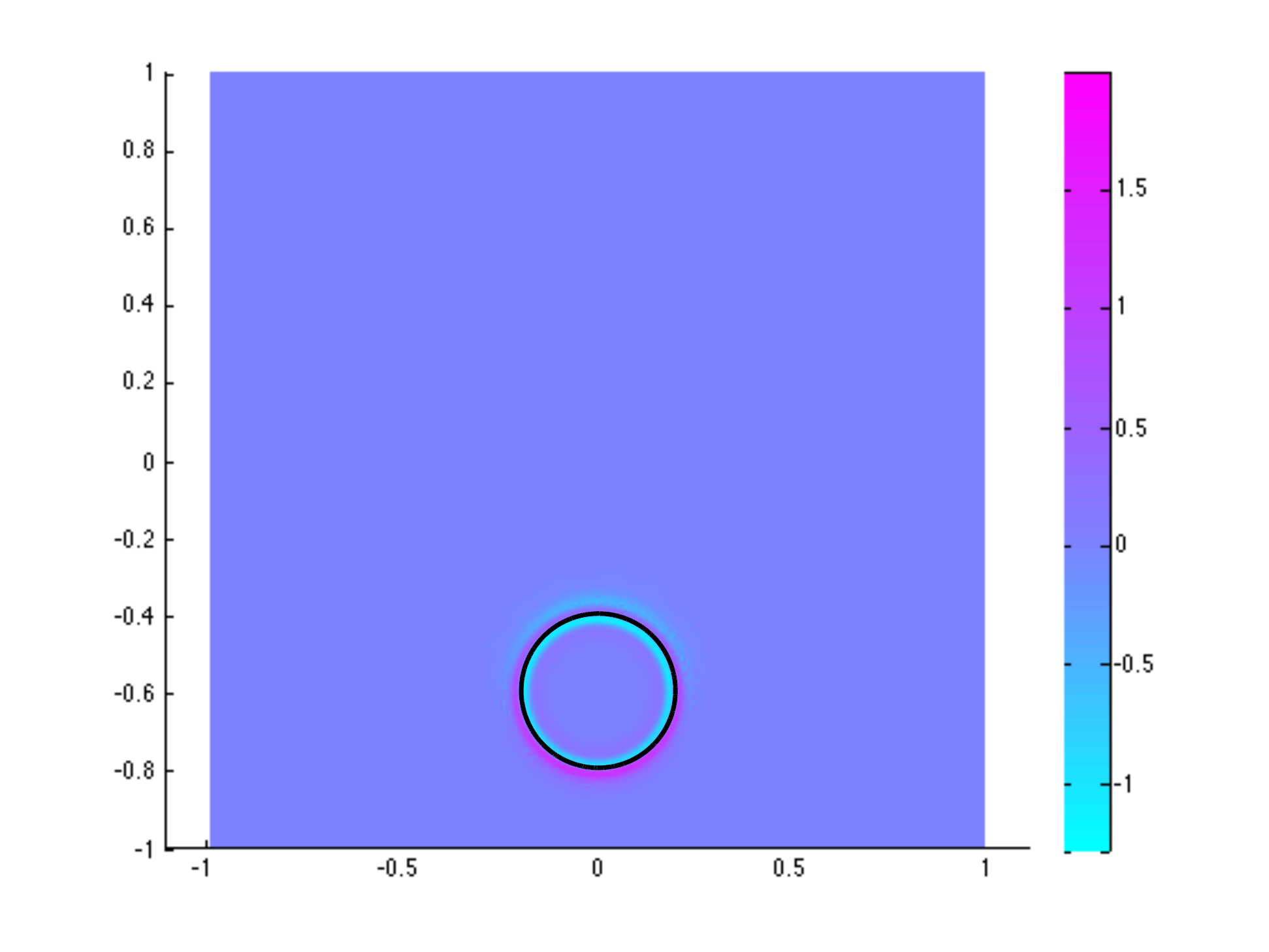}
                \caption*{t=0}
        \end{subfigure}%
\hspace{-4mm}
        \begin{subfigure}[h]{0.275\textwidth}
                \centering
                \includegraphics[width=\textwidth]{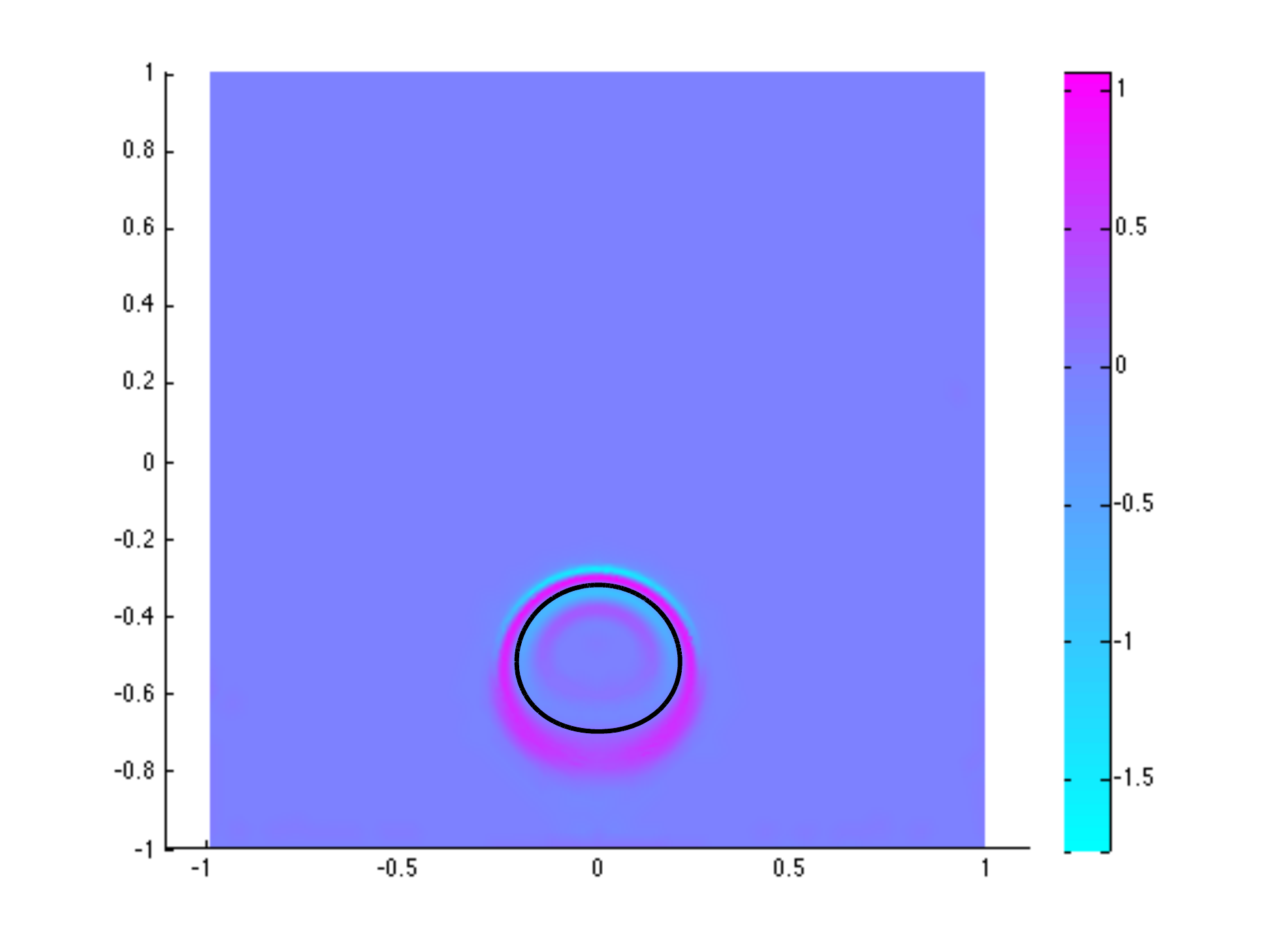}
                \caption*{t=0.125}
        \end{subfigure}%
\hspace{-4mm}
        \begin{subfigure}[h]{0.275\textwidth}
                \centering
                \includegraphics[width=\textwidth]{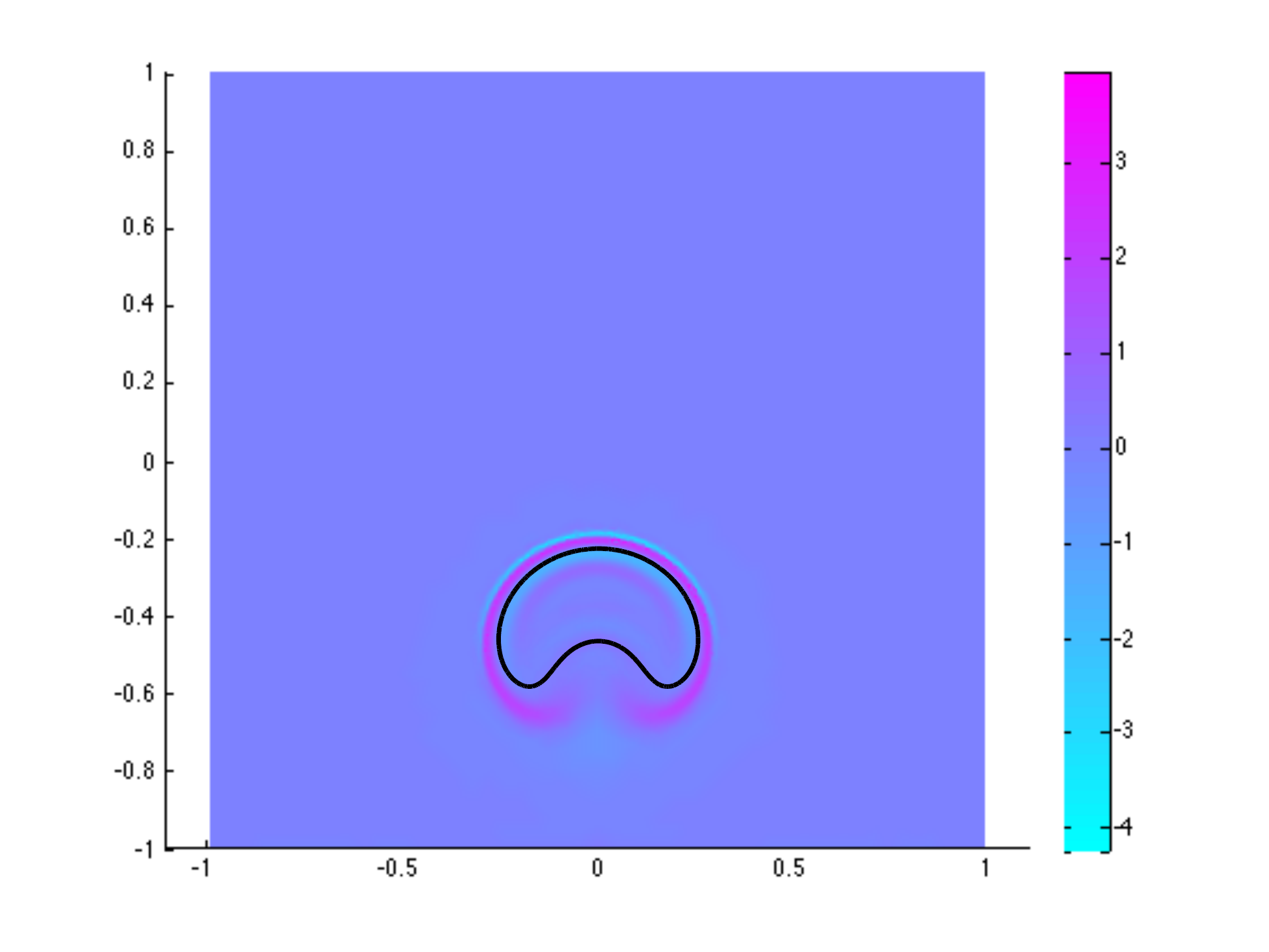}
                \caption*{t=0.225}
        \end{subfigure}%
\hspace{-4mm}
        \begin{subfigure}[h]{0.275\textwidth}
                \centering
                \includegraphics[width=\textwidth]{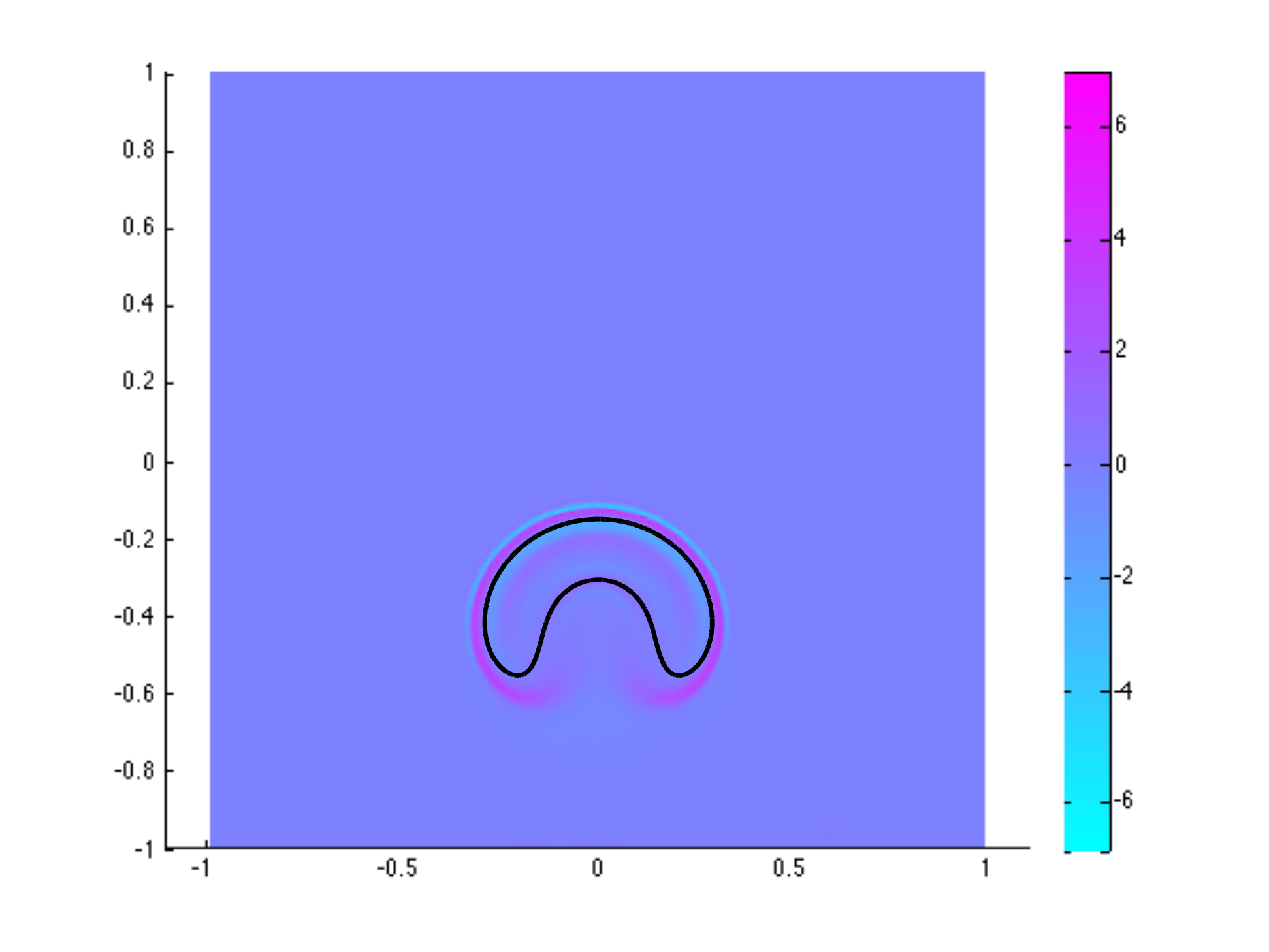}
                \caption*{t=0.3}
        \end{subfigure}
                \caption*{}
        \vspace{-8mm}
        \hspace{-8mm}
        \begin{subfigure}[h]{0.275\textwidth}
                \centering
                \includegraphics[width=\textwidth]{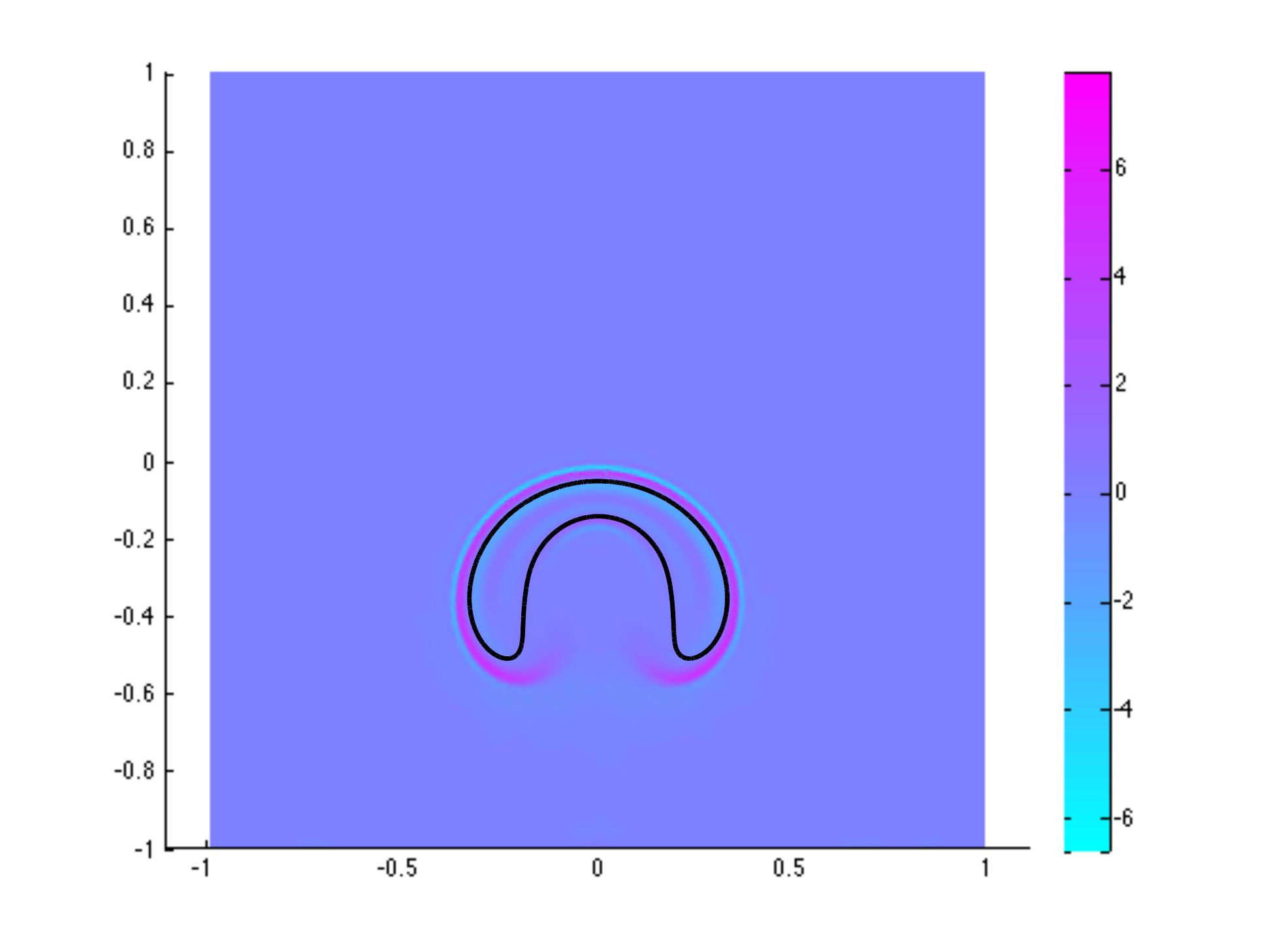}
                \caption*{t=0.4}
        \end{subfigure}%
\hspace{-4.5mm}
        \begin{subfigure}[h]{0.275\textwidth}
                \centering
                \includegraphics[width=\textwidth]{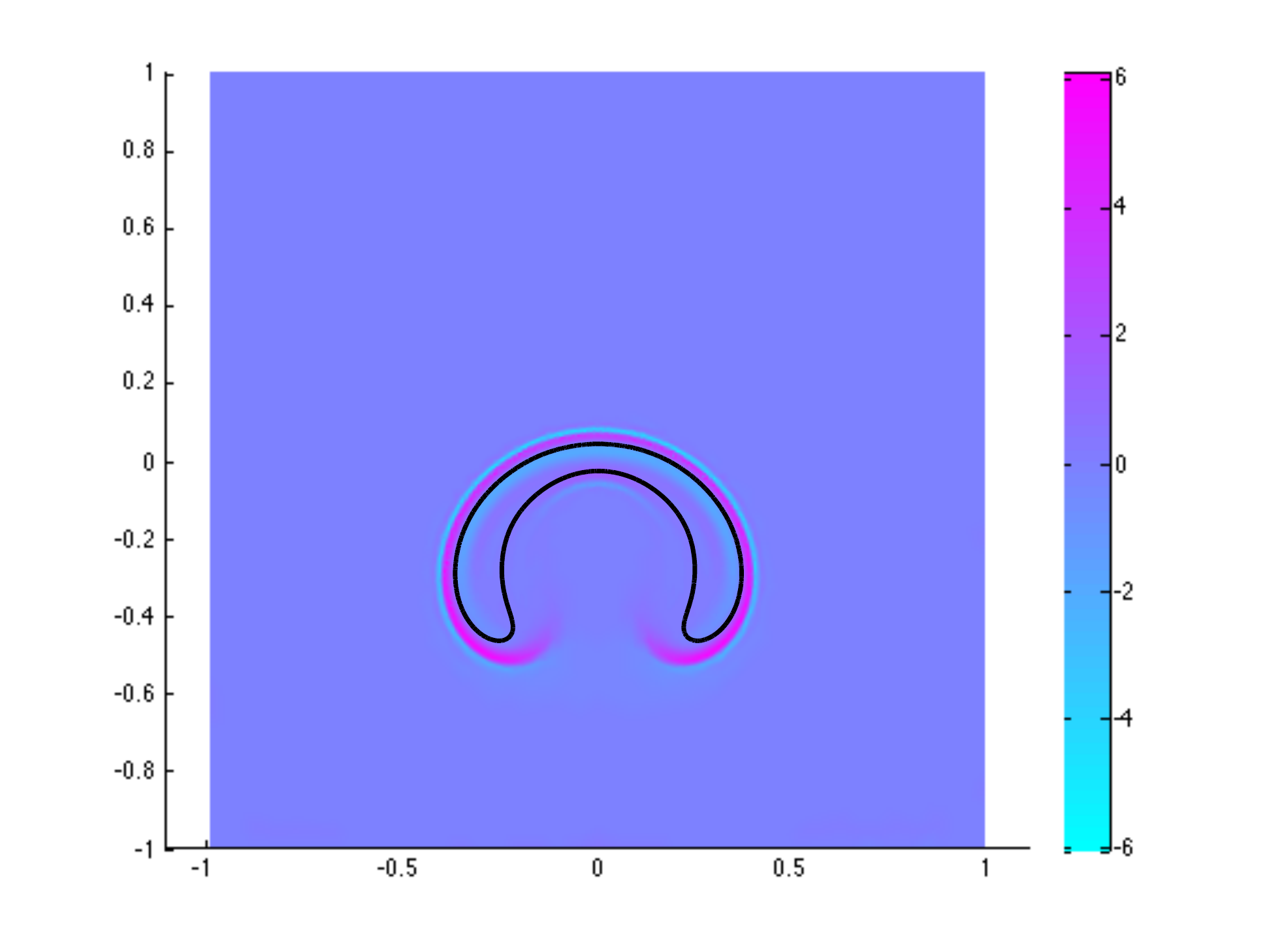}
                \caption*{t=0.5}
        \end{subfigure}%
\hspace{-4.5mm}
        \begin{subfigure}[h]{0.275\textwidth}
                \centering
                \includegraphics[width=\textwidth]{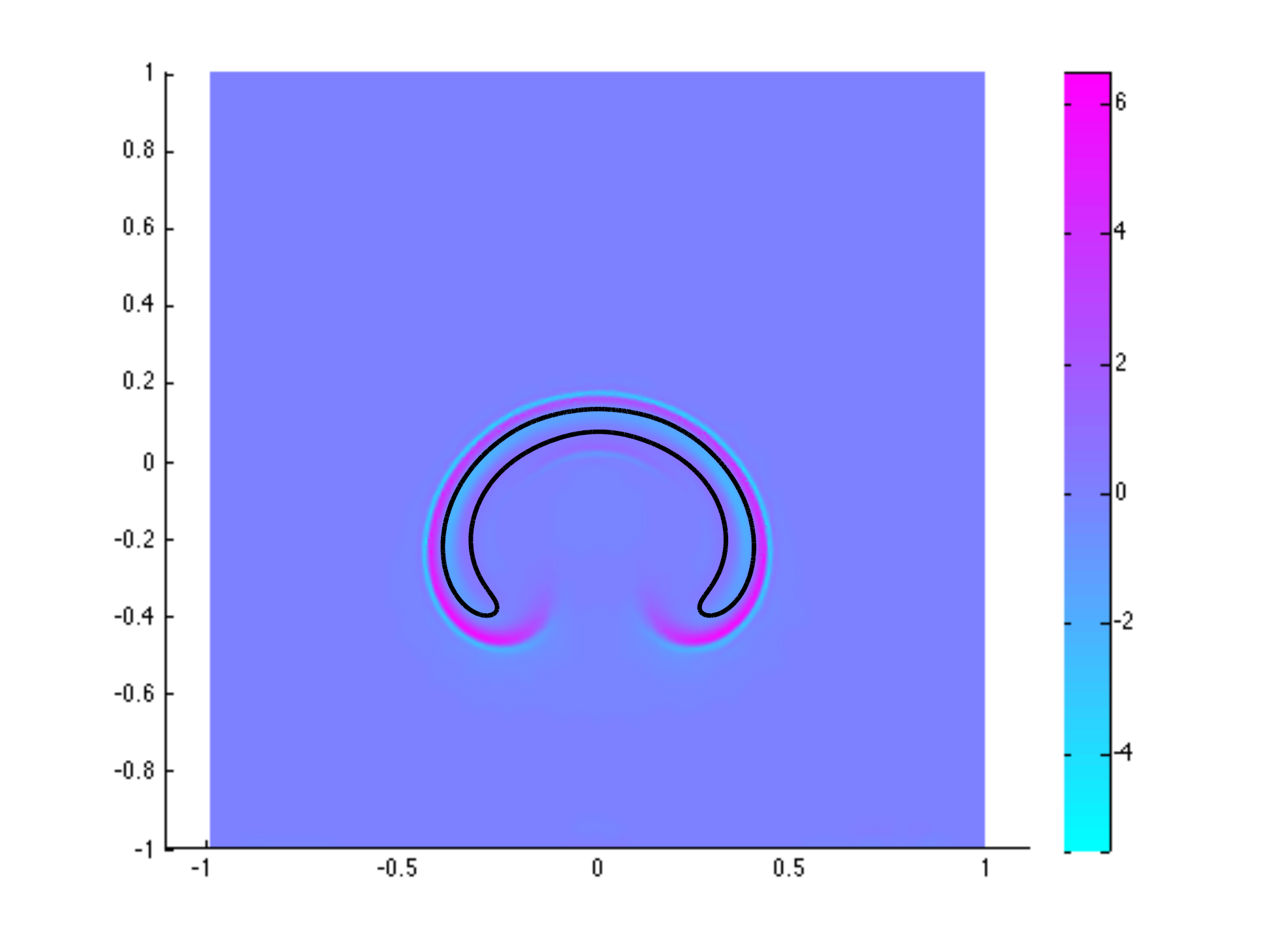}
                \caption*{t=0.6}
        \end{subfigure}
\hspace{-4.5mm}
        \begin{subfigure}[h]{0.275\textwidth}
                \centering
                \includegraphics[width=\textwidth]{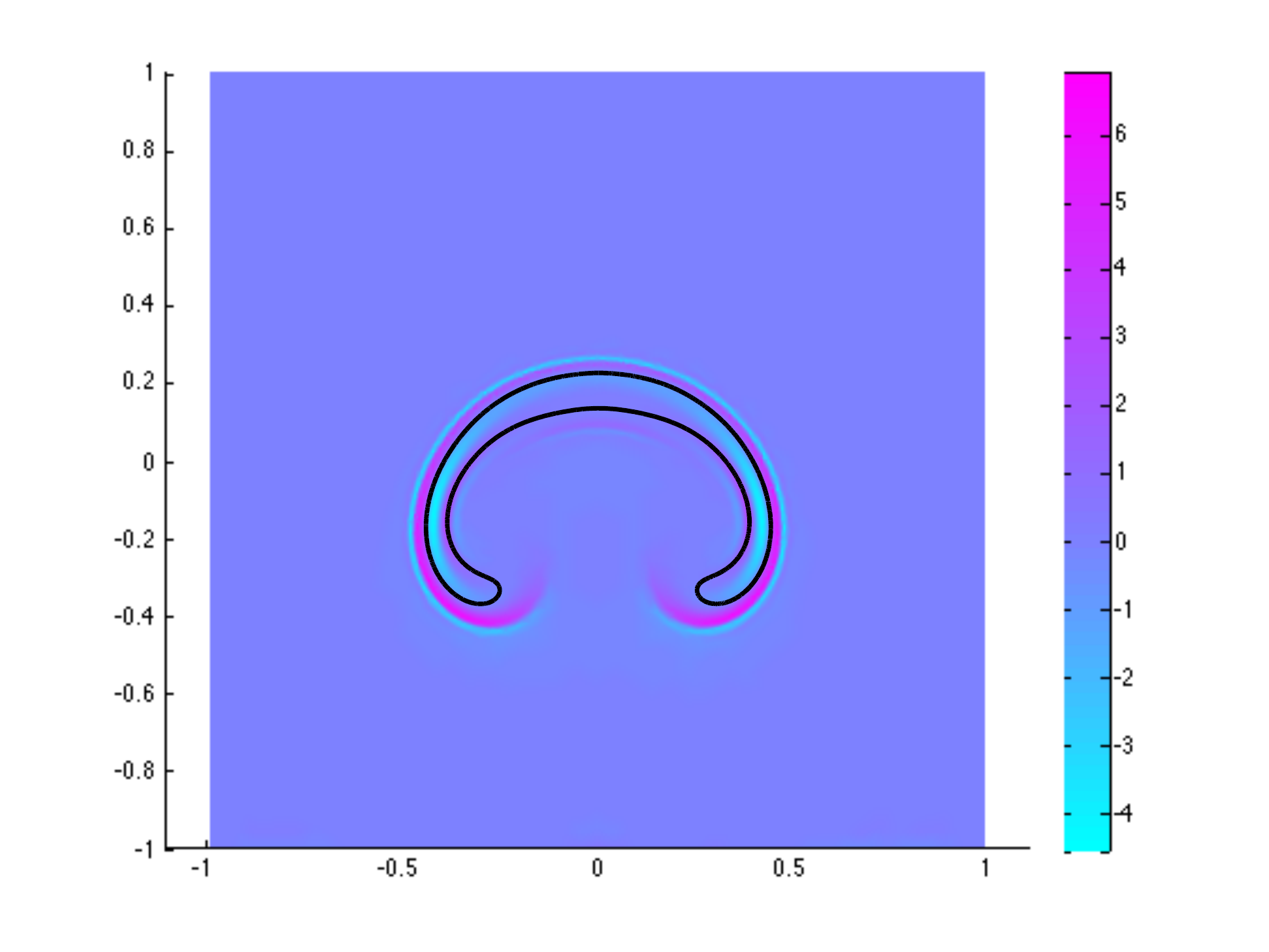}
                \caption*{t=0.7}
        \end{subfigure}
                \caption*{}
        \vspace{-8mm}
\hspace{-8mm}
        \begin{subfigure}[h]{0.275\textwidth}
                \centering
                \includegraphics[width=\textwidth]{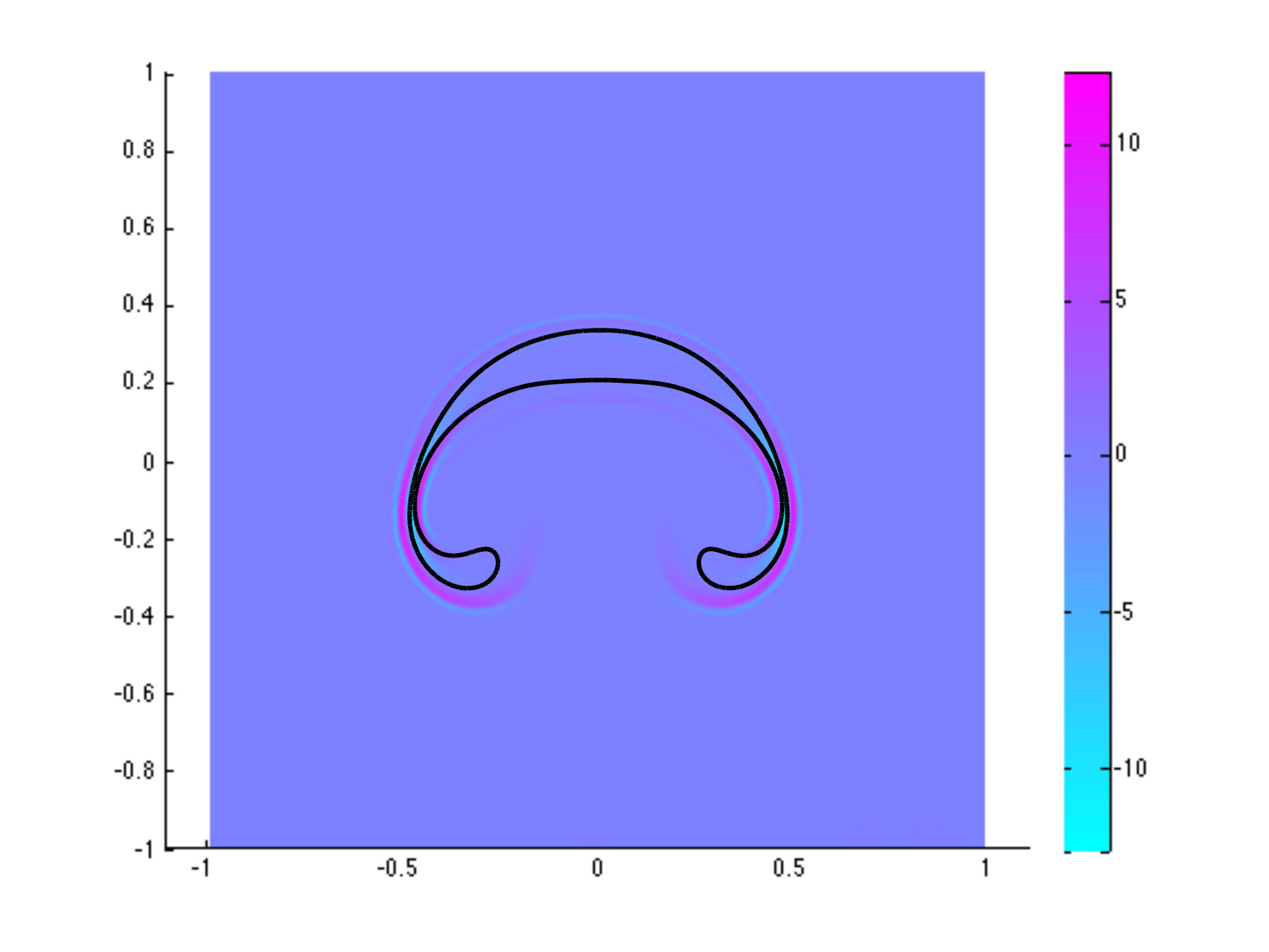}
                \caption*{t=0.825}
        \end{subfigure}%
\hspace{-4mm}
        \begin{subfigure}[h]{0.275\textwidth}
                \centering
                \includegraphics[width=\textwidth]{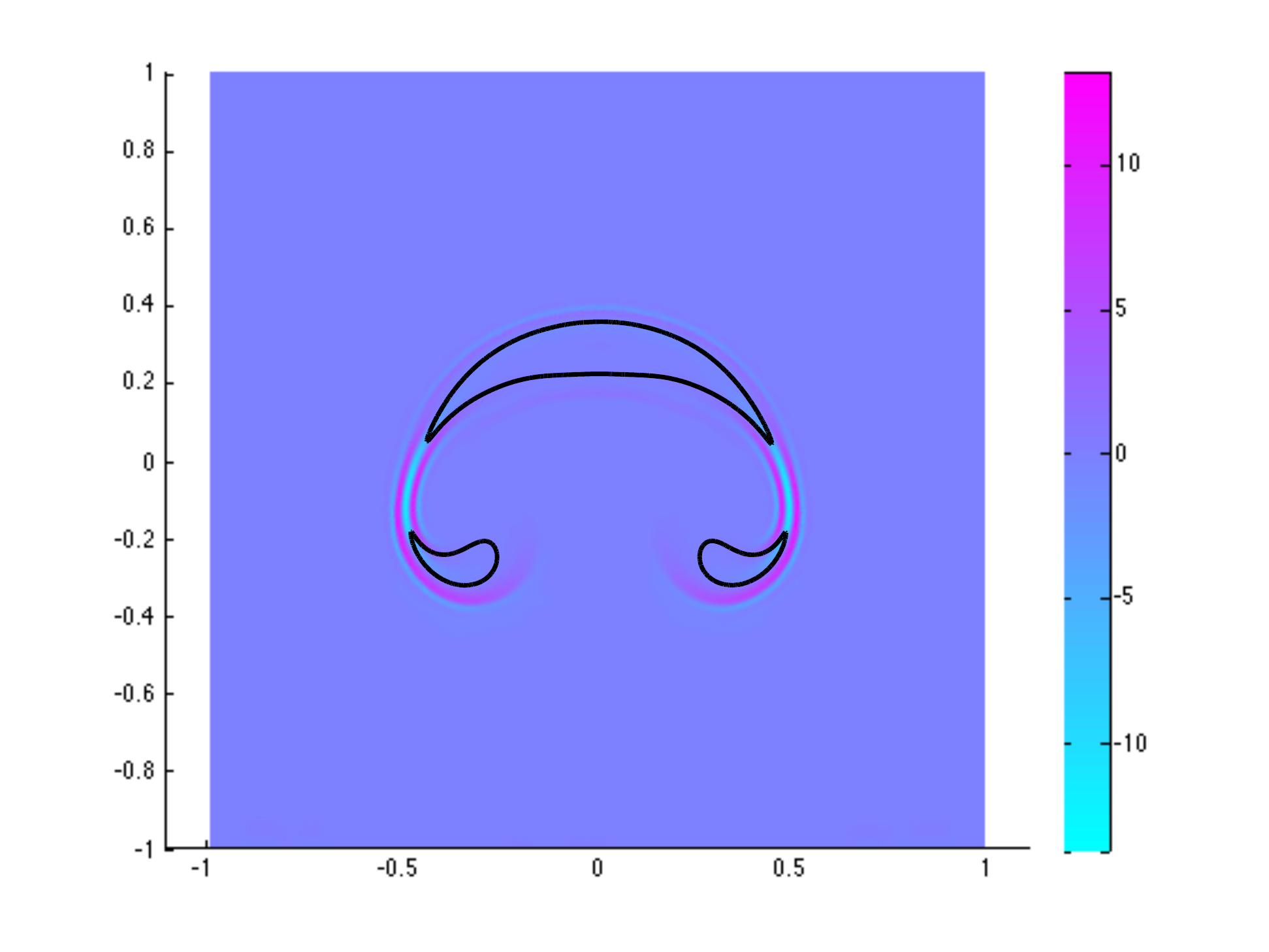}
                \caption*{t=0.85}
        \end{subfigure}%
\hspace{-4mm}
        \begin{subfigure}[h]{0.275\textwidth}
                \centering
                \includegraphics[width=\textwidth]{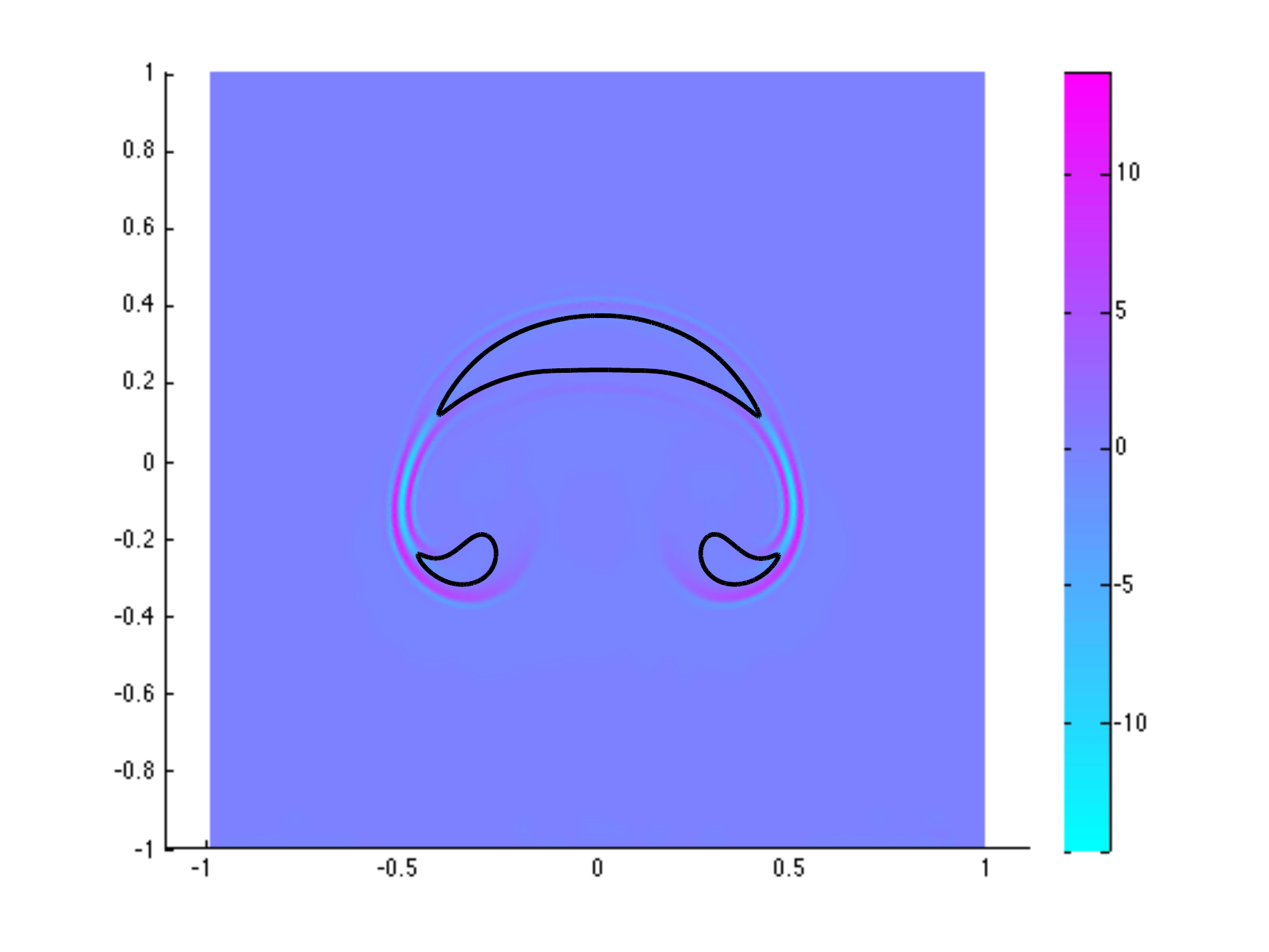}
                \caption*{t=0.875}
        \end{subfigure}%
\hspace{-4mm}
        \begin{subfigure}[h]{0.275\textwidth}
                \centering
                \includegraphics[width=\textwidth]{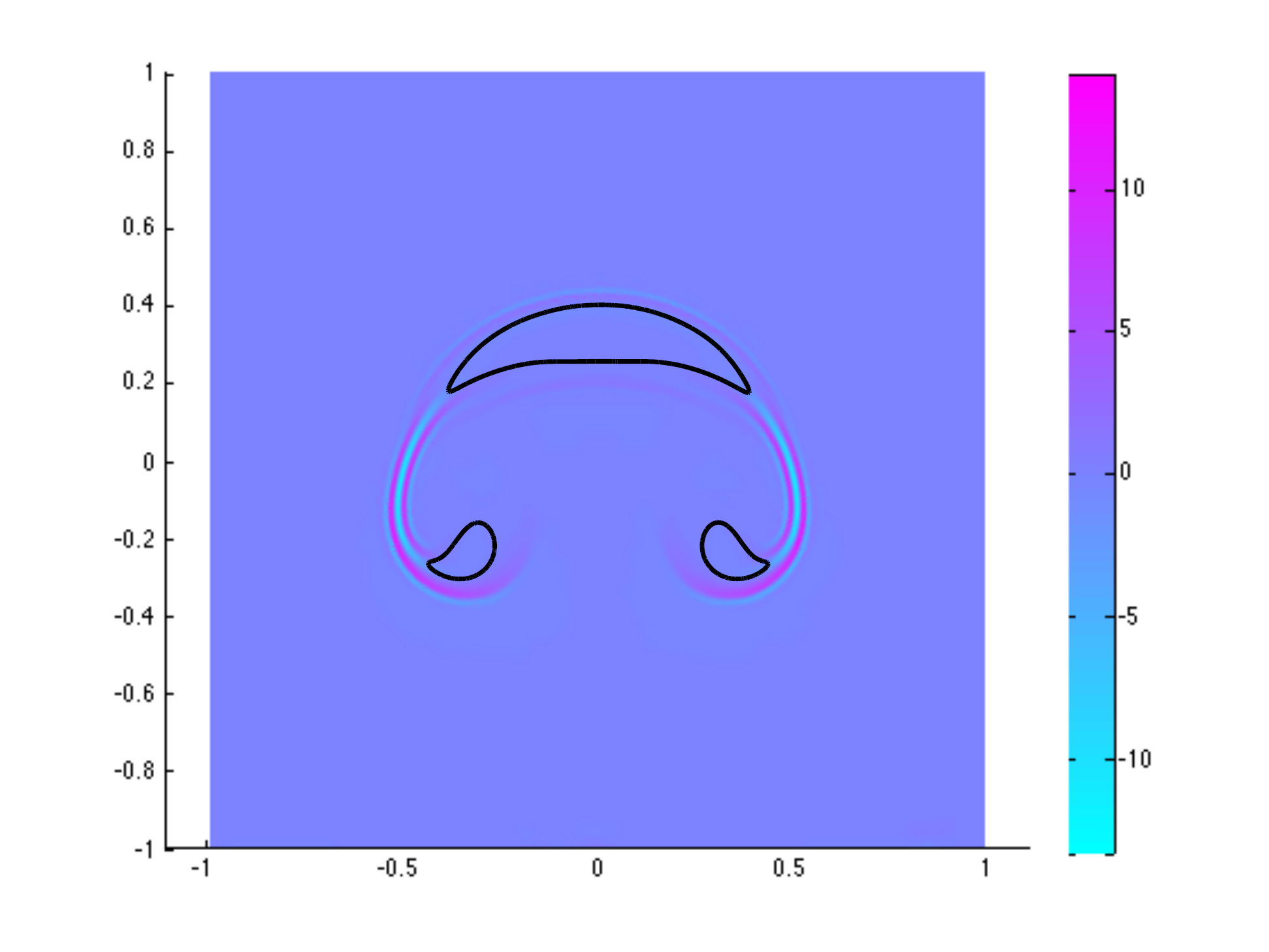}
                \caption*{t=0.915}
        \end{subfigure}%
         \caption{The drop interfaces (black solid line) with $\nabla \cdot \bold{u}$ (with color representing the value distribution) for the rising drop with density ratio 1 : 50, $\rho_1=1, \rho_2=50$, $\epsilon=0.01$, $C=200\epsilon^{2}$, $M=1/(20\epsilon)$, $Pe=1000/\epsilon$, $1/Fr^2=10$ and $\Delta t = 0.00025$.}\label{fig--rising1to50--divv}
\end{figure}
\begin{figure}
\hspace{-1.5mm}
        \begin{subfigure}[h]{0.5\textwidth}
               \includegraphics[width=\textwidth]{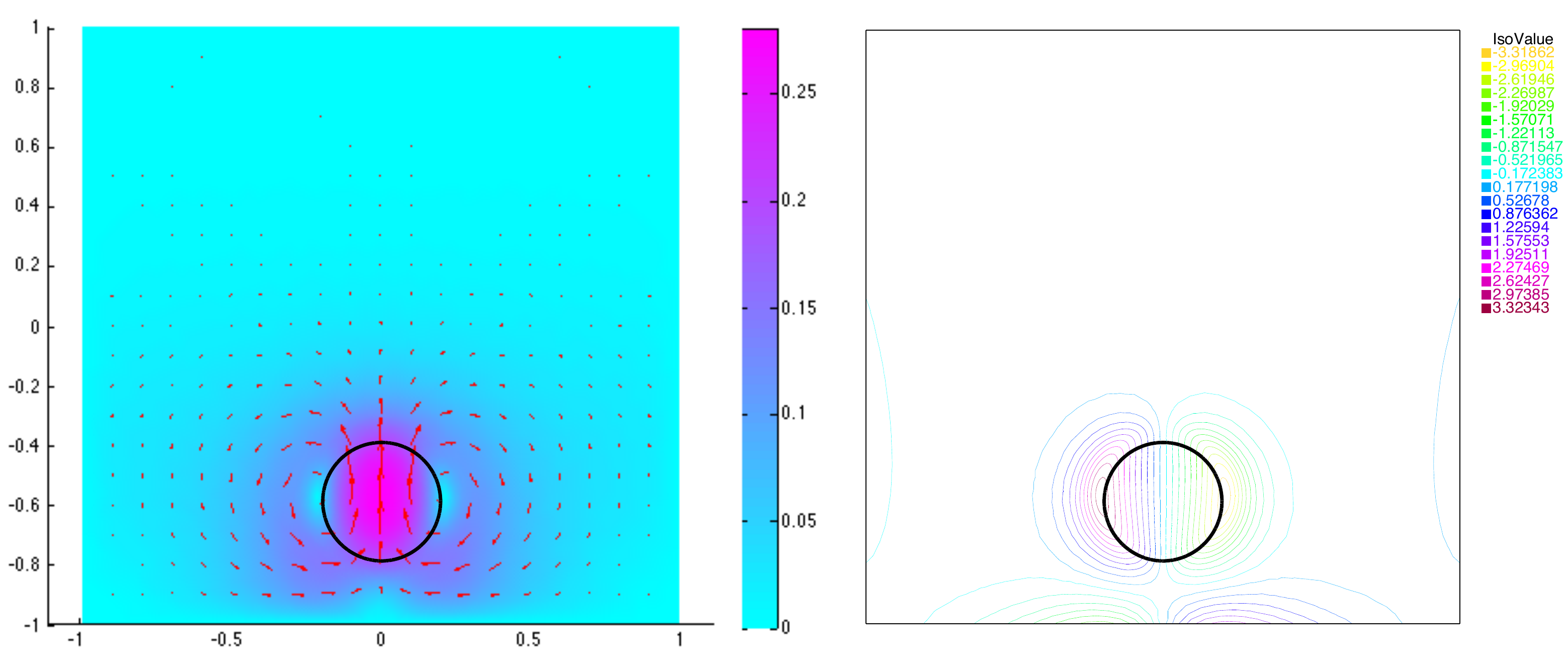}
                \caption*{t=0}
        \end{subfigure}%
\hspace{0mm}
        \begin{subfigure}[h]{0.5\textwidth}
                \includegraphics[width=\textwidth]{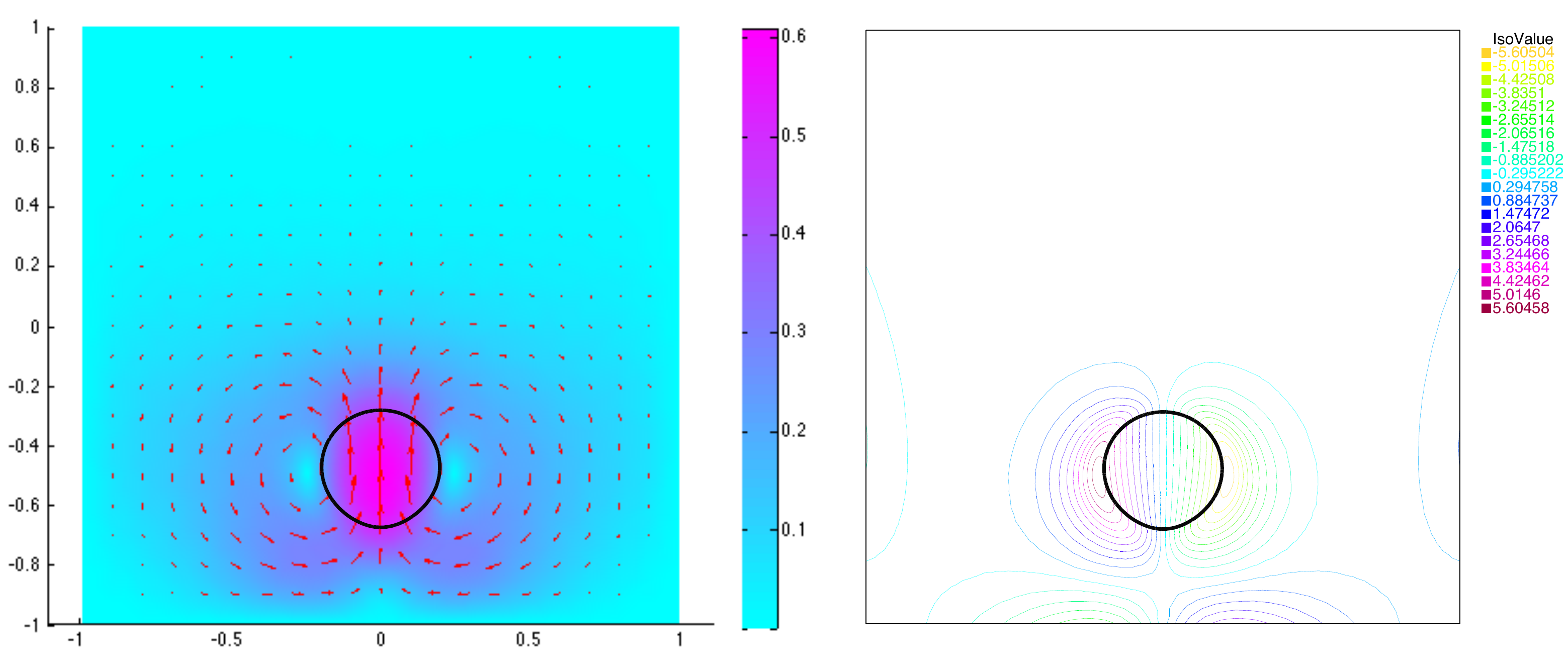}
                \caption*{t=0.4}
        \end{subfigure}%
\hspace{0mm}
        \begin{subfigure}[h]{0.5\textwidth}
               \includegraphics[width=\textwidth]{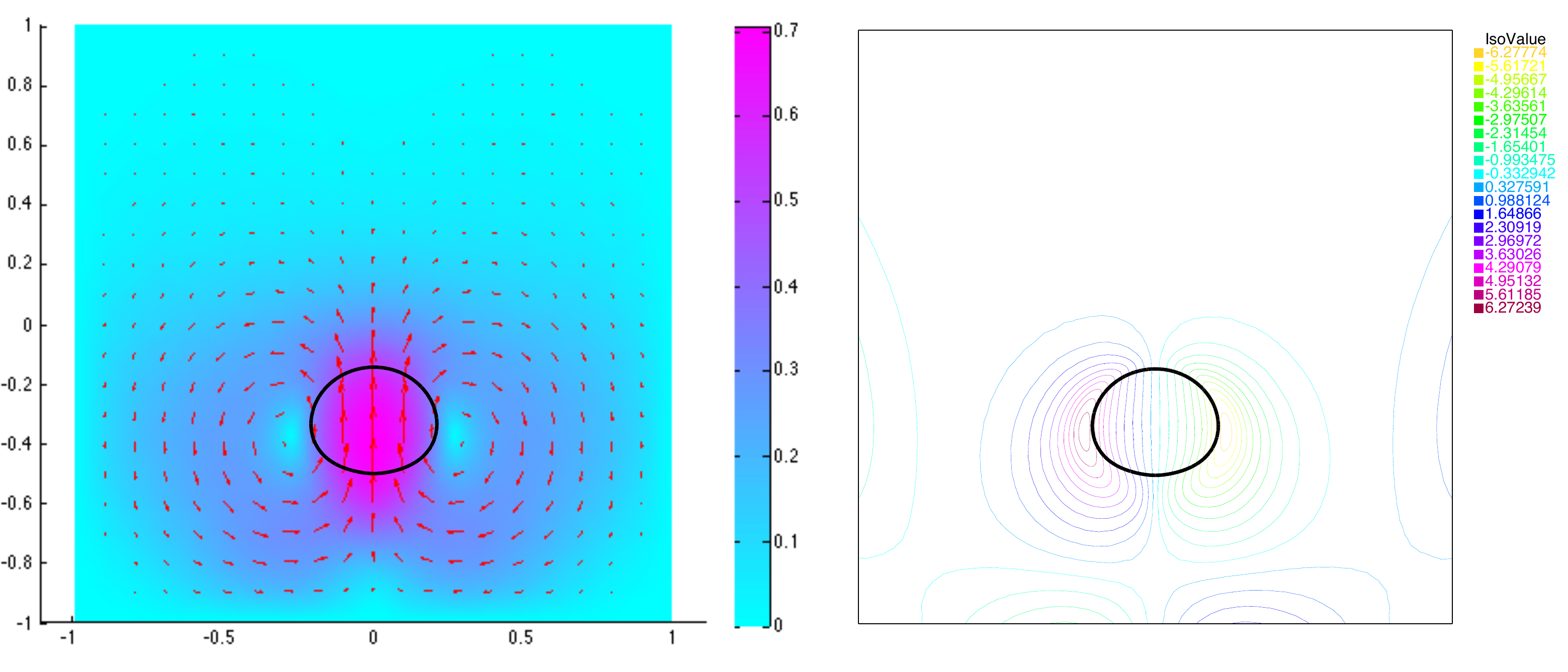}
                \caption*{t=0.7}
        \end{subfigure}%
\hspace{0mm}
        \begin{subfigure}[h]{0.5\textwidth}
                \includegraphics[width=\textwidth]{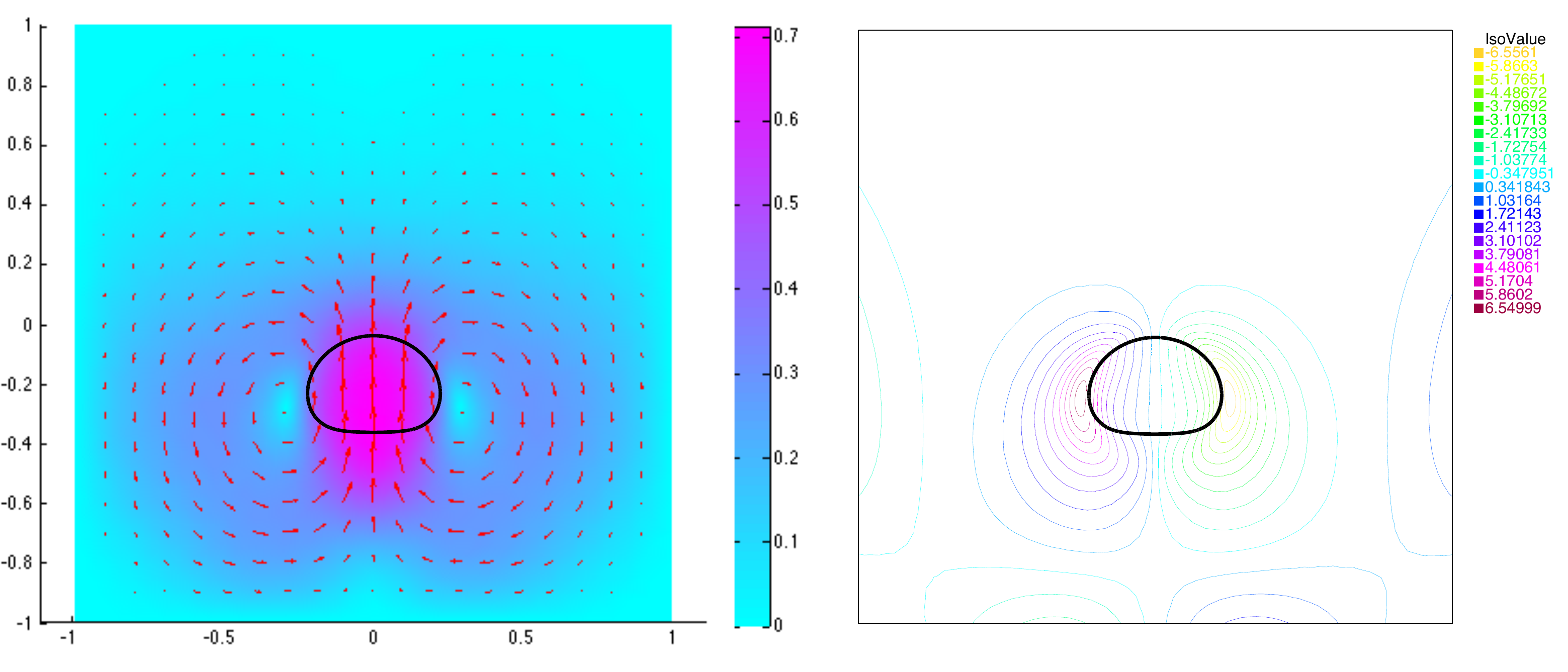}
                \caption*{t=0.9}
        \end{subfigure}%
\hspace{0mm}
        \begin{subfigure}[h]{0.5\textwidth}
               \includegraphics[width=\textwidth]{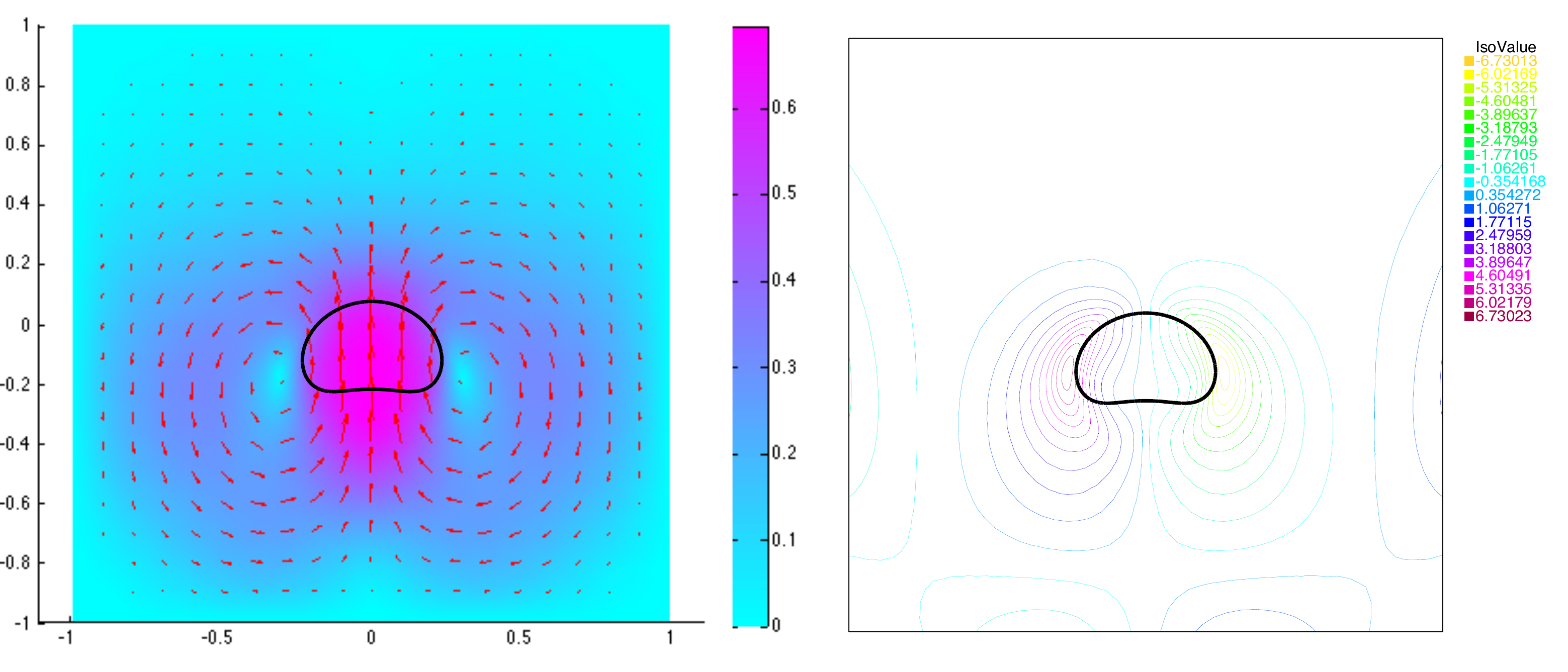}
                \caption*{t=1.1}
        \end{subfigure}%
\hspace{0mm}
        \begin{subfigure}[h]{0.5\textwidth}
                \includegraphics[width=\textwidth]{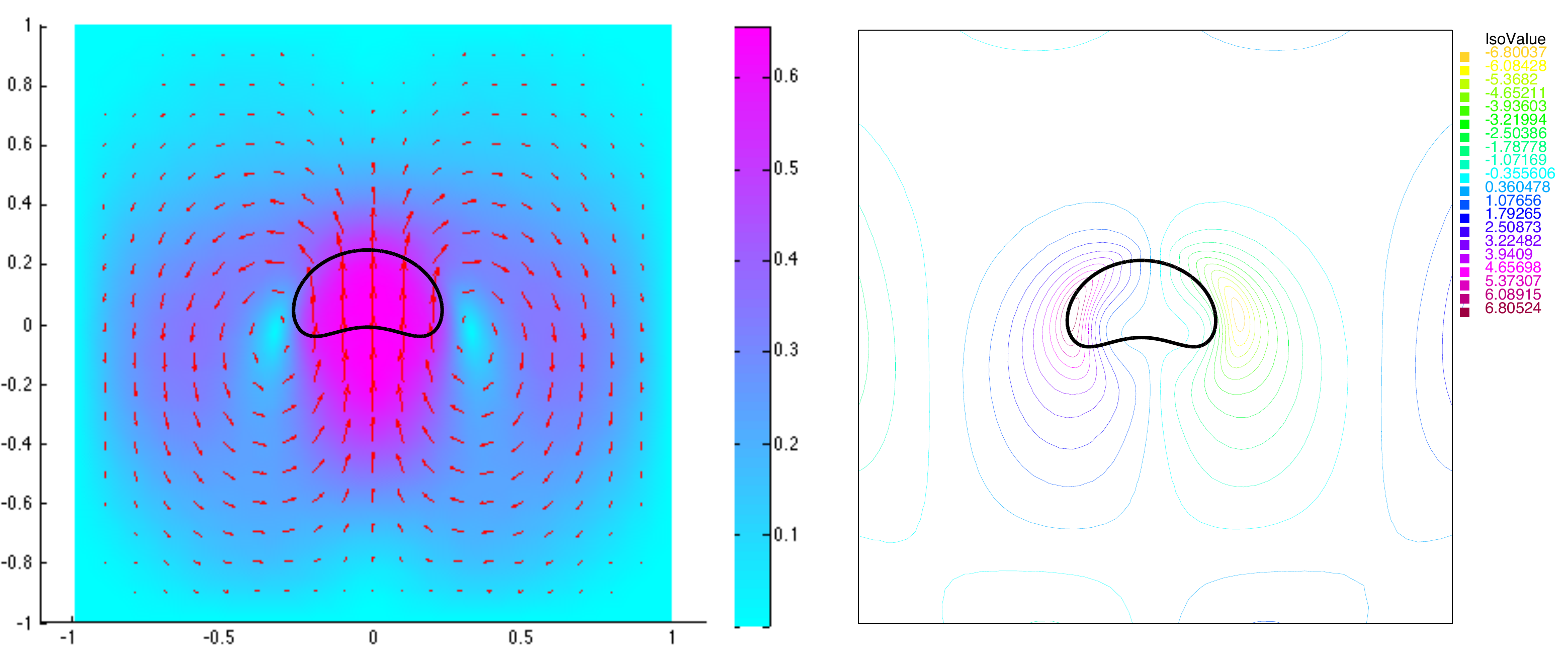}
                \caption*{t=1.4}
        \end{subfigure}%
\hspace{0mm}
        \begin{subfigure}[h]{0.5\textwidth}
               \includegraphics[width=\textwidth]{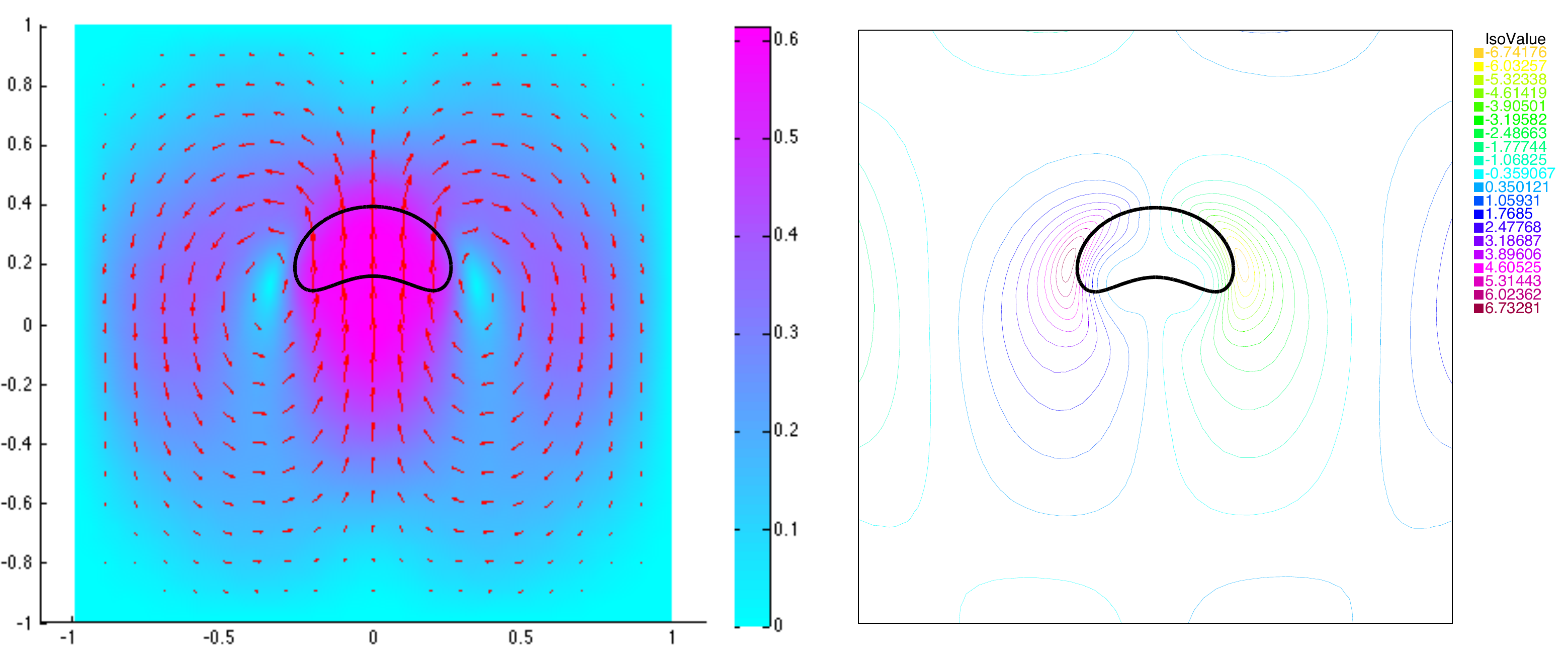}
                \caption*{t=1.7}
        \end{subfigure}%
\hspace{0mm}
        \begin{subfigure}[h]{0.5\textwidth}
                \includegraphics[width=\textwidth]{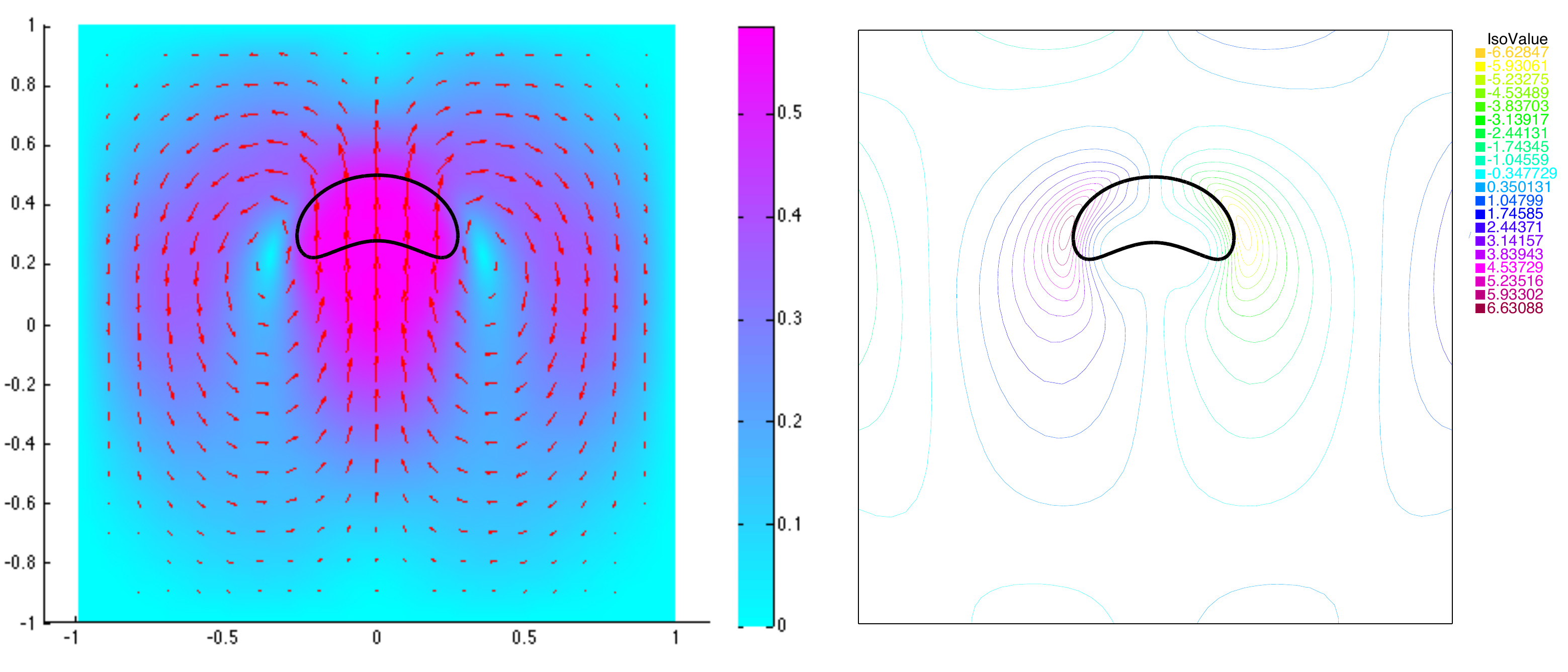}
                \caption*{t=1.9}
        \end{subfigure}%
\hspace{-8mm}
               \caption{The flow field (with arrows representing the velocity vectors and color representing the norm value) and vorticity contours for the rising drop with density ratio  1 : 2, $\rho_1=1, \rho_2=2$,  $\epsilon=0.01$, $C=200\epsilon^{2}$, $M=1/(20\epsilon)$, $Pe=1000/\epsilon$, $1/Fr^2=10$ and $\Delta t = 0.001$.}\label{fig--risdrop1to2--vor}
\end{figure}
\begin{figure}
\hspace{-8mm}
        \begin{subfigure}[h]{0.5\textwidth}
                \includegraphics[width=\textwidth]{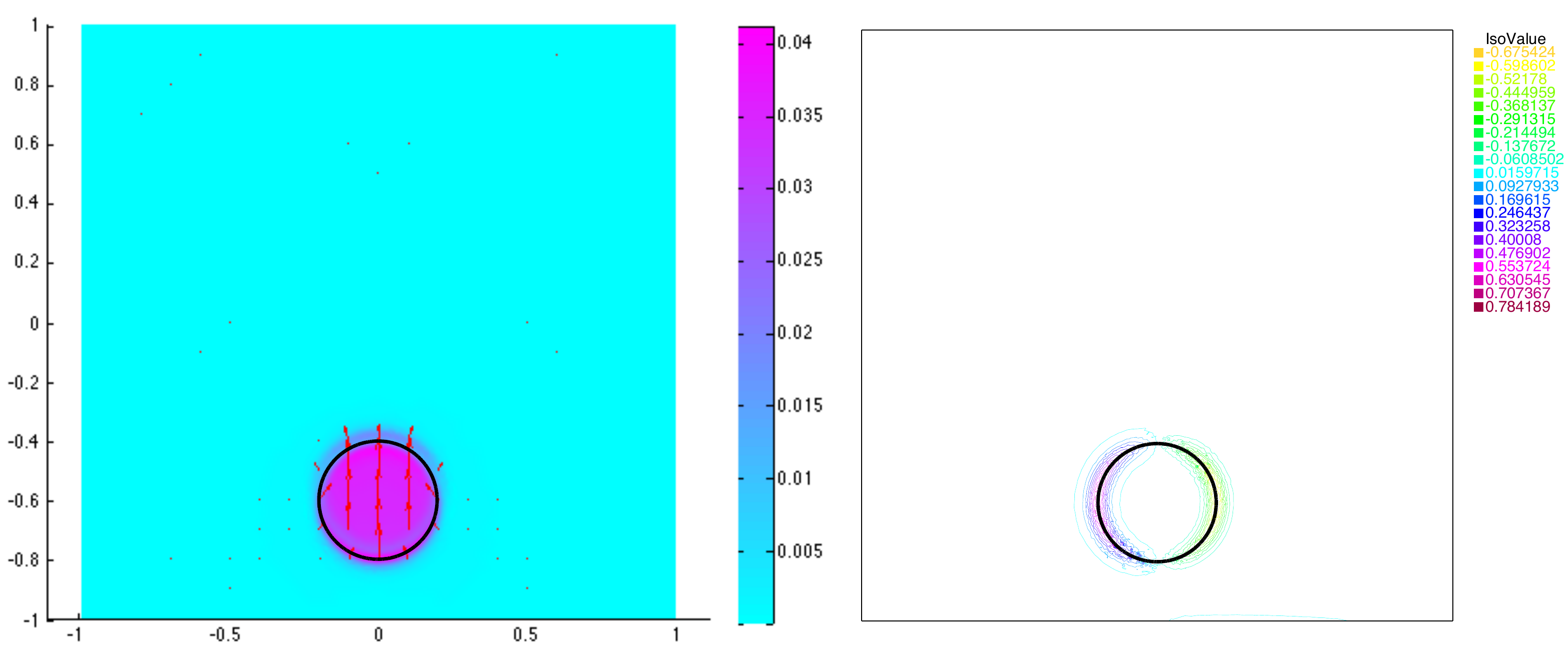}
                \caption*{t=0}
        \end{subfigure}%
\hspace{-8mm}
        \begin{subfigure}[h]{0.5\textwidth}
                \centering
                \includegraphics[width=\textwidth]{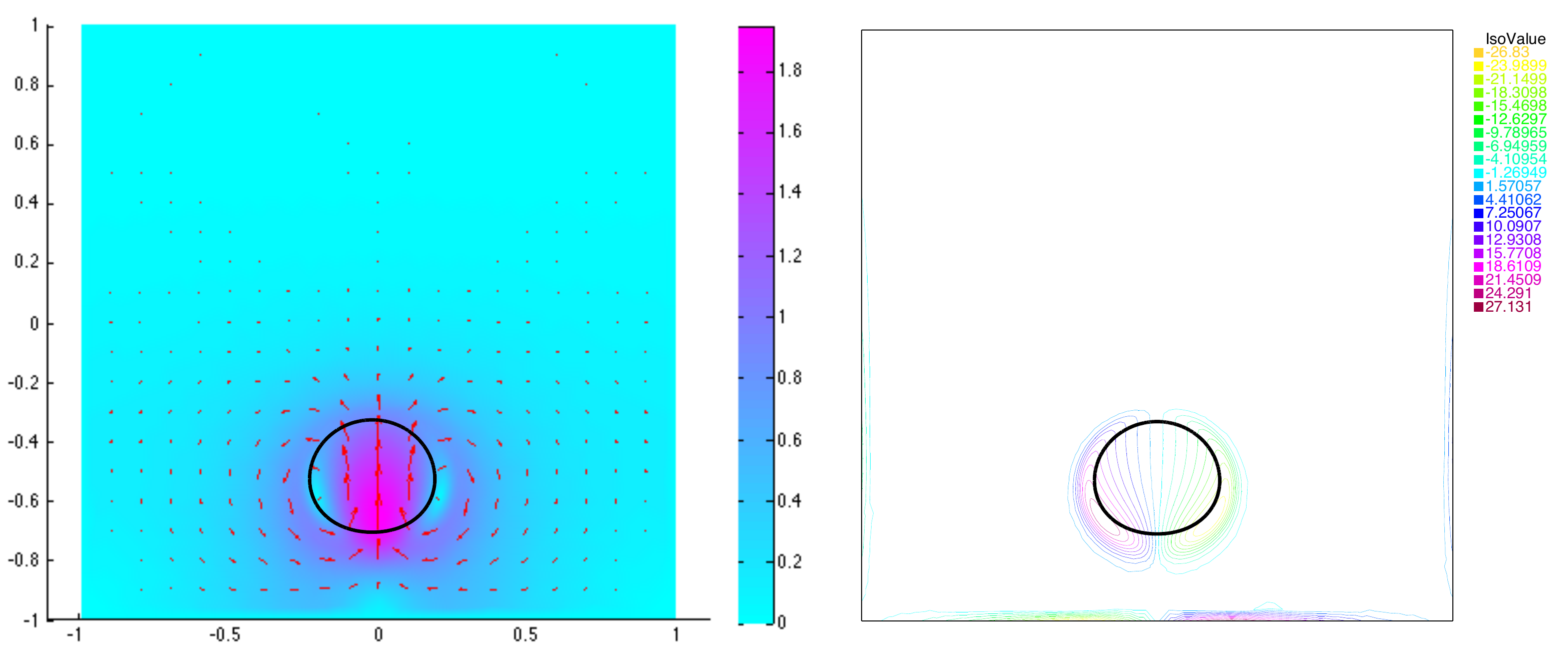}
                \caption*{t=0.125}
        \end{subfigure}%
\hspace{-8mm}
        \begin{subfigure}[h]{0.5\textwidth}
                \centering
                \includegraphics[width=\textwidth]{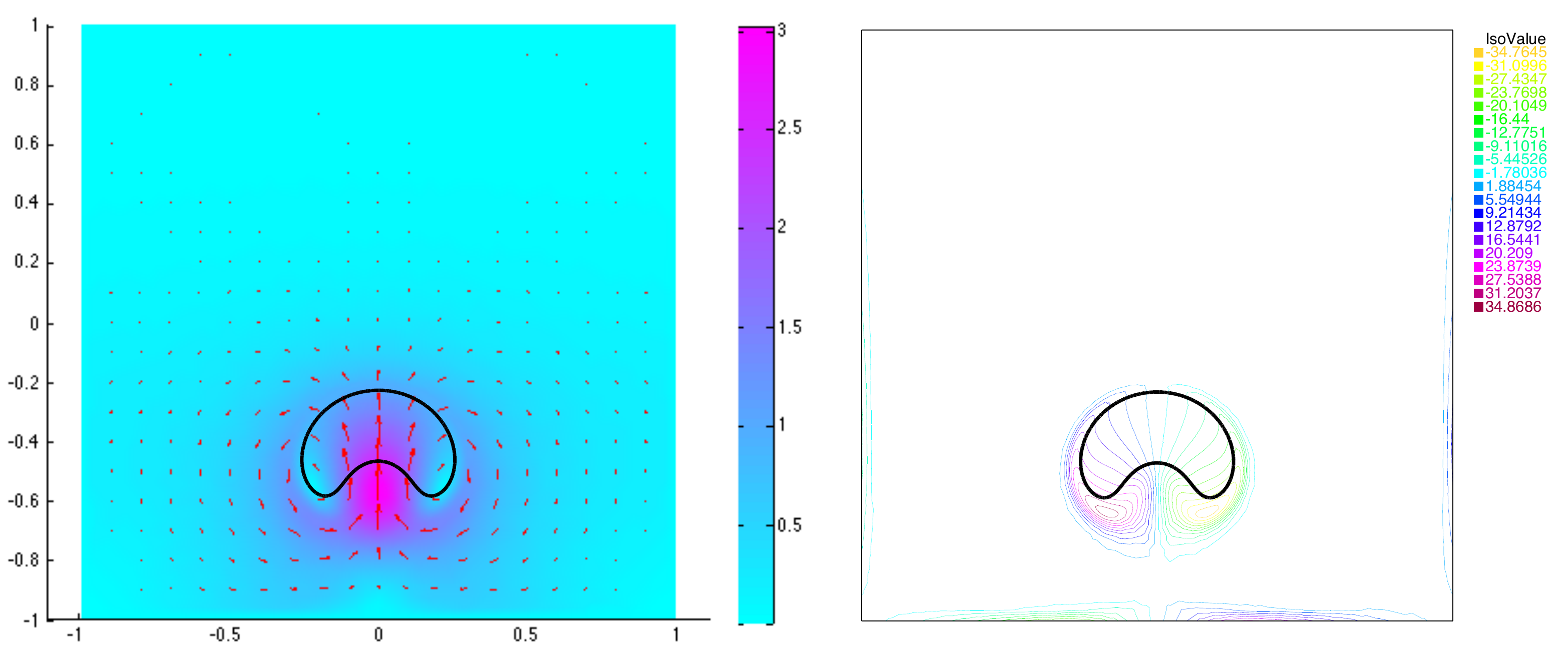}
                \caption*{t=0.225}
        \end{subfigure}%
\hspace{0mm}
        \begin{subfigure}[h]{0.5\textwidth}
                \centering
                \includegraphics[width=\textwidth]{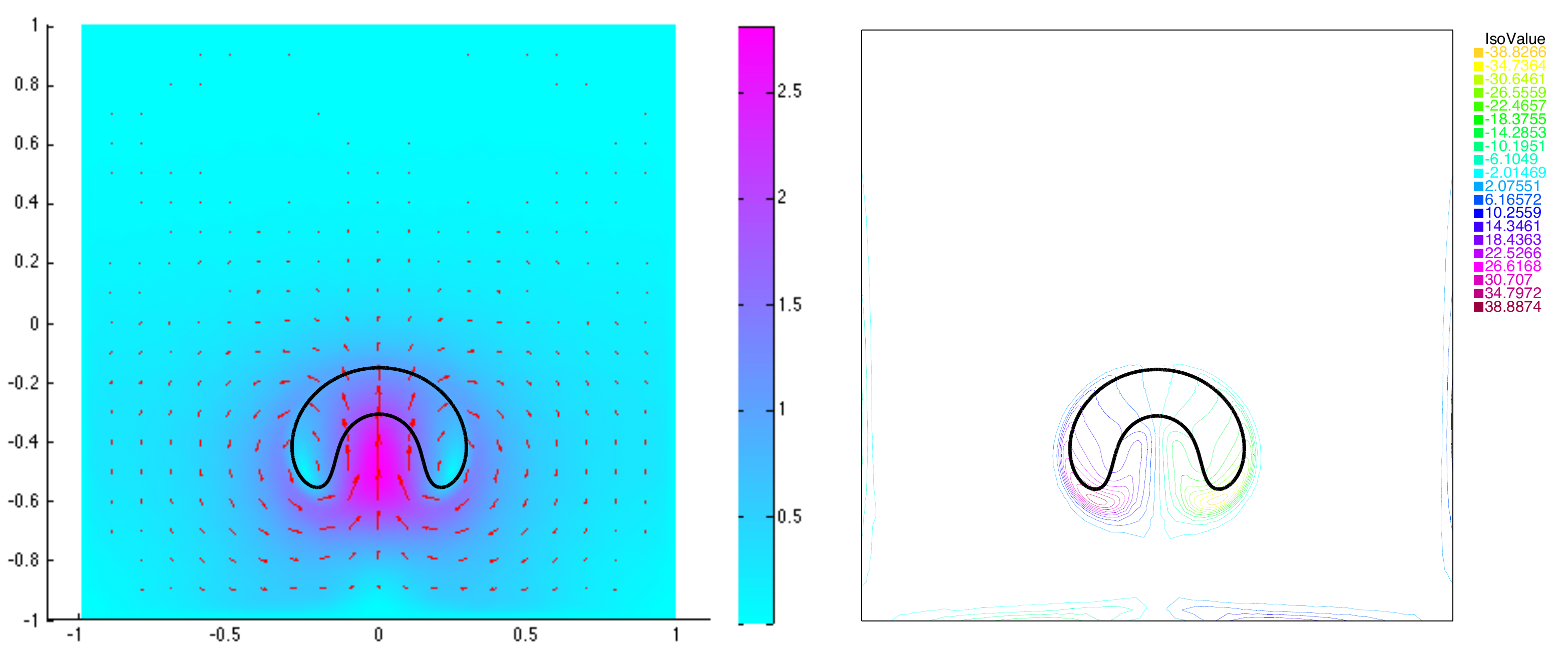}
                \caption*{t=0.3}
        \end{subfigure}%
\hspace{-8mm}
        \begin{subfigure}[h]{0.5\textwidth}
                \centering
                \includegraphics[width=\textwidth]{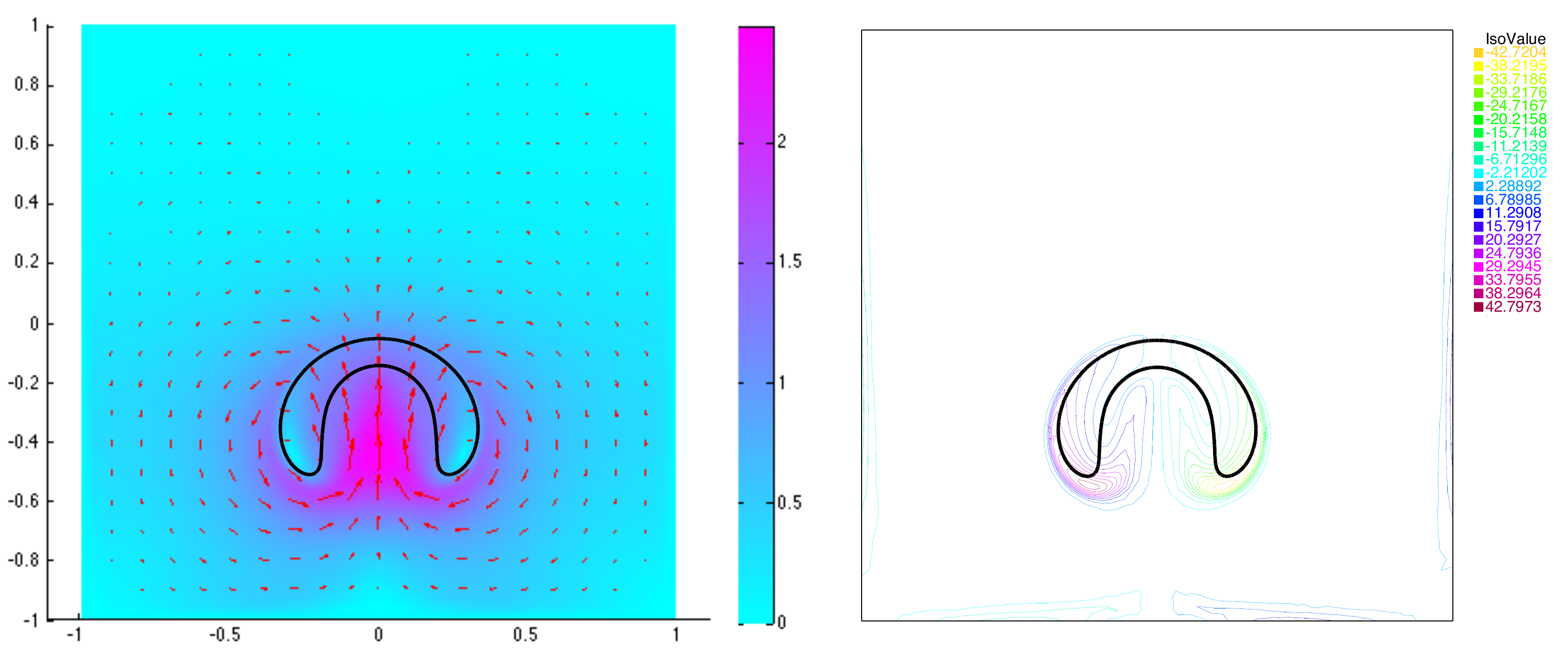}
                \caption*{t=0.4}
        \end{subfigure}%
\hspace{0mm}
        \begin{subfigure}[h]{0.5\textwidth}
                \centering
                \includegraphics[width=\textwidth]{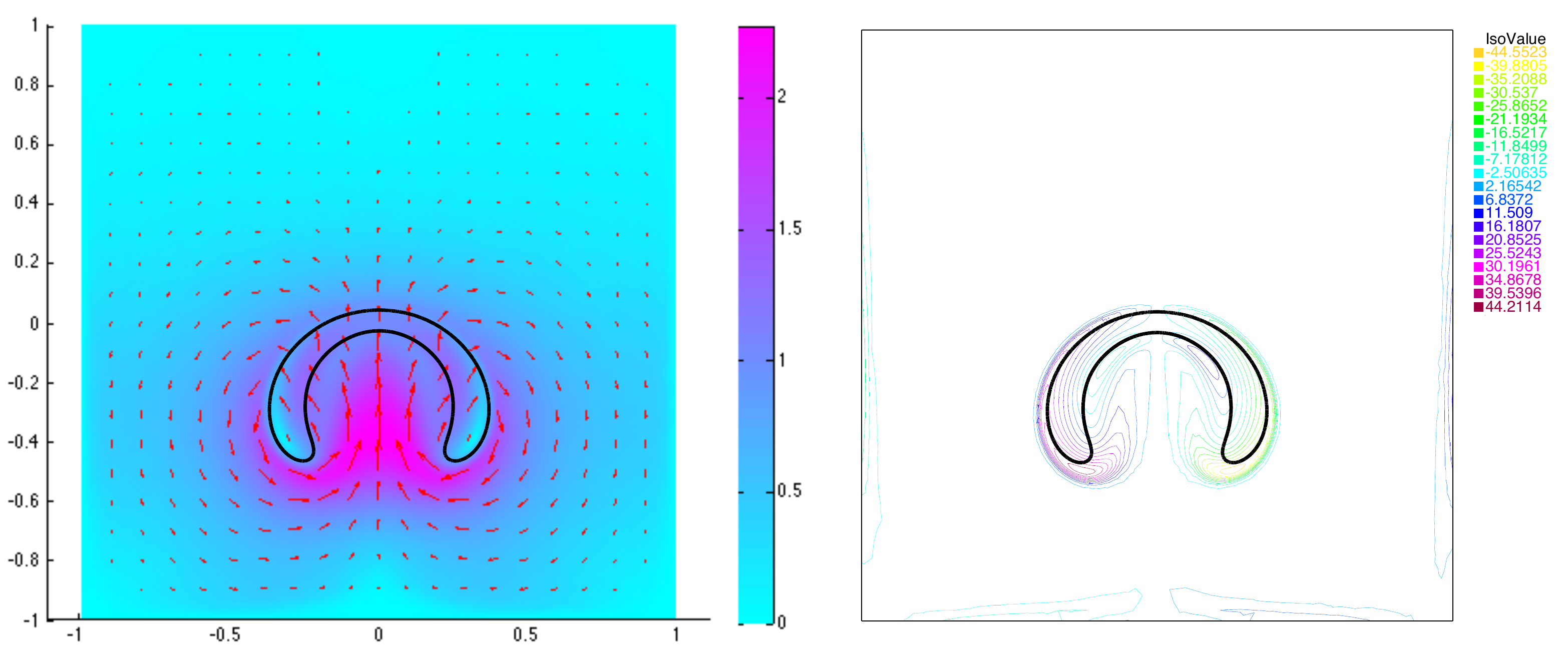}
                \caption*{t=0.5}
        \end{subfigure}%
\hspace{-8mm}
        \begin{subfigure}[h]{0.5\textwidth}
                \centering
                \includegraphics[width=\textwidth]{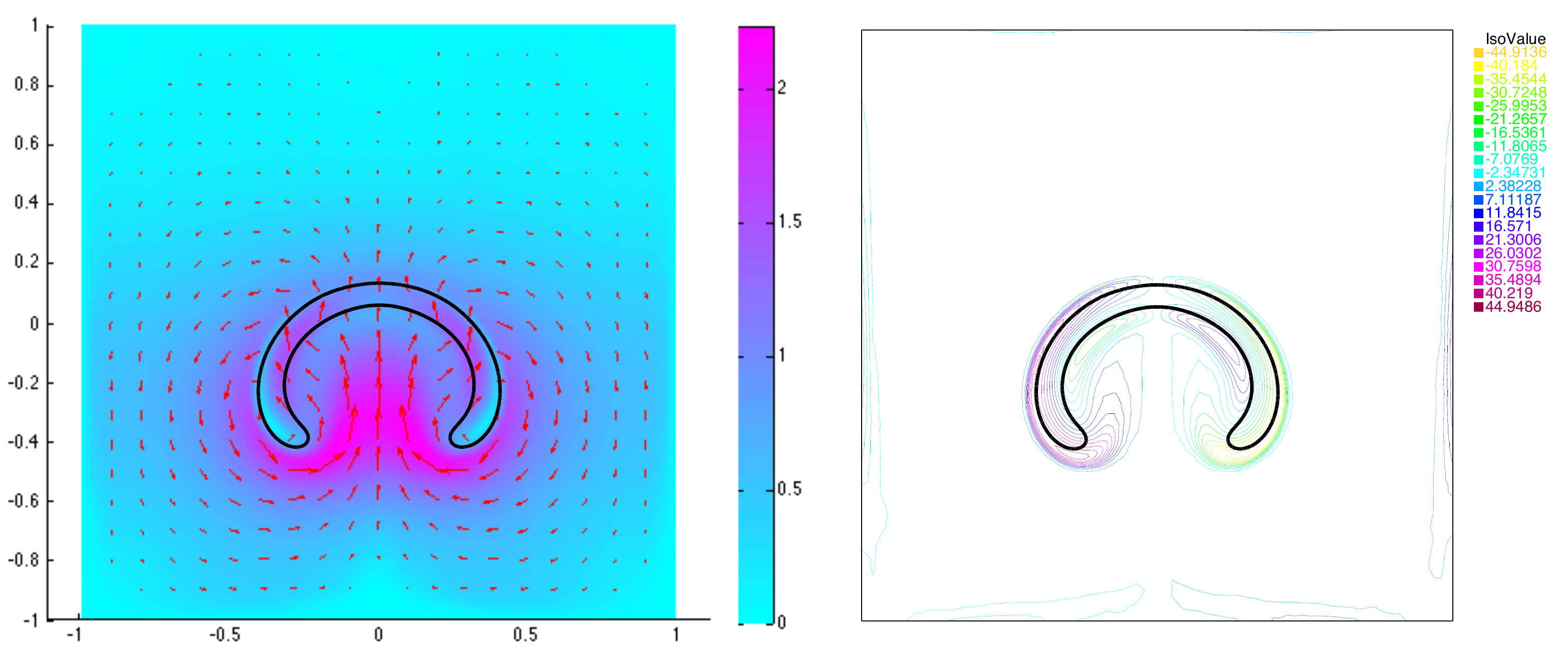}
                \caption*{t=0.6}
        \end{subfigure}%
\hspace{0mm}
        \begin{subfigure}[h]{0.5\textwidth}
                \centering
                \includegraphics[width=\textwidth]{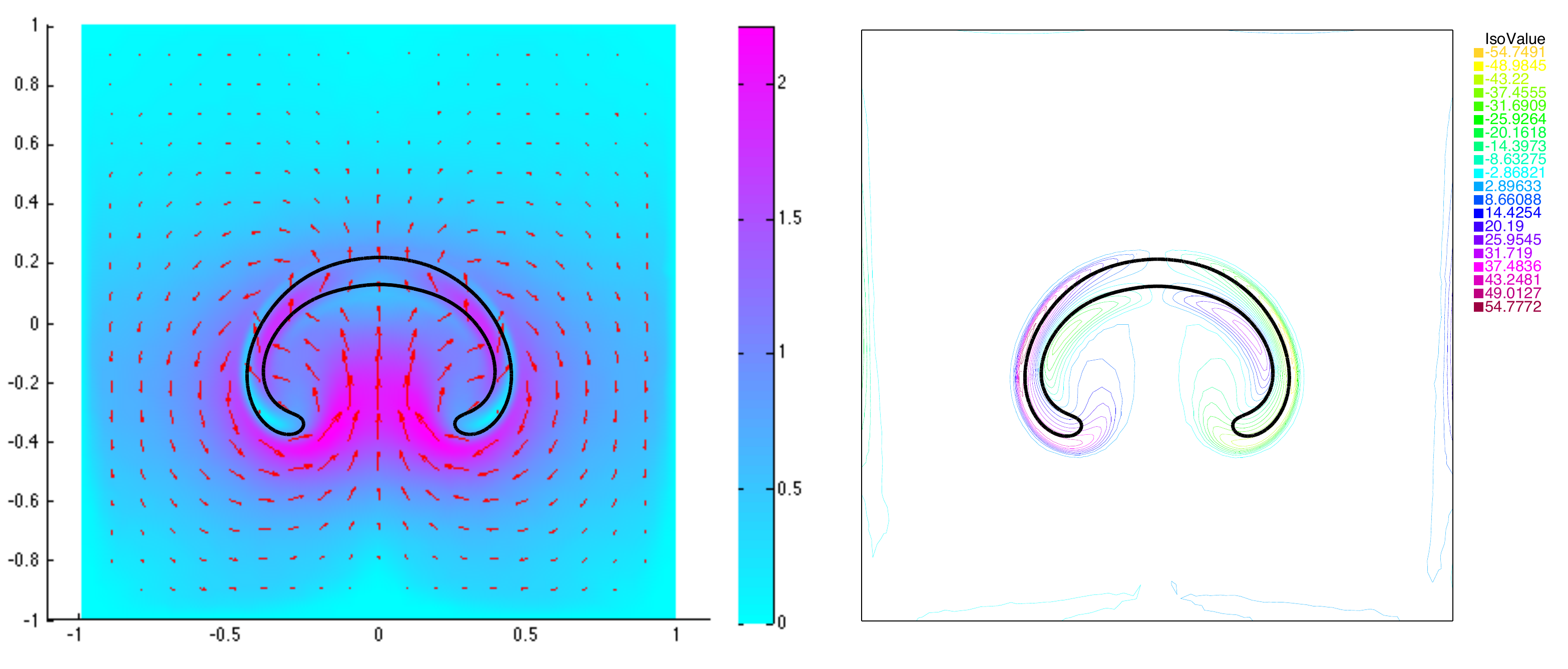}
                \caption*{t=0.7}
        \end{subfigure}%
                               \caption{The flow field (with arrows representing the velocity vectors and color representing the norm value) and vorticity contours for the rising drop with density ratio 1 : 50 from $t=0$ to $t=0.7$. $\rho_1=1, \rho_2=50$, $\epsilon=0.01$, $C=200\epsilon^{2}$, $M=1/(20\epsilon)$, $Pe=1000/\epsilon$, $1/Fr^2=10$ and $\Delta t = 0.00025$.}\label{fig--risdrop1to50--vor1}
\end{figure}
\begin{figure}
\hspace{0mm}
        \begin{subfigure}[h]{0.5\textwidth}
                \centering
                \includegraphics[width=\textwidth]{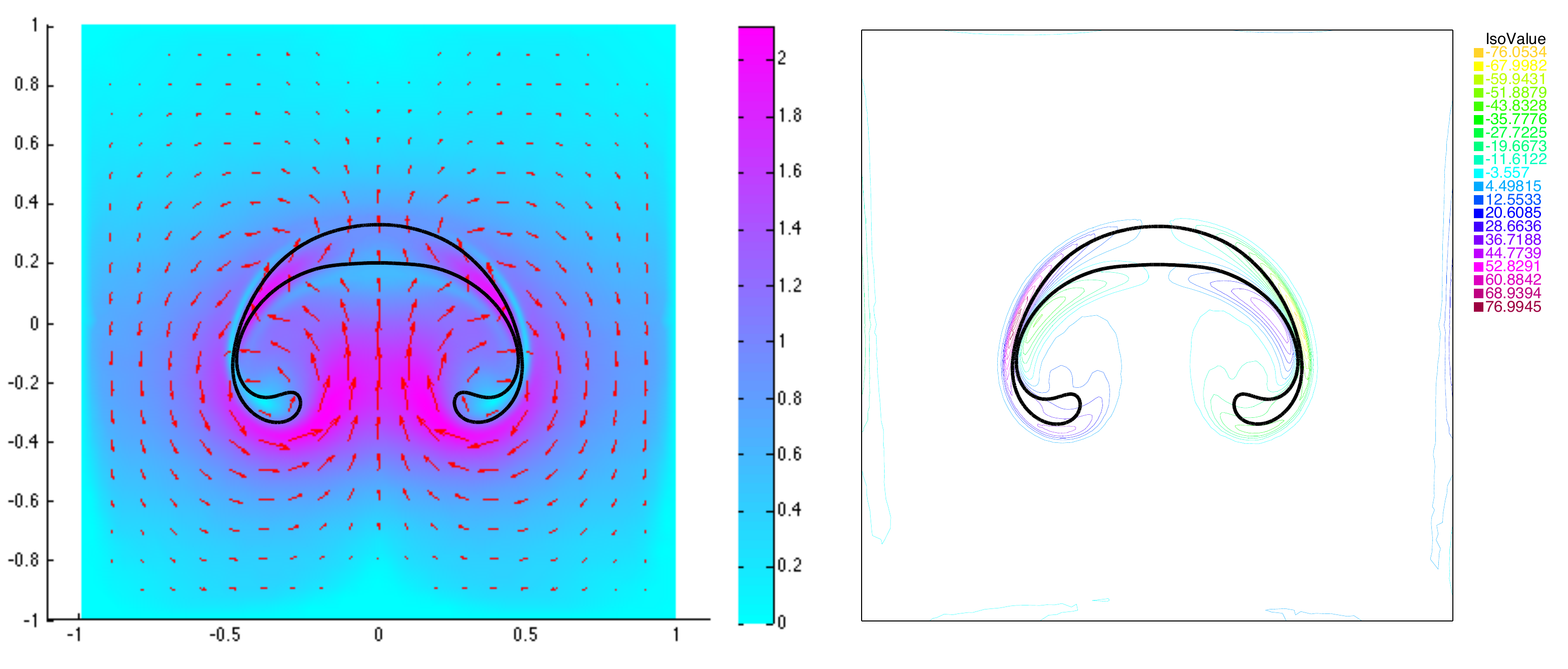}
                \caption*{t=0.825}
        \end{subfigure}%
\hspace{0mm}
        \begin{subfigure}[h]{0.5\textwidth}
                \centering
                \includegraphics[width=\textwidth]{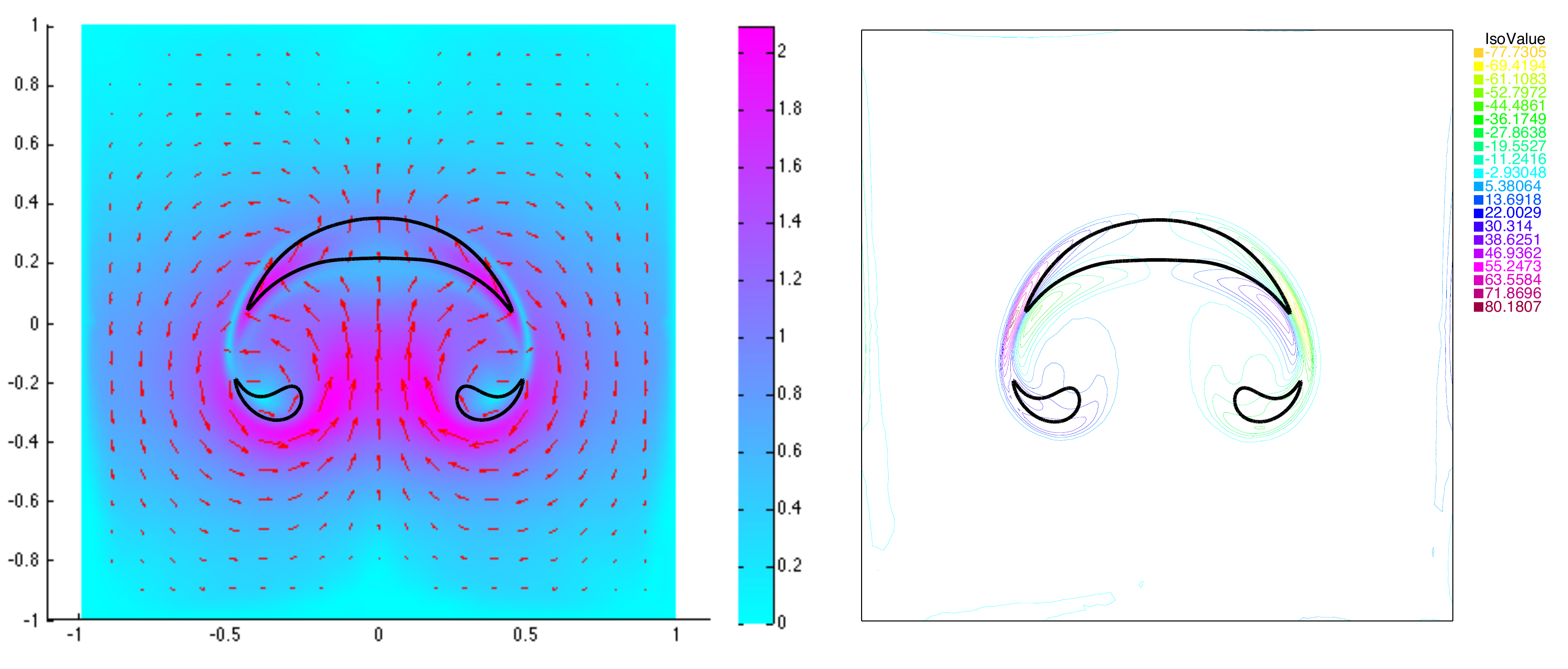}
                \caption*{t=0.85}
        \end{subfigure}%
\hspace{-8mm}
        \begin{subfigure}[h]{0.5\textwidth}
                \centering
                \includegraphics[width=\textwidth]{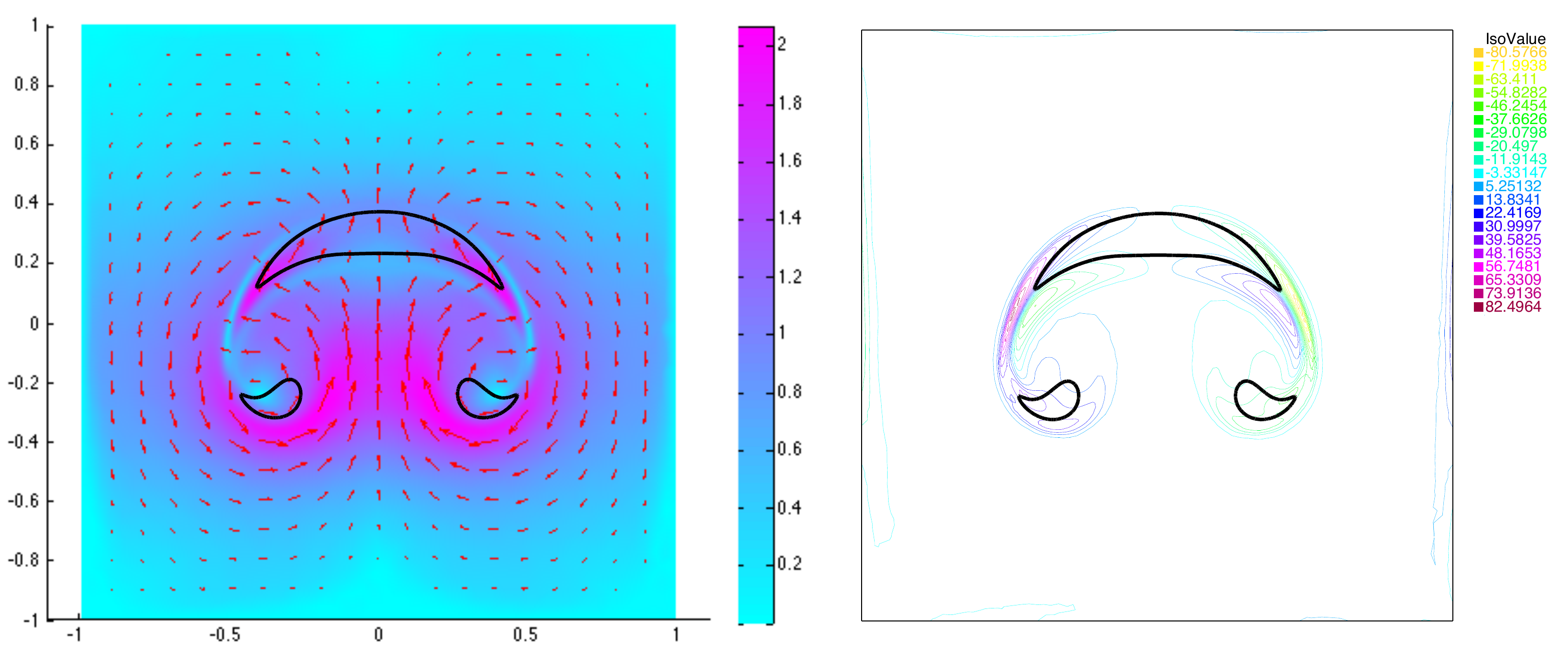}
                \caption*{t=0.875}
        \end{subfigure}%
\hspace{0mm}
        \begin{subfigure}[h]{0.5\textwidth}
                \centering
                \includegraphics[width=\textwidth]{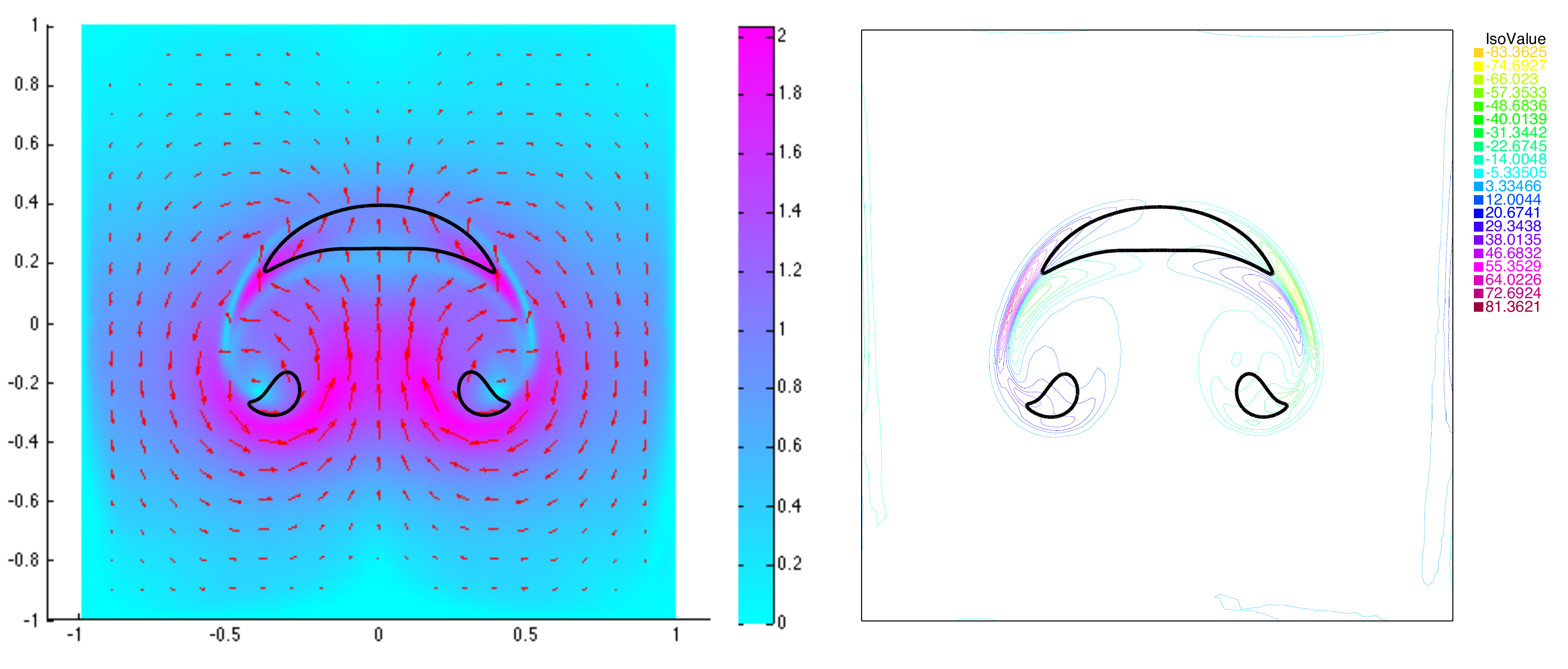}
                \caption*{t=0.915}
        \end{subfigure}%
               \caption{The flow field (with arrows representing the velocity vectors and color representing the norm value) and vorticity contours for the rising drop with density ratio 1 : 50 from $t=0.825$ to $t=0.915$ $\rho_1=1, \rho_2=50$, $\epsilon=0.01$, $C=200\epsilon^{2}$, $M=1/(20\epsilon)$, $Pe=1000/\epsilon$, $1/Fr^2=10$ and $\Delta t = 0.00025$.}\label{fig--risdrop1to50--vor2}
\end{figure}
\begin{figure}
        \begin{subfigure}[h]{0.3\textwidth}
                \includegraphics[width=\textwidth]{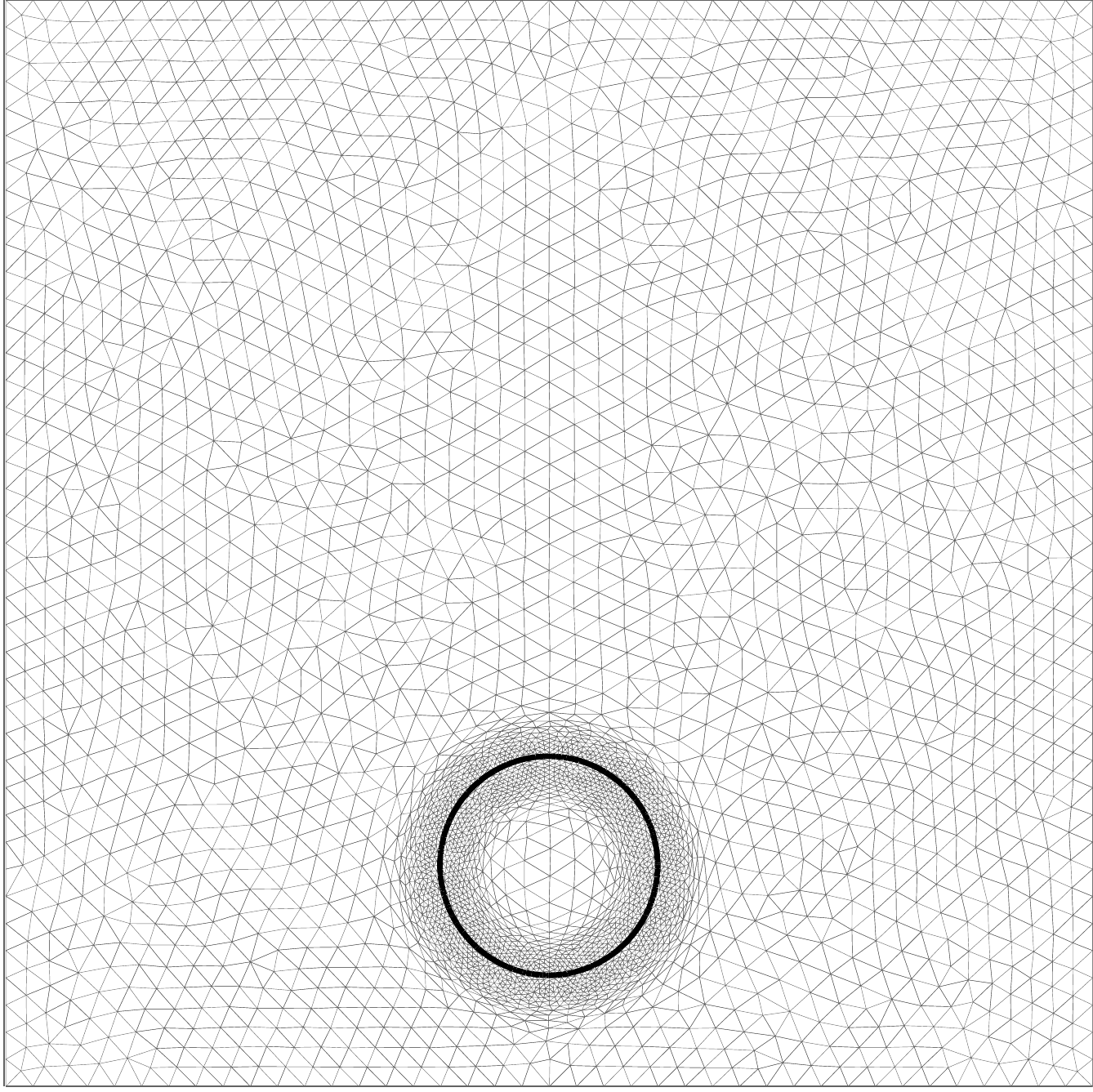}
                \caption*{t=0}
        \end{subfigure}%
\hspace{4mm}
        \begin{subfigure}[h]{0.3\textwidth}
                \includegraphics[width=\textwidth]{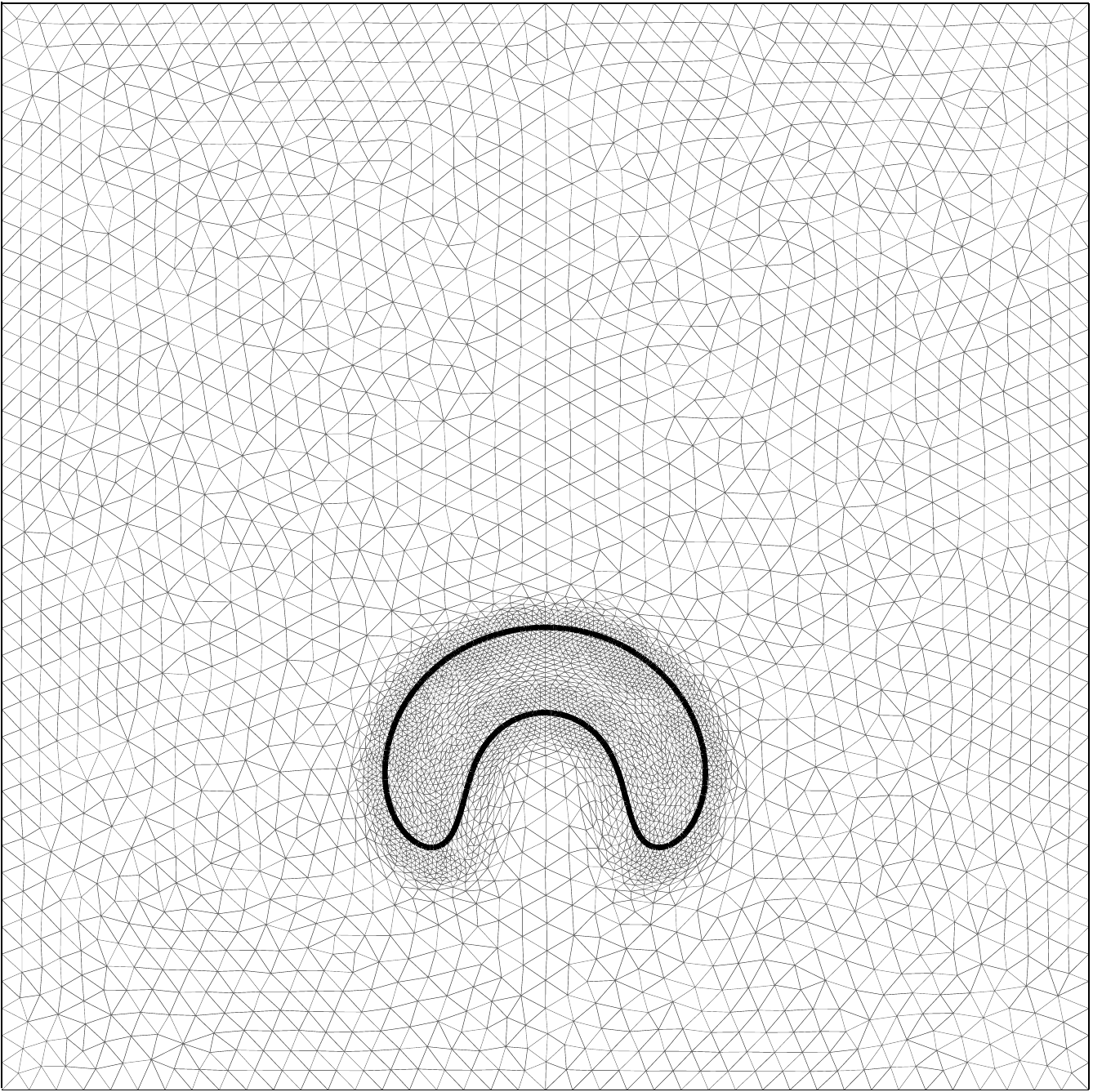}
                \caption*{t=0.3}
        \end{subfigure}%
\hspace{4mm}
        \begin{subfigure}[h]{0.3\textwidth}
                \includegraphics[width=\textwidth]{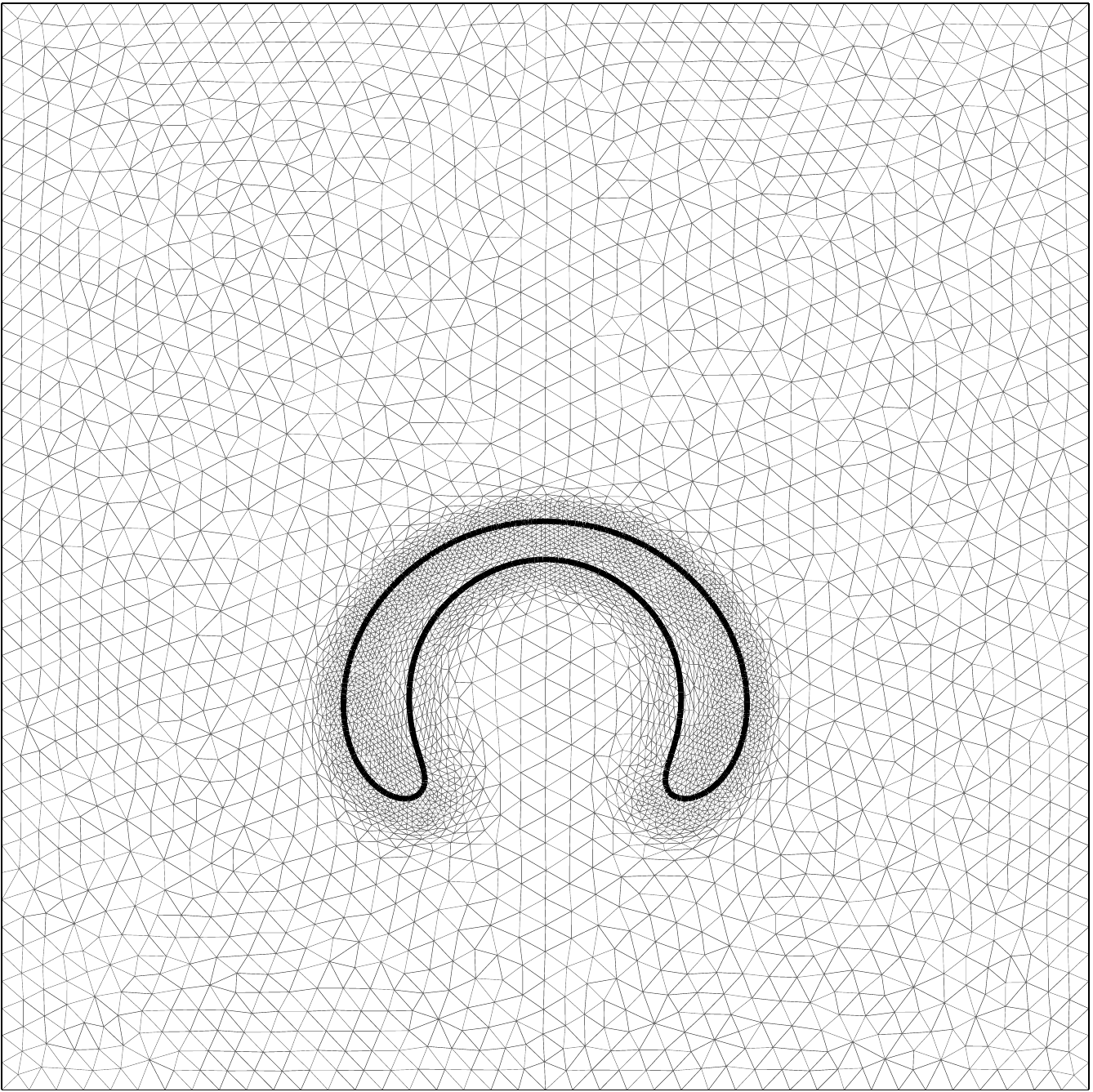}
                \caption*{t=0.5}
        \end{subfigure}%
\hspace{4mm}
\hspace{4mm}
        \begin{subfigure}[h]{0.3\textwidth}
                \includegraphics[width=\textwidth]{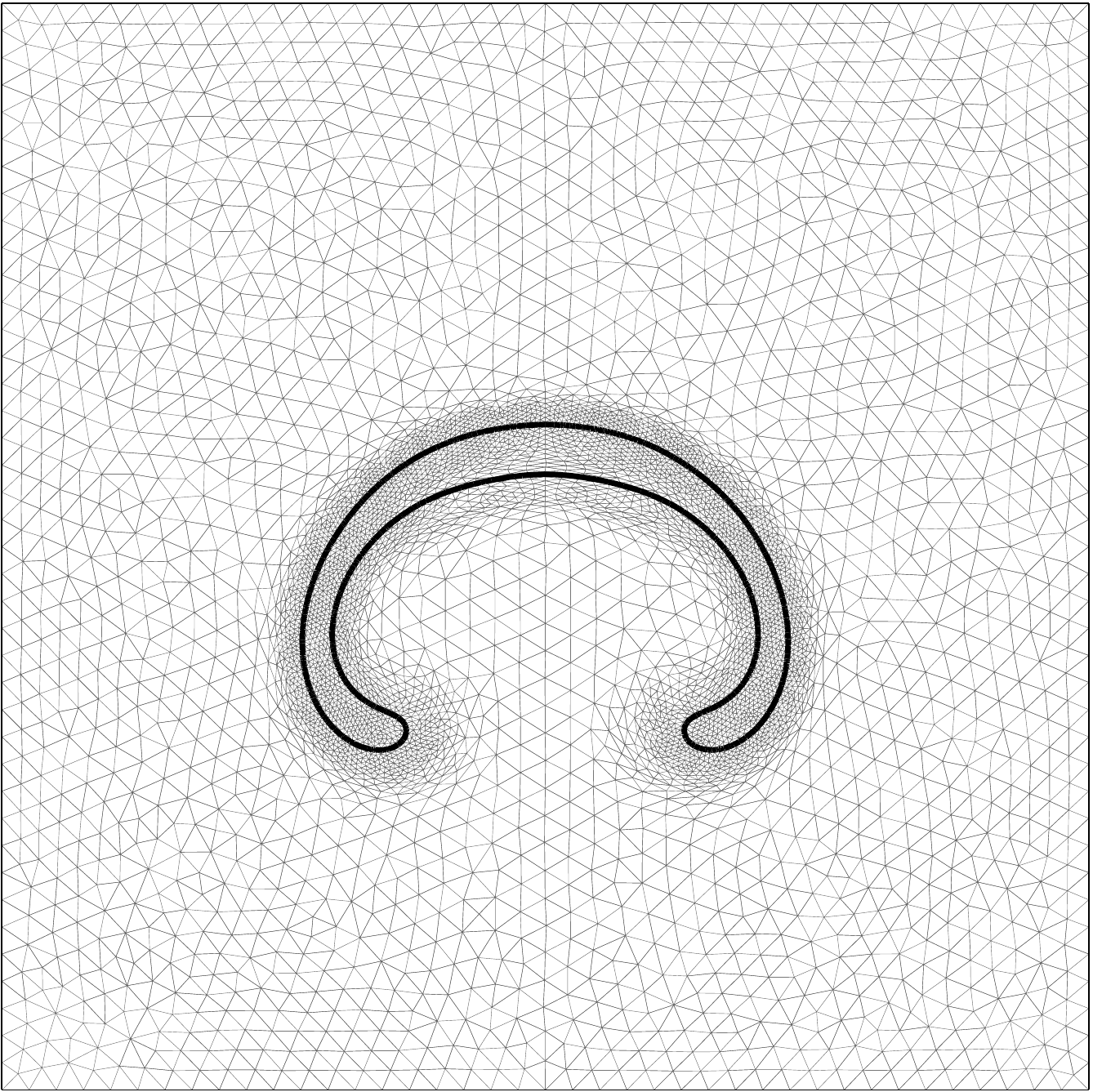}
                \caption*{t=0.7}
        \end{subfigure}%
\hspace{7mm}
        \begin{subfigure}[h]{0.3\textwidth}
                \includegraphics[width=\textwidth]{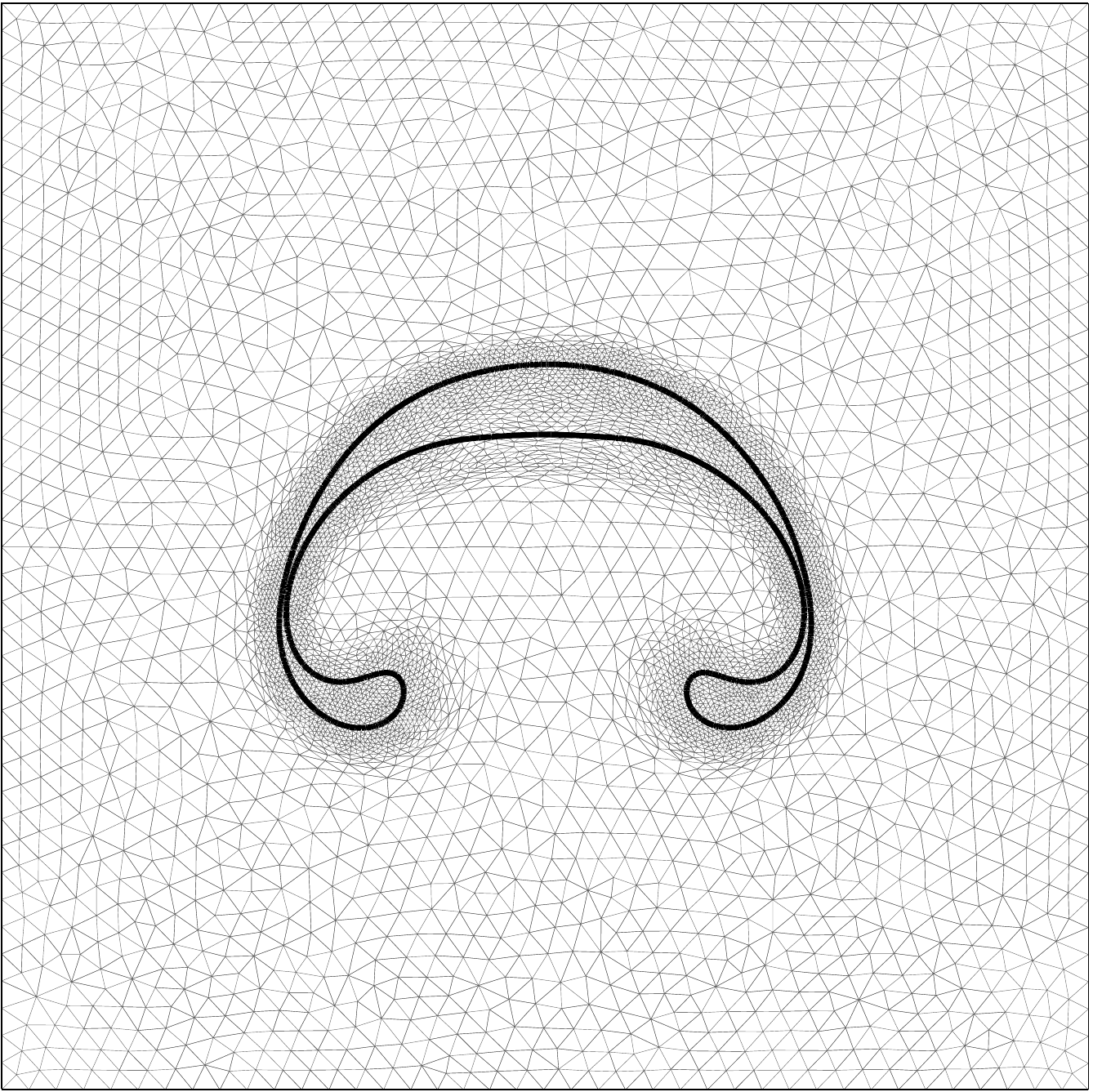}
                \caption*{t=0.825}
        \end{subfigure}%
\hspace{7mm}
        \begin{subfigure}[h]{0.3\textwidth}
                \includegraphics[width=\textwidth]{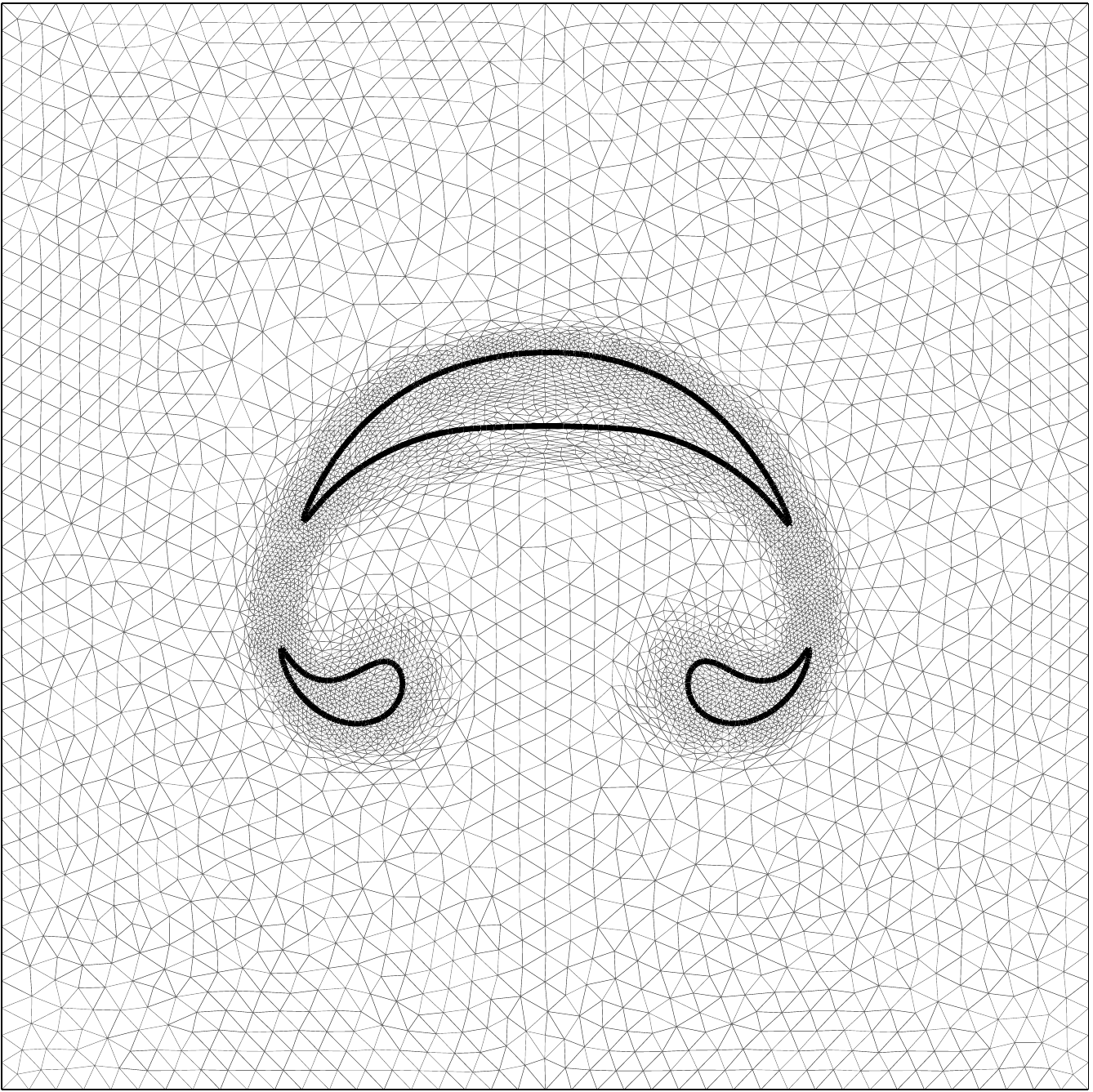}
                \caption*{t=0.85}
        \end{subfigure}%
               \caption{The adaptive for the rising drop with density ratio 1 : 50. $\epsilon=0.01$, $C=200\epsilon^{2}$, $M=1/(20\epsilon)$, $Pe=1000/\epsilon$ and $1/Fr^2=10$.}\label{fig--mesh}
\end{figure}
\newpage
\section{Conclusion}
In this paper, we designed a new numerical method to solve two-phase flow using the quasi-incompressible NSCH system with a variable density. Gravitational forces are incorporated into the system in a thermodynamically consistent way. We reformulated the continuous system that enabled the use of $C^0$ finite elements. We designed a $C^0$ finite element method and a special temporal scheme that ensured that the scheme has an energy law at the discrete level, which is analogous to the one in the continuous level. To our knowledge, this is the first such scheme for the quasi-incompressible NSCH flow system.\\\\
Two examples are computed to test our energy law preserving adaptive numerical scheme for this variable density two-phase model and to see the effects of the density ratio. In the case of kissing drops, increasing the density ratio increases the rate of coalescence because the surface tension also increases. The drop volumes are well-preserved by the numerical scheme. Simulations also confirm that the discrete energy functional is non-increasing, as predicted by our theory. In the example of the rising drop, the effects of the density ratio are much more obvious, where the shape of the drop deforms faster with larger density ratio compared to the case where the density ratio is smaller. Moreover, in the case of the density ratio 1 : 50, the pinch-off event is smoothly captured by the numerical scheme. In the NSCH model, the velocity is not solenoidal near interfaces because fluids of different densities may mix. Our simulations capture this feature. Namely, the numerical results reveal that away from interfaces the fluid is incompressible, while near interfaces waves of expansion and contraction are observed. Increasing the density ratio results in narrower waves with larger magnitudes.\\\\
In future work, we will perform extensive studies of the two-phase problems with more complicated interface dynamics, e.g., the moving contact line problems, where the fluid-fluid interface interacts with solid wall, and the dynamics of interface with Marangoni effects, where surface tension gradients are induced by inhomogeneous temperature distributions or surfactants that can be absorbed at the liquid/gas or liquid/liquid interfaces. Our algorithm may also be implemented using super computers so as to possibly simulate an air bubble in water.
\section*{Acknowledgement}
The authors would like to express their gratitude to Dr. Xiliang Lu for a few early discussions regarding the quasi-incompressible NSCH system. Z. G. was partially supported by the Chinese Scholarship Council for studying at the University of Dundee. P. L. was partially supported by the Fundamental Research Funds for Central Universities (No.06108038 and No.06108137). J. L. acknowledges partial support from the National Science Foundation, Division of Mathematical Sciences, and the National Institutes of Health through grant P50GM76516 for a Center of Excellence in Systems Biology at the University of California, Irvine. 
\bibliographystyle{plain}
\bibliography{ReferenceforPhasefield}
\end{document}